\documentclass[oldversion]{aa}
\usepackage{natbib}
\bibpunct{(}{)}{;}{a}{}{,}
\usepackage{graphicx}
\usepackage{txfonts}
\usepackage{rotating}

\newcommand{\msunyr}{\ensuremath{\mathit{M}_{\odot}{\rm yr}^{-1}}}   % msun/yr
\newcommand{\kms}{\ensuremath{{\rm km\,s^{-1}}}}                   % $\kms$ec
\newcommand{\msun}{\ensuremath{\mathit{M}_{\odot}}}   % msun
\newcommand{\lsun}{\ensuremath{\mathit{L}_{\odot}}}                  % solar luminosity
\newcommand{\rsun}{\ensuremath{\mathit{R}_{\odot}}}                  % solar radius
\newcommand{\mdot}{\ensuremath{\dot{M}}}                             % mass loss rate
\newcommand{\teff}{\ensuremath{\mathit{T}_{\rm eff}}}                % effectieve temperatuur
\newcommand{\vblack}{\ensuremath{v_{\rm black}}}                         %rotational velocity
\newcommand{\vedge}{\ensuremath{v_{\rm edge}}}                         %photospheric velocity

\begin{document}

\title{Detection of high-velocity material from the wind-wind collision zone of Eta Carinae across the 2009.0 periastron passage\thanks{Based on observations made with ESO Telescopes at the La Silla Paranal Observatory under programme IDs 381.D-0262, 282.D-5043, and 383.D-0240; with the {\it Hubble Space Telescope} Imaging Spectrograph ({\it HST}/STIS) under programs 9420 and 9973; and with the 1.6m telescope of the OPD/LNA (Brazil).}}
\author{J. H. Groh\inst{1}\fnmsep \thanks{\email{jgroh@mpifr.de}}
\and
K.~E. Nielsen\inst{2,3}
\and
A. Damineli\inst{4}
\and
T.~R. Gull\inst{2}
\and
T.~I. Madura\inst{5}
\and
D.~J. Hillier\inst{6}
\and 
M. Teodoro\inst{4}
\and
T. Driebe\inst{1}
\and
G. Weigelt\inst{1}
\and
H. Hartman\inst{8}
\and
F. Kerber\inst{9}
\and 
A.~T. Okazaki\inst{7}
\and
S.~P. Owocki\inst{5}
\and
F. Millour\inst{1}
\and
K. Murakawa\inst{1}
\and
S. Kraus\inst{10}
\and
K.-H. Hofmann\inst{1}
\and
D. Schertl\inst{1}
}
\institute{
%1
Max-Planck-Institut f\"ur Radioastronomie, Auf dem H\"ugel 69, D-53121 Bonn, Germany
%2
\and Astrophysics Science Division, Code 667, NASA Goddard Space Flight Center, Greenbelt, MD 20771, USA 
%3
\and  Department of Physics, IACS, Catholic University of America, Washington DC 20064, USA
%4
\and Instituto de Astronomia, Geof\'{\i}sica e Ci\^encias  Atmosf\'ericas, Universidade de S\~ao Paulo, Rua do Mat\~ao 1226, Cidade Universit\'aria, 05508-090, S\~ao Paulo, SP, Brazil
%5
\and Bartol Research Institute, University of Delaware, Newark, DE 19716, USA
%6
\and Department of Physics and Astronomy, University of Pittsburgh, 3941 O'Hara Street, Pittsburgh, PA 15260, USA
%7
\and Faculty of Engineering, Hokkai-Gakuen University, Toyohira-ku, Sapporo 062-8605, Japan
%8
\and Lund Observatory, Lund University, Box 43, 221 00 Lund, Sweden
%9
\and
ESO, Karl-Schwarzschild-Strasse 2, 85748 Garching, Germany
\and
Department of Astronomy, University of Michigan, 500 Church Street, Ann Arbor, MI 48103, USA
}

\authorrunning{Groh et al. }
\titlerunning{High-velocity material in Eta Car across periastron}

\date{Received  / Accepted }

\abstract{

We report near-infrared spectroscopic observations of the Eta Carinae massive binary system during 2008--2009 using the CRIRES spectrograph mounted on the 8\,m UT\,1 Very Large Telescope (VLT Antu). We detect a strong, broad absorption wing in \ion{He}{i} $\lambda$10833 extending up to $-1900~\kms$ across the 2009.0 spectroscopic event. Analysis of archival {\it Hubble Space Telescope}/Space Telescope Imaging Spectrograph ultraviolet and optical data reveals the presence of a similar high-velocity absorption (up to $-2100~\kms$) in the ultraviolet resonance lines of \ion{Si}{iv} $\lambda\lambda$1394, 1403 across the 2003.5 event. Ultraviolet resonance lines from low-ionization species, such as \ion{Si}{ii} $\lambda\lambda$1527, 1533 and \ion{C}{ii} $\lambda\lambda$1334, 1335, show absorption only up to $-1200~\kms$, indicating that the absorption with velocities $-1200$ to $-2100~\kms$ originates in a region markedly faster and more ionized than the nominal wind of the primary star. Seeing-limited observations obtained at the 1.6\,m OPD/LNA telescope during the last four spectroscopic cycles of Eta Carinae (1989--2009) also show high-velocity absorption in \ion{He}{i} $\lambda$10833 during periastron. Based on the large OPD/LNA dataset, we determine that material with velocities more negative than $-900~\kms$ is present in the phase range $0.976 \leq \phi \leq 1.023$ of the spectroscopic cycle, but absent in spectra taken at $\phi \leq 0.947$ and $\phi \geq 1.049$. Therefore, we constrain the duration of the high-velocity absorption to be 95 to $206~\mathrm{days}$ (or 0.047 to 0.102 in phase). We suggest that the high-velocity absorption component originates from shocked gas in the wind-wind collision zone, at distances of 15 to 45~AU in the line-of-sight to the primary star. With the aid of three-dimensional hydrodynamical simulations of the wind-wind collision zone, we find that the dense high-velocity gas is in the line-of-sight to the primary star only if the binary system is oriented in the sky so that the companion is behind the primary star during periastron, corresponding to a longitude of periastron of $\omega \sim 240\degr-270\degr$. We study a possible tilt of the orbital plane relative to the Homunculus equatorial plane and conclude that our data are broadly consistent with orbital inclinations in the range $i=40\degr-60\degr$.}

\maketitle

\section{\label{intro}Introduction} 
 
Eta Carinae offers a unique opportunity to study the evolution of the most massive stars and their violent, giant outbursts during the Luminous Blue Variable (LBV) phase. Eta Car is located at a distance of $2.3 \pm 0.1$ kpc \citep{walborn73,hillier92,dh97,smith06} in the Trumpler 16 OB cluster in Carina and presents itself as a very luminous central object  ($L_{\star}\geq 5\times 10^6~\lsun$, \citealt{dh97}) enshrouded in massive ejecta of $\sim12-20~\msun$ (the Homunculus nebula; \citealt{smith03b}). 

The optical and near-infrared spectra of Eta Car present low-ionization permitted and forbidden lines mainly of \ion{H}{i}, \ion{Fe}{ii}, \ion{N}{ii}, [\ion{Fe}{ii}], and [\ion{Ni}{ii}] \citep{thackeray53}. Since 1944, high-ionization forbidden lines, such as [\ion{Fe}{iii}], [\ion{Ar}{iii}], and [\ion{Ne}{iii}], appeared in the optical spectrum of Eta Car \citep{gaviola53,humphreys08,feast01}. High-spatial resolution imaging and spectroscopy have shown that the narrow emission component of the low- and high-ionization forbidden lines arise in the ejecta \citep{davidson1995,nielsen05,gull09}, in condensations known as the Weigelt blobs \citep{weigelt86,hofmann88,zethson01,hartman05}. It was noticed that at some epochs the high-ionization forbidden lines become weaker and eventually vanish, and then recover their normal strength after several months \citep{gaviola53, whitelock83, zanella84, damineli96, damineli98, damineli00,damineli08_period,damineli08_multi}. Throughout this paper, we refer to these epochs when the high-ionization lines disappear as a spectroscopic event, and to the time interval between consecutive spectroscopic events as a spectroscopic cycle. 

Extensive monitoring of the optical spectrum of Eta Car showed that the spectroscopic events repeat periodically every 5.54 yr  \citep{damineli96,damineli00,damineli08_period,damineli08_multi}, leading to the suggestion that Eta Car is a binary system \citep{damineli97} consisting of two very massive stars, Eta Car~A (primary) and Eta Car~B (secondary), amounting to at least $110~\msun$ \citep{hillier01}. The spectroscopic events are related to the periastron passage of Eta Car~B, and the binary scenario is supported by numerous multi-wavelength observations from X-ray \citep{corcoran97,pittard98,pc02,ishibashi99,corcoran01,corcoran05,hamaguchi07,henley08}, ultraviolet \citep{smith04,iping05}, optical \citep{vg03,vg06,lajus03,lajus09, lajus10, sd04,nielsen07,damineli08_period,damineli08_multi}, near-infrared \citep{feast01,whitelock04}, and radio wavelengths \citep{duncan03,abraham05a}. Although most orbital parameters of Eta Car are uncertain, the wealth of multi-wavelength works mentioned above supports a high eccentricity ($e\sim0.9$) and an orbital period of $2022.7 \pm 1.3$~d \citep{damineli08_period,lajus10}.

Significant advancement on obtaining the properties of Eta Car~A has been achieved in recent years, confirming that it is an LBV star with a high mass-loss rate $2.5\times10^{-4}$ to $10^{-3}~\msunyr$)  and a wind terminal velocity in the range 500 to 600~$\kms$ \citep{davidson1995,hillier01,pc02,smith03,hillier06}. Several observational works have suggested that the wind of Eta Car~A is latitude-dependent \citep{smith03,vb03,weigelt07}, with the polar wind presenting the higher densities and faster velocities. \citet{smith03}, based on H$\alpha$ absorption profiles obtained with HST/STIS during 1998--2000, found evidence for material with velocities of up to $-1200~\kms$ during most of the spectroscopic cycle but only in the polar wind of Eta Car~A. \citet{smith03} also suggested that the wind of Eta Car~A became roughly spherical across the 1998.0 spectroscopic event, with a terminal velocity around 600~\kms. Extremely massive material moving faster than $3000~\kms$ was discovered by \citet{smith08b} in distant ejecta far from the Homunculus nebula, and is thought to be related to the Giant Eruption and not (at least directly) to Eta Car~B.

Little is known about Eta Car~B, since it has never been observed directly. The role of Eta Car~B in the giant outbursts and on the long-term evolution of Eta Car~A is not yet understood. Earlier analysis of the ionization of the ejecta around Eta Car have suggested an O-type or WR nature for Eta Car~B \citep{verner05,teodoro08}.  \citet{mehner10} showed that a broad range of luminosities ($10^5$ to $10^6~\lsun$) and effective temperatures ($36\,000$ to $41\,000~\mathrm{K}$) of Eta Car~B are able to account for the relatively high ionization stage of the ejecta, adding further uncertainty to the evolutionary status and exact position of Eta Car~B in the HR diagram. 

X-ray observations require that Eta Car~B must have a wind terminal velocity on the order of $3000~\kms$ \citep{pc02,okazaki08,parkin09}. X-ray studies have also shown that a strong and variable wind-wind collision zone is present between Eta Car~A and Eta Car~B \citep{hamaguchi07,henley08}. The wind of Eta Car~B likely influences geometry and ionization of the dense wind of Eta Car~A, since numerical simulations have suggested that Eta Car~B creates a cavity in the wind of Eta Car~A \citep{pc02,okazaki08,parkin09}. The extended outer interacting wind structure has been shown to produce broad ($\sim400~\kms$), time-variable forbidden line emission \citep{gull09}.

Spectroscopic observations of the near-infrared \ion{He}{i} $\lambda$10833\footnote{Vacuum wavelengths and heliocentric velocities are adopted in this paper. The spectra presented here have not been corrected by the systemic velocity of $-8~\kms$ of Eta Car \citep{smith04_vel}.} line in Eta Car have shown evidence for a brief appearance of fast-moving material up to $\sim1500~\kms$ during previous periastron passages \citep{damineli98,gdj07,damineli08_multi}. Ultraviolet observations with the International Ultraviolet Explorer (IUE) obtained during the 1980 spectroscopic event suggested that the resonance lines of \ion{Si}{iv} $\lambda\lambda$1393,1403 show absorptions up to $-1240~\kms$ \citep{viotti89}. X-rays observations of high-ionization silicon and sulfur emission lines detected increased line widths (1000 to 1500~\kms) during the 2003.5 spectroscopic event, suggesting that these lines come from the inner, hotter part of the wind-wind collision zone \citep{henley08} or from jets \citep{behar07}. A clear detection of very high-velocity material ($v>1500~\kms$) coming directly from Eta Car~B or from the wind-wind collision zone has been elusive so far in the ultraviolet, optical, and infrared, in particular because the flux of Eta Car~A is several orders of magnitude larger than Eta Car~B \citep{hillier06}. The relative flux between Eta Car~B and Eta Car~A increases towards the ultraviolet, but the absorptions of the wind of Eta Car~A likely modifies -- and masks -- the UV spectrum of Eta Car~B. 

While previous spectroscopic observations of \ion{He}{i} $\lambda$10833 used only moderate spectral resolving power ($R\simeq7000$) and seeing--limited spatial resolution ($1\farcs5$), we monitored Eta Car across the 2009.0 spectroscopic event at much higher spectral ($R\simeq90\,000$) and spatial ($0\farcs3$) resolutions in the near-infrared, where the strong, unblended \ion{He}{i} $\lambda$10833 line is present. We obtained spectroscopic observations with the highest spectral and spatial resolutions obtained so far for longslit spectroscopy of Eta Car in the near-infrared. We combined our data with multi-wavelength diagnostics from the ultraviolet to the near-infrared. The goal of this paper is to characterize the origin, formation region, physical conditions, and the phase interval when the high-velocity material can be detected in \ion{He}{i} $\lambda$10833, to possibly constrain the orbital parameters of the Eta Car binary system. In particular, we aim at constraining the longitude of periastron $\omega$, given that some previous studies have determined $\omega$ around $240\degr-270\degr$, i.e., Eta Car~B is behind Eta Car~A during periastron (e.g., \citealt{damineli97,pc02,nielsen07,corcoran05,iping05,henley08,okazaki08,parkin09}), while others have found $\omega$ $\sim50\degr-90\degr$, i.e., Eta Car~B is in front of Eta Car~A during periastron (e.g., \citealt{abraham05a,abraham07,abraham09,diego05,diego09,kashi07b,kashi08a,kashi09b}). Sideways orbital orientations (i.e., $\omega=0\degr$ or $180\degr$) have also been suggested in the literature (e.g., \citealt{ishibashi99,smith04}).

This paper is organized as follows. Sect. \ref{obs} describes the near-infrared spectroscopic data obtained at the 8\,m Very Large Telescope (VLT) during 2008--2009, the archival ultraviolet and optical {\it HST}/STIS from 2002--2003, and the archival and new near-infrared observations recorded at the 1.6\,m OPD/LNA telescope. Section \ref{highvel} reports the detection of a strong high-velocity ($-1900~\kms$), broad absorption wing in \ion{He}{i} $\lambda$10833 during the 2009.0 periastron passage. A similar high-velocity absorption component is also seen in the ultraviolet resonance lines in spectra obtained with {\it HST}/STIS during 2003.5 and in the OPD/LNA data. In Sect. \ref{timescale} we derive the timescale for the presence of the high-velocity absorption component. In Sect. \ref{disc} we discuss possible scenarios to explain the presence of high-velocity gas in Eta Car during periastron, and argue that our observations provide direct detection of high-velocity material flowing from the wind-wind collision zone in the Eta Car binary system.

\section{\label{obs} Observations}

The multi-wavelength spectroscopic observations used to investigate the presence of high-velocity gas are summarized in Table \ref{tab1}, \ref{tab2}, and \ref{tab3}. Since we are interested in studying material with velocities well above the terminal speed of the wind of Eta Car~A, in this paper the designation ``high-velocity'' means velocities more negative than $-900~\kms$.

Throughout this paper we use the ephemeris obtained by \citet{damineli08_period}\footnote{The \citet{damineli08_period} definition of phase zero is shifted by $-0.002$ in phase (or $-4$ d) from the date of X-ray minimum: JD(X-ray minimum)=2\,450\,799.8 + $2024(E-10)$ \citep{corcoran05}.} to calculate the phases $\phi$ across a given spectroscopic cycle $E$:  JD(phase zero)=2\,452\,819.8 + $2022.7 (E-11)$. Note that phase zero is defined according to the disappearance of the narrow component of \ion{He}{i} $\lambda6678$ and does not necessarily coincide exactly with the time of periastron passage. However, given the high orbital eccentricity of the Eta Car system, the periastron passage is expected to occur close to the phase zero of the spectroscopic cycle. We adopt the nomenclature described by \citet{gd04} to label the spectroscopic cycles of Eta Car, so that cycle \#1 started after the event observed in 1948 by Gaviola. Therefore, phase zero of the 1992.5, 1998.0, 2003.5, and 2009.0 spectroscopic events corresponds to $\phi=9.0$, $\phi=10.0$, $\phi=11.0$, and $\phi=12.0$, respectively. 

\subsection{VLT/CRIRES high-resolution near-infrared spectroscopy}

Spatially resolved spectra of Eta Car were recorded with the CRyogenic high-resolution InfraRed Echelle Spectrograph ({\it CRIRES}; \citealt{kaufl04}) mounted on the 8-m VLT Unit Telescope 1.  Spectra were obtained in 2008 May 05 ($\phi$=11.875), 2008 Dec 26 ($\phi$=11.991), 2009 Jan 08 
($\phi$=11.998), 2009 Feb 09 ($\phi$=12.014) and 2009 Apr 03 ($\phi$=12.040), and are summarized in Table \ref{tab1}. The air-mass ($<$1.4) and seeing ($<$ 0\farcs8 in the $V$ band) during the observations led to an excellent performance of the multi-conjugate adaptive-optics system 
and to an image quality with $0\farcs23-0\farcs28$ FWHM at 20587 {\AA} and $0\farcs31-0\farcs35$ FWHM at 10833 {\AA} for a standard star. All data were recorded with a 31\arcsec$\times$ 0\farcs2 slit at position angle
PA=325\degr, with a plate scale of 0\farcs085/pixel. To avoid saturation and to enhance the signal-to-noise ratio, 
55 exposures of 1 second were co-added.  Normal and interleaved grating 
settings (see the CRIRES User Manual) were used to obtain the desired spectral coverage and some 
wavelength overlap around \ion{He}{i} $\lambda$10833 and \ion{He}{i} $\lambda$20587. 

The spectra were reduced in the standard way using IRAF and other custom IDL data reduction routines specific for CRIRES developed by one of us (JHG). The spectral resolving power is estimated to be R$\sim$90\,000 using unresolved calibration lamp lines. The spectra were extracted by averaging 3 pixels in the spatial direction, centered on the central source of Eta Car, thus corresponding to a $0\farcs26 \times 0\farcs20$ spatial region. The  spectra were corrected to the heliocentric rest frame and normalized to continuum. The error in the continuum normalization is estimated to be less than 5\%. Using as reference the Th-Ar spectral line database from \citet{kerber08}, an error in the wavelength calibration of $\simeq0.5~\kms$ was achieved. 

\begin{figure*}
\resizebox{1.02\hsize}{!}{\includegraphics{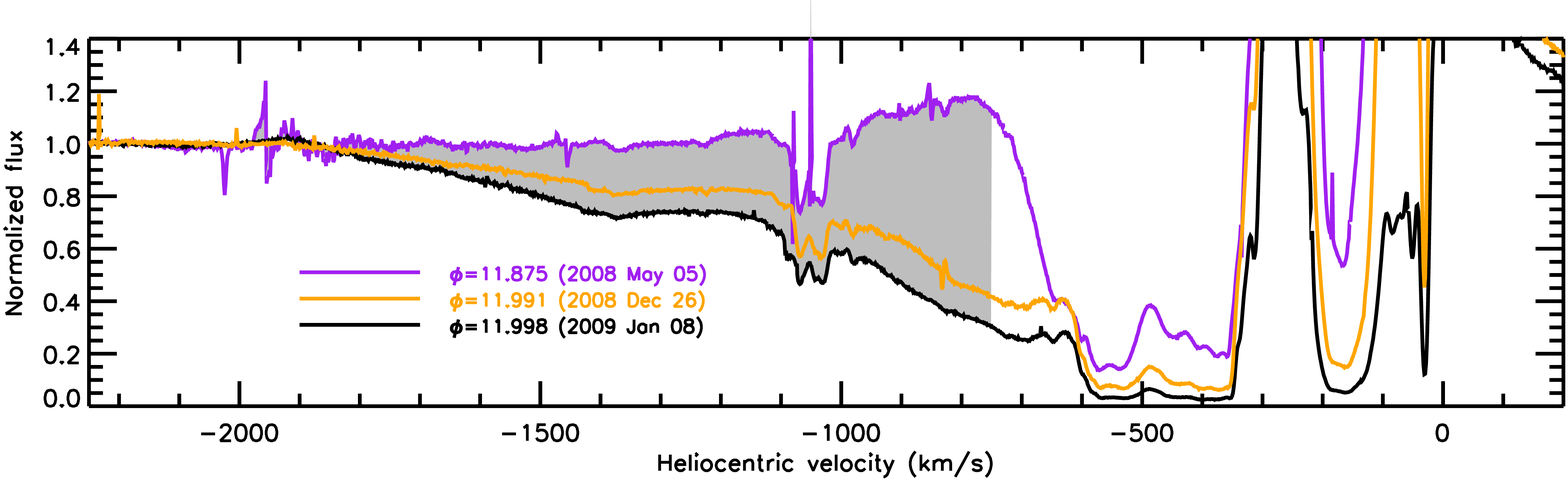}}\\
\resizebox{1.02\hsize}{!}{\includegraphics{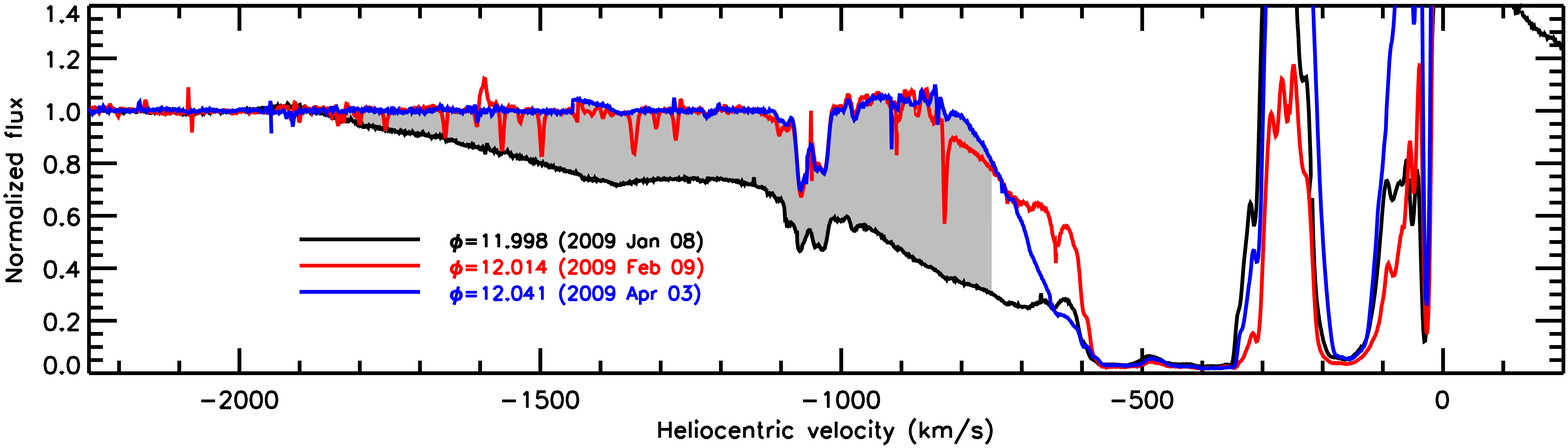}}\\
\caption{\label{fig1}Continuum-normalized \ion{He}{i} $\lambda$10833 CRIRES spectrum from the inner $0\farcs26 \times 0\farcs20$ spatial region around the central source of Eta Car obtained before/during (top panel) and during/after (bottom panel) the 2009.0 spectroscopic event. The grey region corresponds to the excess absorption due to the high-velocity material in Eta Car during the 2009.0 spectroscopic event. The 2009 Jan 08 spectrum is repeated in both panels for clarity. Note that the many narrow absorption features blueward of $\sim-800~\kms$ seen in the Feb 2009 spectrum (red line), except the $-1050~\kms$ feature, are residuals from the removal of telluric lines. The broad emission between $-600$ and $-1500~\kms$ seen in the 2008 May 05 spectrum is due to electron scattering. }
\end{figure*}

\begin{table}
\caption{\label{tab1} Summary of VLT/CRIRES Eta Car spectroscopic observations used in this paper} 
\centering
\begin{tabular}{l c c c}
\hline\hline
Date & $\phi$ & JD & High-velocity absorption\\  
\hline
2008 May 05 & 11.875 & 2454591.53 & No \\
2008 Dec 26 & 11.991 & 2454826.82 & Yes\\
2009 Jan 08 & 11.998 & 2454839.80 & Yes\\
2009 Feb 09 & 12.014 & 2454871.77 & Yes\\
2009 Apr 03 & 12.040 & 2454924.61 & No \\
\hline
\end{tabular}
\end{table} 

\subsection{OPD/LNA medium-resolution near-infrared spectroscopy}

\begin{table}
\centering
\caption{\label{tab2} Summary of OPD/LNA Eta Car spectroscopic observations used in this paper} 
\begin{tabular}{l c c c}
\hline\hline
Date & $\phi$ & JD & High-velocity absorption?\\
\hline
1989 Mar 28 & 8.4263 & 2447614.50 & No \\
1990 Jan 25 & 8.5761 & 2447917.50 & No \\
1990 Jun 16 & 8.6463  & 2448059.50  & No \\
1990 Dec 30 & 8.7442  & 2448257.50  & No\\
1991 Jan 29 & 8.7590  & 2448287.50  & No\\
1991 May 25 & 8.8164  & 2448403.50  & No\\
1992 Mar 20 & 8.9647  & 2448703.50  &Yes \\
1992 May 27 & 8.9976 & 2448769.50 & Yes \\
1992 Jun 02 & 9.0006 & 2448775.50 & Yes \\
1992 Jun 03 & 9.0011 & 2448776.50 & Yes \\
1992 Jul 16 & 9.0228 & 2448820.40 & Yes \\
1997 Jul 22 & 9.9285 & 2450652.39 & No \\
1998 Jan 20 & 10.0181 & 2450833.70 & Yes \\
1998 May 07 & 10.0713 & 2450941.31 & No \\
1998 May 12 & 10.0738 & 2450946.27 & No \\
1998 Jun 14 & 10.0897 & 2450978.53 & No \\
1998 Jul 08 & 10.1020 & 2451003.40 & No \\
1999 Apr 26 & 10.2460 & 2451294.56 & No \\
1999 Jul 02 & 10.2795 & 2451362.41 & No \\
2000 Dec 12 & 10.5407 & 2451890.76 & No \\
2001 Jun 09 & 10.6296 & 2452070.46 & No \\
2001 Jun 10 & 10.6301 & 2452071.41 & No \\
2001 Jun 12 & 10.6311 & 2452073.43 & No \\
2002 Feb 14 & 10.7528 & 2452319.70 & No \\
2002 Feb 19 & 10.7552 & 2452324.62 & No \\
2002 Apr 29 & 10.7898 & 2452394.48 & No \\
2002 Apr 30 & 10.7903 & 2452395.46 & No \\
2002 Jul 20 & 10.8303 & 2452476.39 & No \\
2002 Nov 04 & 10.8829 & 2452582.76 & No \\
2002 Dec 21 & 10.9061 & 2452629.76 & No \\ 
2003 May 13 & 10.9771 & 2452773.41 & Yes \\
2003 May 15 & 10.9781 & 2452775.44 & Yes \\
2003 May 25 & 10.9830 & 2452785.38 & Yes \\
2003 Jun 17 & 10.9944 & 2452808.45 & Yes \\
2003 Jun 22 & 10.9969 & 2452813.45 & Yes \\
2003 Jun 24 & 10.9979 & 2452815.43 & Yes \\
2003 Jun 26 & 10.9989 & 2452817.38 & Yes \\
2003 Jun 28 & 10.9998 & 2452819.36 & Yes \\
2003 Jun 29 & 11.0003 & 2452820.38 & Yes \\
2003 Jul 03 & 11.0023 & 2452824.37 & Yes \\
2003 Jul 04 & 11.0028 & 2452825.33 & Yes \\
2003 Jul 08 & 11.0048 & 2452829.46 & Yes \\
2003 Jul 09 & 11.0053 & 2452830.40 & Yes \\
2003 Aug 11 & 11.0216 & 2452863.38 & No \\
2003 Aug 13 & 11.0226 & 2452865.39 & No \\
2003 Dec 15 & 11.0835 & 2452988.67 & No \\
2004 Feb 29 & 11.1212 & 2453064.83 & No \\
2004 Apr 30 & 11.1516 & 2453126.41 & No \\
2005 Mar 25 & 11.3143 & 2453455.47 & No \\
2005 Jul 26 & 11.3751 & 2453578.37 & No \\
2006 Mar 14 & 11.4893 & 2453809.47 & No \\
2006 Jun 06 & 11.5308 & 2453893.43 & No \\
2006 Aug 11 & 11.5634 & 2453959.38 & No \\
2007 Jun 28 & 11.7221 & 2454280.43 & No \\
2007 Jun 29 & 11.7226 & 2454281.39 & No \\
2007 Jun 30 & 11.7231 & 2454282.39 & No \\
2007 Aug 02 & 11.7394 & 2454315.36 & No \\
2008 May 19 & 11.8833 & 2454606.40 & No \\
2008 Dec 05 & 11.9819 & 2454805.74 & Yes \\
2008 Dec 07 & 11.9828 & 2454807.70 & Yes \\
2008 Dec 09 & 11.9838 & 2454809.70 & Yes \\
2009 Jan 08 & 11.9986 & 2454839.60 & Yes \\
2009 Feb 19 & 12.0193 & 2454881.57 & No \\
2009 Apr 20 & 12.0494 & 2454942.46 & No \\  
\hline
\end{tabular}
\end{table} 

We use a large amount of archival near-infrared medium-resolution spectroscopic observations, obtained during 1992--2004 \citep{damineli98,damineli08_multi,gdj07}, and gathered additional spectra of Eta Car during 2004--2009 with the 1.6-m telescope of the Brazilian Observatorio Pico dos Dias/Laboratorio Nacional de Astrofisica (OPD/LNA). The data reduction has been performed using standard near-infrared spectroscopic techniques as described in \citet{gdj07} and \citet{damineli08_multi}. These spectra cover the \ion{He}{i} $\lambda$10833 line with a moderate spectral resolving power ($R\simeq7000$) and seeing--limited spatial resolution ($\sim 1\farcs5$), but have excellent time-sampling of the order of days around the 2003.5 spectroscopic event. The 2009.0 event was covered with a time sampling of the order of a month. The time evolution of the emission component of the \ion{He}{i} $\lambda$10833 line in the OPD/LNA dataset from the 2003.5 event has been extensively discussed in \citet{gdj07} and \citet{damineli08_multi}. Table \ref{tab2} lists the OPD/LNA dataset used in this paper.

\subsection{{\it HST}/STIS ultraviolet and optical spectroscopy}

\begin{table}
\caption{\label{tab3} Summary of {\it HST}/STIS Eta Car spectroscopic observations used in this paper} 
\centering
\begin{tabular}{l c c c}
\hline\hline
Date & $\phi$ & JD & High-velocity absorption?\\
\hline
2002 Jan 20 & 10.738 & 2452240 & No \\
2002 Jul 04 & 10.820 & 2452460 & No \\
2003 Feb 13 & 10.930 & 2452683 & No \\
2003 Jun 01 & 10.984 & 2452792 & Yes \\
2003 Jun 22 & 10.995 & 2452813 & Yes \\
2003 Jul 05 & 11.001 & 2452825 & Yes \\
2003 Jul 29 & 11.013 & 2452851 & Yes \\
\hline
\end{tabular}
\end{table} 

Spectroscopic observations of Eta Car were obtained with the Hubble Space Telescope Imaging Spectrograph ({\it HST}/STIS) from 1998 Jan 1 until 2004 Mar 6. In this paper we use ultraviolet and optical archival data obtained across the 2003.5 spectroscopic event, which was covered by the HST Eta Car Treasury project\footnote{The reduced data are available online at http://etacar.umn.edu}. These spectra have been extensively described in previous works (e.\,g., \citealt{davidson05, hillier06,nielsen07, gull09}), and here we summarize their main characteristics. Ultraviolet observations were conducted using the MAMA echelle grating E140M, providing a spectral resolving power of 30\,000 throughout the 1170--1700~{\AA} spectral range. An aperture of $0\farcs2 \times 0\farcs2$ or $0\farcs3 \times 0\farcs2$ was used, and the data reduction was accomplished using STIS GTO IDL CALSTIS software. A 7-pixel (0\farcs0875) extraction centered on the central source was adopted to minimize contamination from the extended ejecta. We refer to \citet{nielsen05} and \citet{hillier06} for further details on the observations and for an extensive analysis of the {\it HST}/STIS ultraviolet spectrum of Eta Car, in particular the dataset obtained in 2002 Jul 04. For the optical range, the G430M and G750M gratings were used, yielding a resolving power of 6000--8000 across the 1640--10\,100~{\AA} spectral range. Spectra were extracted using 6 half-pixels (0\farcs152). Table \ref{tab3} summarizes the ({\it HST}/STIS data used in this paper.

\section{\label{highvel} High-velocity gas (up to 2000~km\,s$^{-1}$) in Eta Car during the spectroscopic events}

Herein we discuss the behavior of key line profiles across the two most recent spectroscopic events: 2003.5 ($\phi = 11.0$)
and 2009.0 ($\phi = 12.0$). While the ideal case would be that all observations were accomplished across
the same spectroscopic event, availability of instrumentation did not permit such. We note that the 1998.0 and 2003.5 
spectroscopic events were nearly identical in behavior in X-rays \citep{corcoran05}, but that the duration of the X-ray minimum was substantially shorter in the 2009.0 spectroscopic event (M. Corcoran et al. 2010, in prep.). However, optical spectroscopy indicated that the H$\alpha$ line profile was flat-topped in 2003.5 but not in the 1998.0 spectroscopic event \citep{davidson05}. Our understanding of the changes in spectroscopic profiles presented here suggests that while there may be small changes in the wind profiles due to the secular variability (Sect. \ref{highvelopd}), the major changes we see are due to the periodic variations caused by the binary nature of Eta Car.

\subsection{\label{highvelcrires}Detection across the 2009.0 spectroscopic event by VLT/CRIRES}

The \ion{He}{i} $\lambda$10833 absorption line profile, as displayed in Figure \ref{fig1}, evolved considerably across the 2009.0 
spectroscopic event. In all recorded spectra a relatively broad ($\sim100~\kms$), blueshifted emission feature is seen at $-250~\kms$, and is due to gas in the equatorial plane of the Homunculus \citep{smith02, teodoro08}. 

The spectrum obtained at $\phi=11.875$, well before periastron, shows strong \ion{He}{i} $\lambda$10833 absorption extending from $\sim-150~\kms$ up to an edge with velocity $\vedge \simeq -750~\kms$, with the strongest absorption occurring at $\vblack\simeq-580~\kms$. This value of $\vblack$ from $\ion{He}{i} \lambda$10833 is remarkably similar to those derived from UV resonance lines such as \ion{C}{ii} $\lambda$1335 and \ion{Mg}{ii} $\lambda$1240 \citep[$-600 \pm 50 ~\kms$,][]{hillier01}. This range of velocities is consistent with \ion{He}{i} $\lambda$10833 being formed in the wind of Eta Car~A. Eta Car~B might significantly influence the \ion{He}{i} $\lambda$10833 line profile through photoionization of the outer parts of the wind of Eta Car~A and, in this case, the amount of emission and absorption seen in \ion{He}{i} $\lambda$10833 will strongly depend on the orbital parameters of the system. Nevertheless, at phases sufficiently far from periastron such as at $\phi=11.875$, there is no evidence for additional velocity fields in our line-of-sight, such as one would expect in the case that the absorption was formed in the wind of Eta Car~B or in high-velocity material from the wind-wind collision zone.

The profiles from $\phi=11.991$ and $\phi=11.998$ show very different \ion{He}{i} $\lambda$10833 absorptions compared to the one from $\phi=11.875$. The low-velocity absorption strengthened, becoming nearly saturated from $-40$ to $-580~\kms$, with the exception of the equatorial ejecta emission at $-250~\kms$. A broad, high-velocity absorption ranging from $-580$ to $-1900~\kms$ appeared by $\phi=11.991$ and strengthened by $\phi=11.998$. We would not expect this high-velocity absorption from the velocity field of the wind of Eta Car~A, as seen in the spectrum from $\phi=11.875$. Therefore, the \ion{He}{i} $\lambda$10833 absorption line profile strongly indicates that, in addition to the wind of Eta Car~A, at least one more velocity structure is crossing our line-of-sight to Eta Car. That high-velocity absorption is transient; as the spectrum observed at $\phi=12.014$ shows, it has faded considerably and is present only up to $-900~\kms$. By $\phi=12.041$, the \ion{He}{i} $\lambda$10833 profile is quite similar to that recorded at $\phi=11.875$, but the $-40$ to $-580~\kms$ absorption is saturated, indicating a higher column density of \ion{He}{i}.

\begin{figure}
\resizebox{\hsize}{!}{\includegraphics{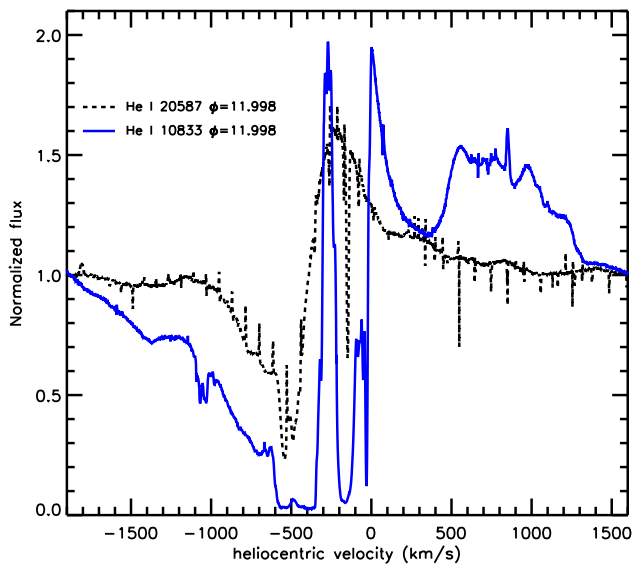}}
\caption{\label{fig2}Comparison between the continuum-normalized \ion{He}{i} $\lambda$10833 (blue) and \ion{He}{i} $\lambda$20587 (black) spectral lines from the inner $0\farcs26 \times 0\farcs20$ spatial region around the central source of Eta Car. The many narrow spikes around \ion{He}{i} $\lambda$20587 are residuals from the removal of telluric lines. Note that a $-146~\kms$ absorption component is present in \ion{He}{i} $\lambda$20587 and might also be present in \ion{He}{i} $\lambda$10833.  The emission feature from $500$ to $1300~\kms$ is due to a blend of \ion{Fe}{ii} lines.}
\end{figure}

The \ion{He}{i} $\lambda$10833 and \ion{He}{i} $\lambda$20587 line profiles, recorded at $\phi=11.998$, demonstrate that the high-velocity absorption component is much stronger in \ion{He}{i} $\lambda$10833 than in the \ion{He}{i} $\lambda$20587 line (Fig. \ref{fig2}). While noticeable from $-600$ to $-1000~\kms$, the absorption is much weaker in the range of $-1100$ to $-1600~\kms$. The \ion{He}{i} $\lambda$10833 line (2s\,\element[][3]{S} -- 2p\,\element[][3]{P}) absorption originates from the metastable triplet state, while the \ion{He}{i} $\lambda$20587 line (2s\,\element[][1]{S} -- 2p\,\element[][1]{P}) originates from the metastable 2s\,\element[][1]{S} state. The population of the 2s\,\element[][1]{S} energy level can be increased through photo-excitation from 584~{\AA} UV photons, which would cause increased \ion{He}{i} $\lambda$20587 absorption. The much weaker high-velocity absorption in the \ion{He}{i} $\lambda$20587 line profile indicates that photo-excitation by hard UV photons is negligible in the region responsible for the high-velocity absorption. The oscillator strength of \ion{He}{i} $\lambda$10833 is about 5 times higher than that of  \ion{He}{i} $\lambda$20587 and, thus, a much stronger \ion{He}{i} $\lambda$10833 as seen in Figure \ref{fig2} is exactly what would be expected.

The high-velocity absorption is not seen in H$\alpha$ or any other hydrogen lines during periastron, since these lines show an edge velocity in the range $-800$ to $-1000~\kms$ \citep{weis05,stahl05,davidson05}, implying that hydrogen is predominantly ionized (H$^0$/H$^{+} \lesssim 10^{-5}$) in the region responsible for the \ion{He}{I} high-velocity absorption. In addition, the absence of high-velocity H$\alpha$ absorption in the range $-1000$ to $-2000~\kms$ during periastron indicates that such material is either helium rich, or it places an upper limit on the electron density ($n_\mathrm{e}$) in that region. Although detailed radiative transfer calculations are needed, we estimate an upper limit of $n_\mathrm{e}\leq10^{10}$ cm$^{-3}$.

\subsection{The OPD/LNA datasets from the 2003.5 and 2009.0 spectroscopic events \label{highvelopd}}

\begin{figure}
\resizebox{\hsize}{!}{\includegraphics{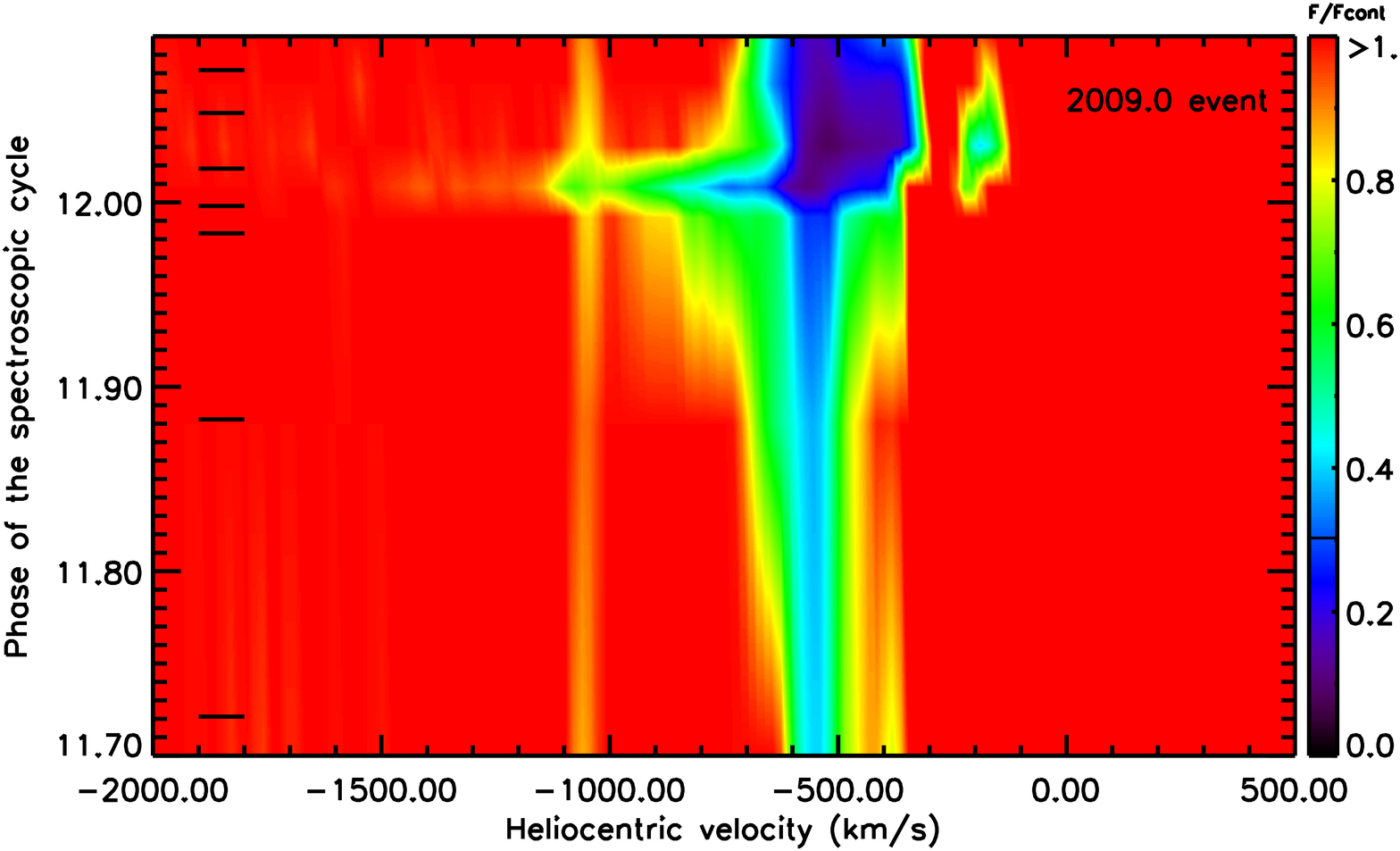}}
\resizebox{\hsize}{!}{\includegraphics{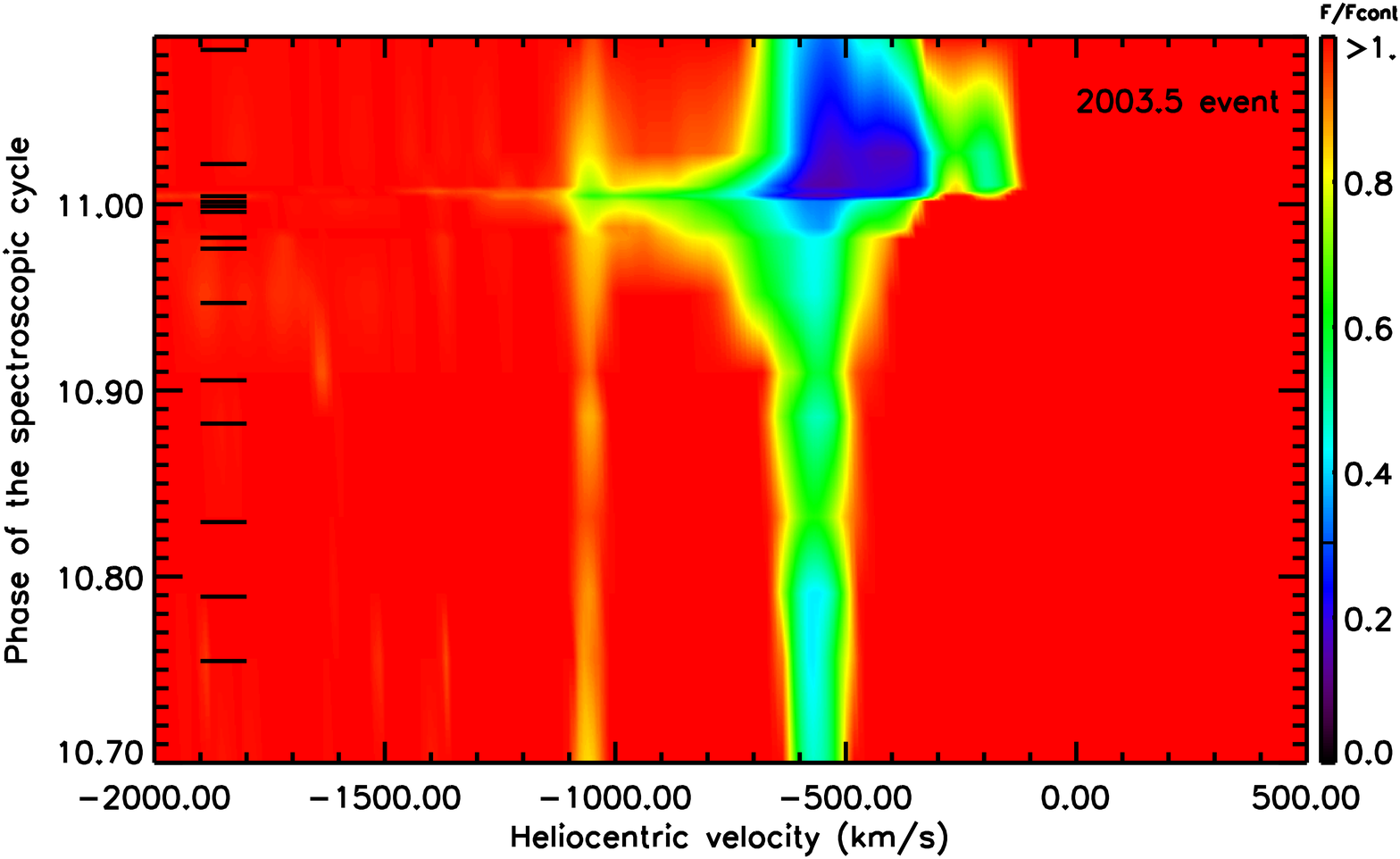}}
\caption{\label{lna_2003_2009} Evolution of the \ion{He}{i} $\lambda$10833 line as a function of orbital phase, for the 2003.5 (bottom panel) and 2009.0 spectroscopic events (upper panel). The spectra were interpolated in phase for visualization purposes, and the continuum-normalized flux is color-coded linearly between 0 (black) and $\geq1$ (red) to emphasize the absorption structure.  The black horizontal tick marks on the left correspond to the observed phases. The feature running vertically at $-1050~\kms$ is probably formed outside the Homunculus nebula and is not relevant for the purpose of this paper.}
\end{figure}

\begin{figure}
\resizebox{\hsize}{!}{\includegraphics{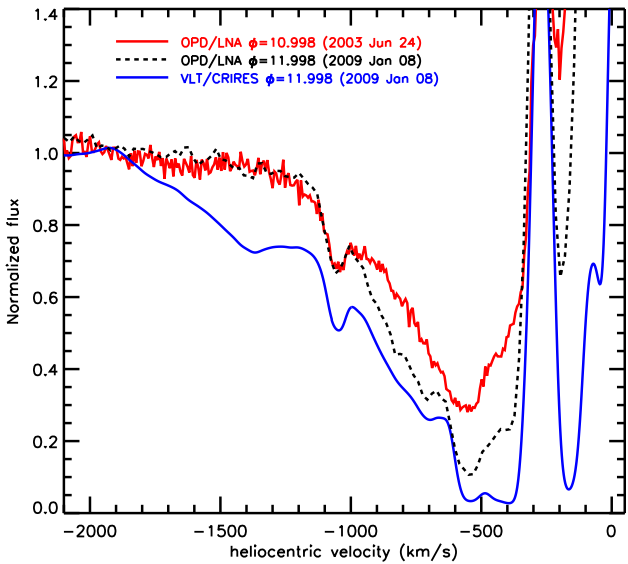}}
\caption{\label{fig3}Comparison between continuum-normalized \ion{He}{i} $\lambda$10833 line profiles obtained at different spatial resolutions and similar orbital phases (but different cycles). The dashed black line shows the spectrum from \citet{damineli08_multi} obtained with the 1.6-m telescope of the Brazilian OPD/LNA, which has $R\simeq9000$, and aperture and spatial resolution of roughly $1\farcs5$.  The CRIRES spectrum (solid blue line) has an aperture and spatial resolution of $\sim0\farcs3$ and was convolved with a Gaussian to match the spectral resolution of the OPD/LNA spectrum. }
\end{figure}

As mentioned before, ground-based data of Eta Car have shown a weak absorption wing of up to $-1500~\kms$ associated with \ion{He}{i} $\lambda$10833 during previous spectroscopic events \citep{damineli98,gdj07,damineli08_multi}. The new spectra obtained across the 2009.0 event showed again the presence of such high-velocity absorption during a brief time interval, which strongly suggests that it is related to the periastron passage of Eta Car~B. Figure \ref{lna_2003_2009} displays the temporal evolution of the \ion{He}{i} $\lambda$10833 absorption during the 2003.5 and 2009.0 spectroscopic events, while Figure \ref{fig3} presents a comparison between the \ion{He}{i} $\lambda$10833 line profiles at a similar phase during the 2003.5 and 2009.0 spectroscopic cycles ($\phi=10.998$ and $\phi=11.998$, respectively), and the VLT/CRIRES spectrum obtained simultaneously to the OPD/LNA data at $\phi=11.998$. 

First, comparing spectra obtained with OPD/LNA at  similar orbital phases close to periastron, we notice that the high-velocity \ion{He}{i} $\lambda$10833 absorption changes from cycle to cycle, and is stronger in the spectrum from the 2009.0 event. This trend in the cycle-to-cycle variability was also noticed in optical \ion{He}{i} lines \citep{gd04}, but for low velocities within the line. Whether this is due to changes in the continuum or in the line-absorbing region, or both, still remains to be seen. Secular variability from one (or both) stellar winds in the Eta Car system cannot be excluded.

Second, we notice that the  \ion{He}{i} $\lambda$10833 high-velocity absorption is stronger and extends to higher velocities in the VLT/CRIRES than in the OPD/LNA spectrum. This is not due to the different spectral resolutions, and we argue that the small aperture and better spatial resolution of the VLT/CRIRES instrument ($\sim0\farcs3$) spatially resolves extended scattered continuum and \ion{He}{i} $\lambda$10833 line emission, which are otherwise included in the OPD/LNA spectra, since  they have a lower spatial resolution ($1\farcs5$) than the VLT/CRIRES data. 

\subsection{\label{highvelstis}Detection across the 2003.5 spectroscopic event by {\it HST}/STIS}

\subsubsection{Ultraviolet resonance lines}

\begin{figure}
\resizebox{\hsize}{!}{\includegraphics{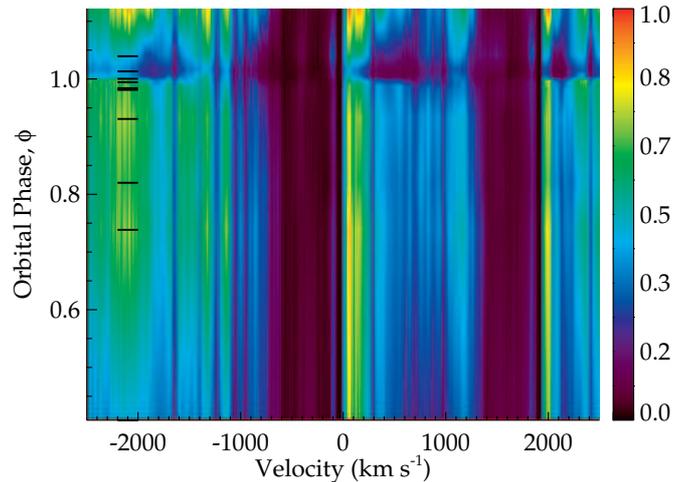}}
\caption{\label{uv1} Similar to Fig. \ref{lna_2003_2009}, but showing the evolution of the \ion{Si}{iv} $\lambda$$\lambda$1394, 1402 lines as a function of orbital phase. The velocity scale refers to \ion{Si}{iv} $\lambda$1394. In this particular Figure no flux scaling was applied to illustrate the global decrease in the UV flux close to the spectroscopic event. The spectra were interpolated in phase for visualization purposes and intensity color-coded between the minimum flux (black) and max flux  (red) seen across the spectroscopic cycle. The black horizontal tick marks on the right correspond to the observed phases.}
\end{figure}

We analyzed the archival {\it HST}/STIS ultraviolet spectra of Eta Car searching for evidence of high-velocity absorption in UV resonance lines. The time variability of the UV spectrum across the spectroscopic cycle is complex and will be analyzed in detail elsewhere (Nielsen et al. 2010, in prep.). To illustrate the variability of the UV spectrum of Eta Car around selected UV resonance lines, we computed color-coded intensity plots by stacking the spectra in velocity space as a function of phase, and by interpolating in phase the flux between the observations (Figures \ref{uv1} and \ref{uv2}).

We found that some of the UV resonance lines show evidence for high-velocity absorption during the 2003.5 spectroscopic event, but two other effects influence the interpretation of the spectra. First, due to the increasingly high line density to shorter wavelengths, the ultraviolet continuum level of Eta Car cannot be accurately determined \citep{hillier01,hillier06}. Second, the line profiles of UV resonance lines are extremely affected by strong blending primarily due to \ion{Fe}{ii} lines \citep{hillier01,nielsen05,hillier06}. In addition, the UV flux begins decreasing about six months before the spectroscopic event \citep{smith04,nielsen05} and recovers well after minimum. Similar to what is seen in the optical \citep{nielsen07,damineli08_multi}, the emission and absorption of lines from \ion{Fe}{ii} strongly increase after $\phi=11.0$ in the UV. Both the change in continuum level and the blending due to \ion{Fe}{ii} lines hamper the precise determination of the strength and maximum velocity of the high-absorption component across the spectroscopic event. Therefore, we focus here the evolution of the spectra leading up to $\phi=11.0$. 

We scaled the UV spectra obtained at $\phi=10.820$ (2002 Jul 04), $\phi=10.930$ (2003 Feb 13), and $\phi=10.984$ (2003 Jun 01) to approximately match the continuum of the spectrum from $\phi=10.995$ (2003 Jun 22). As reference, we used a ``control" spectral interval located sufficiently far ($>10000~\kms$) from the rest wavelength of the line of interest. Our purpose for scaling the spectra was to separate changes in the line profile due to the high-velocity absorption from shifts in the continuum level, which otherwise could be misinterpreted as increases in absorption strength. We verified that the ``control" regions, which are also affected by \ion{Fe}{ii} lines, do not change appreciably in the scaled spectra as a function of phase before $\phi=11.0$. Thus, we investigate the temporal behavior of the remaining relative changes in the scaled line profiles, which are intrinsic to the UV resonance lines.

% Figure 6 available electronically only 
\onlfig{6}{
\begin{figure}
\resizebox{\hsize}{!}{\includegraphics{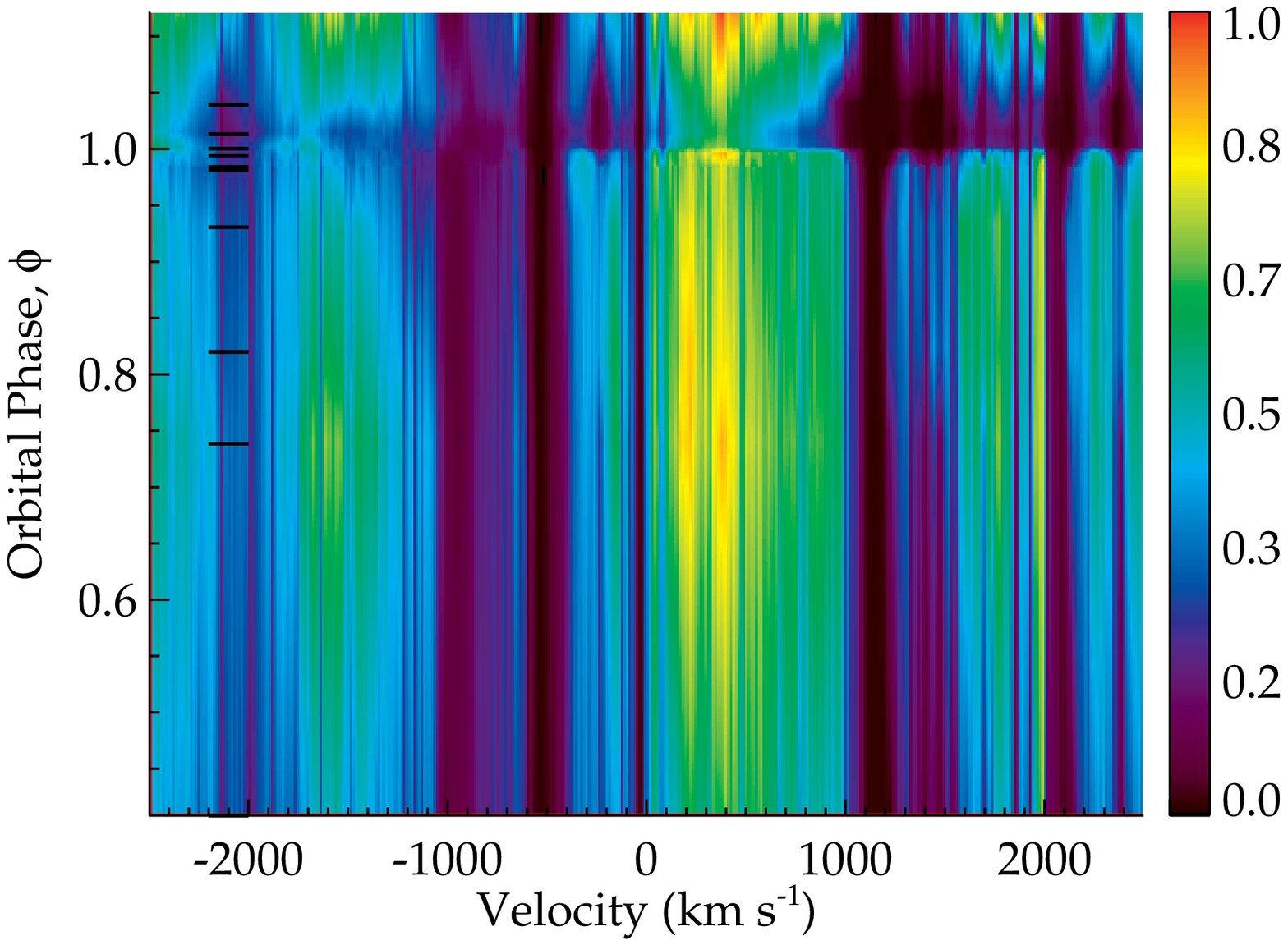}}
\resizebox{\hsize}{!}{\includegraphics{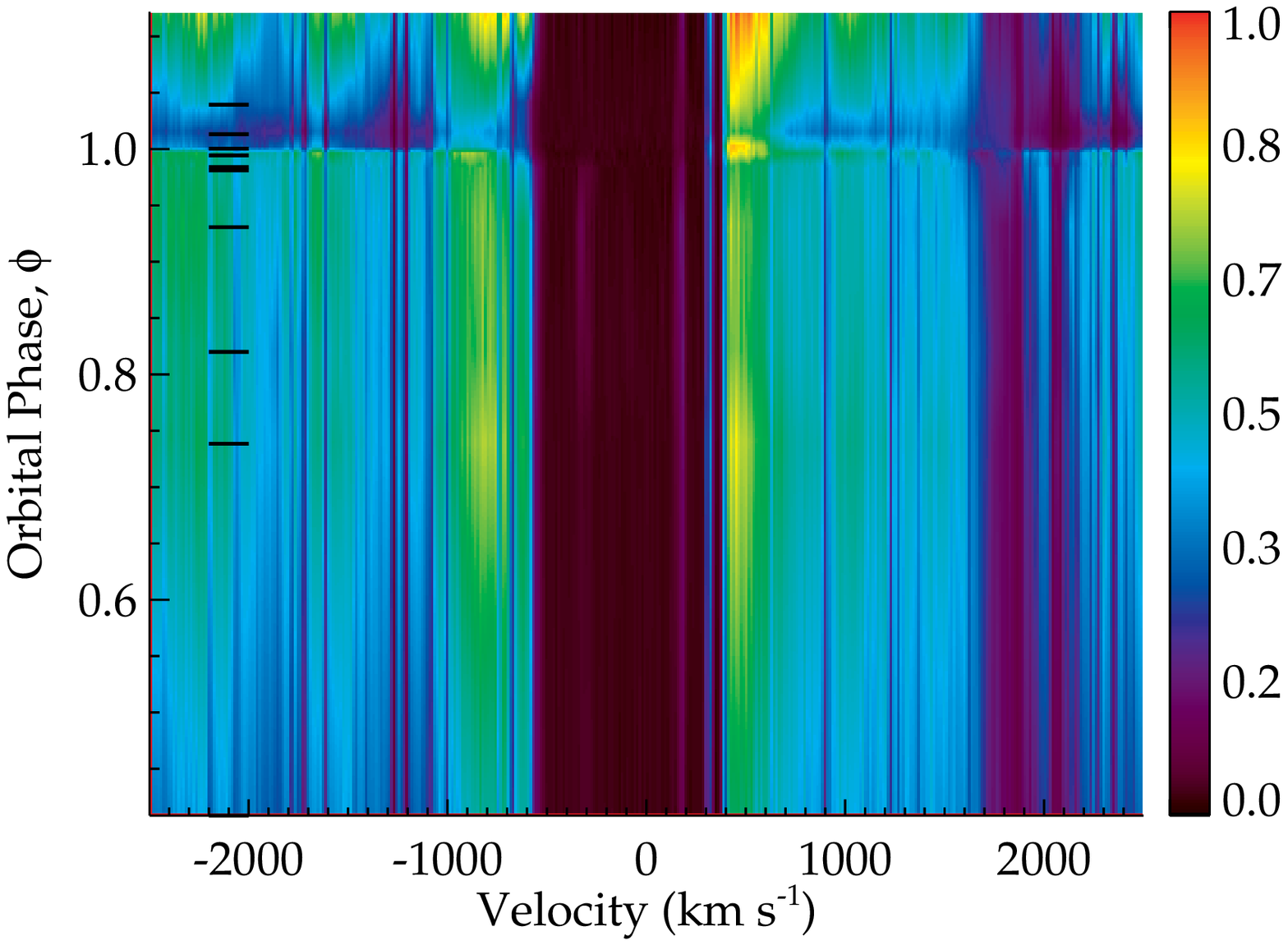}}
\resizebox{\hsize}{!}{\includegraphics{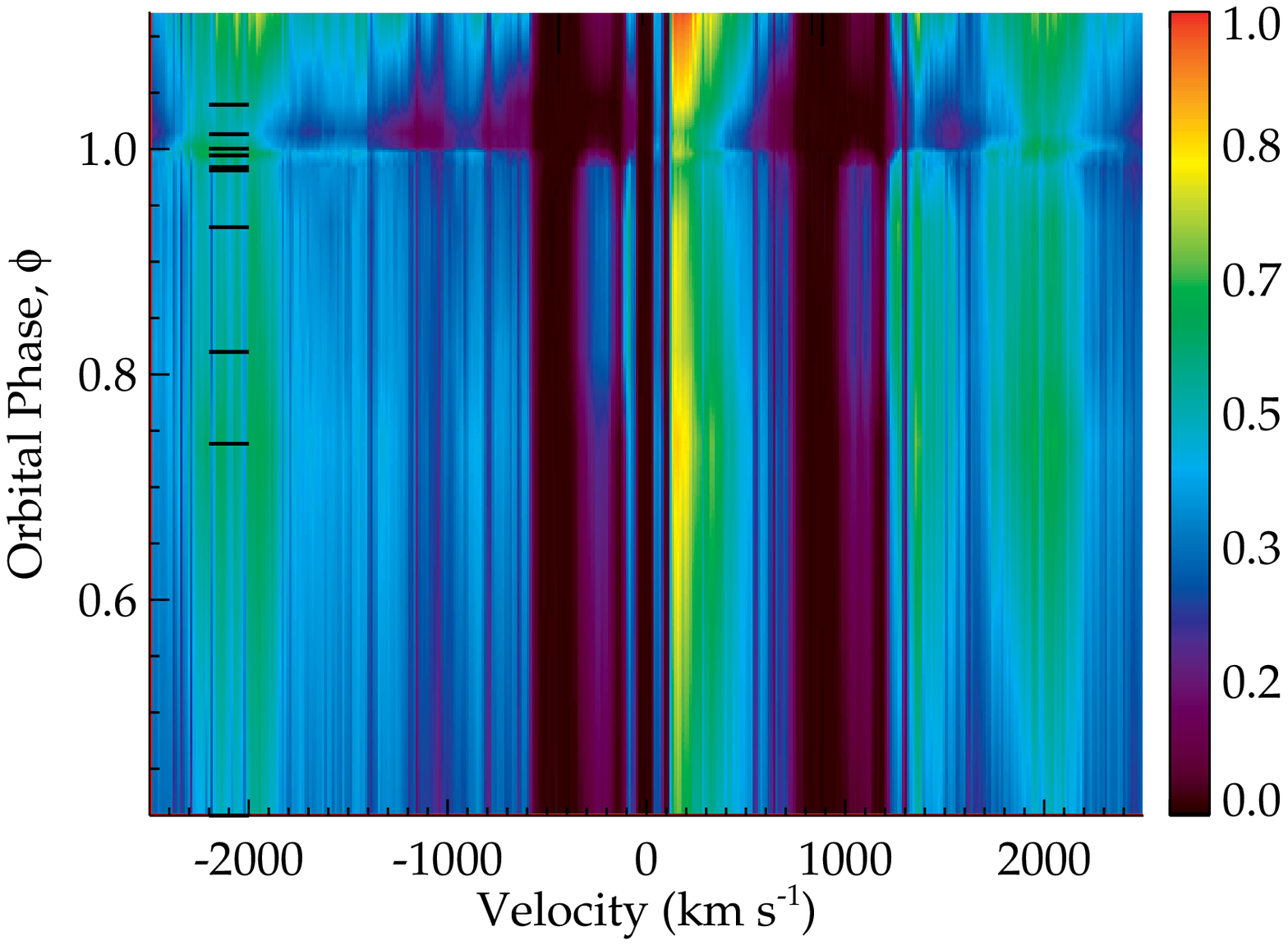}}
\caption{\label{uv2} Similar to Fig. \ref{uv1}, but for \ion{C}{iv} $\lambda$1548, 1551 (upper panel),  \ion{C}{ii} $\lambda\lambda$1334, 1335 (middle panel), \ion{Si}{ii} $\lambda\lambda$1526, 1533 (bottom panel).}
\end{figure}
}% end of onlfig

\begin{figure*}
\resizebox{1.00\hsize}{!}{\includegraphics{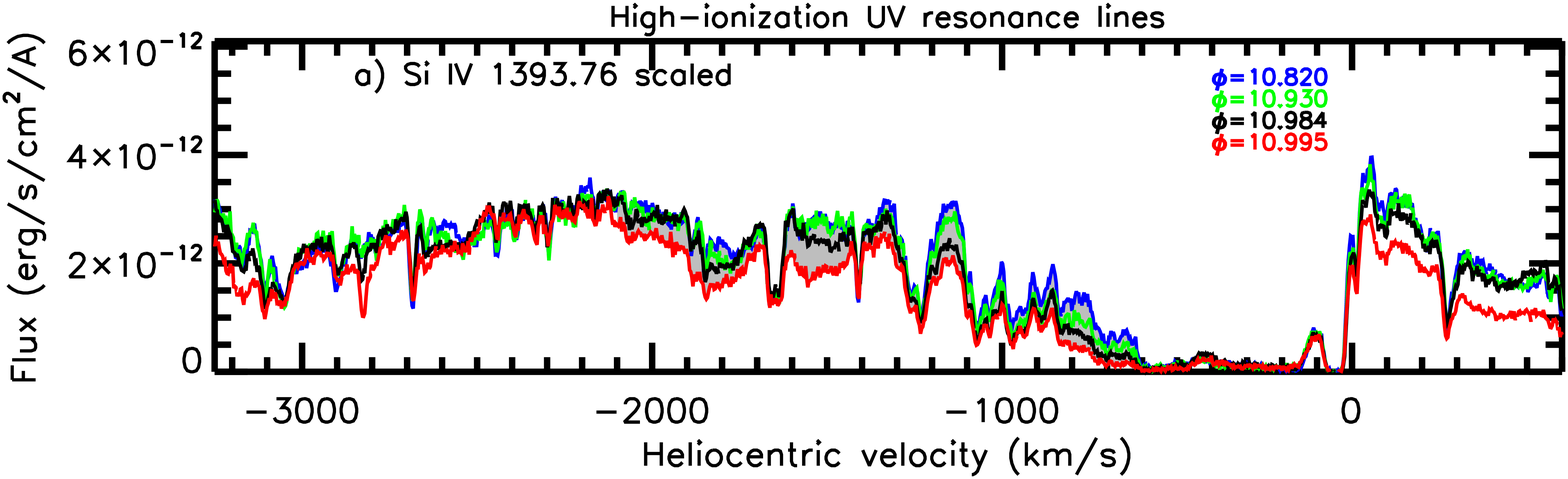} \includegraphics{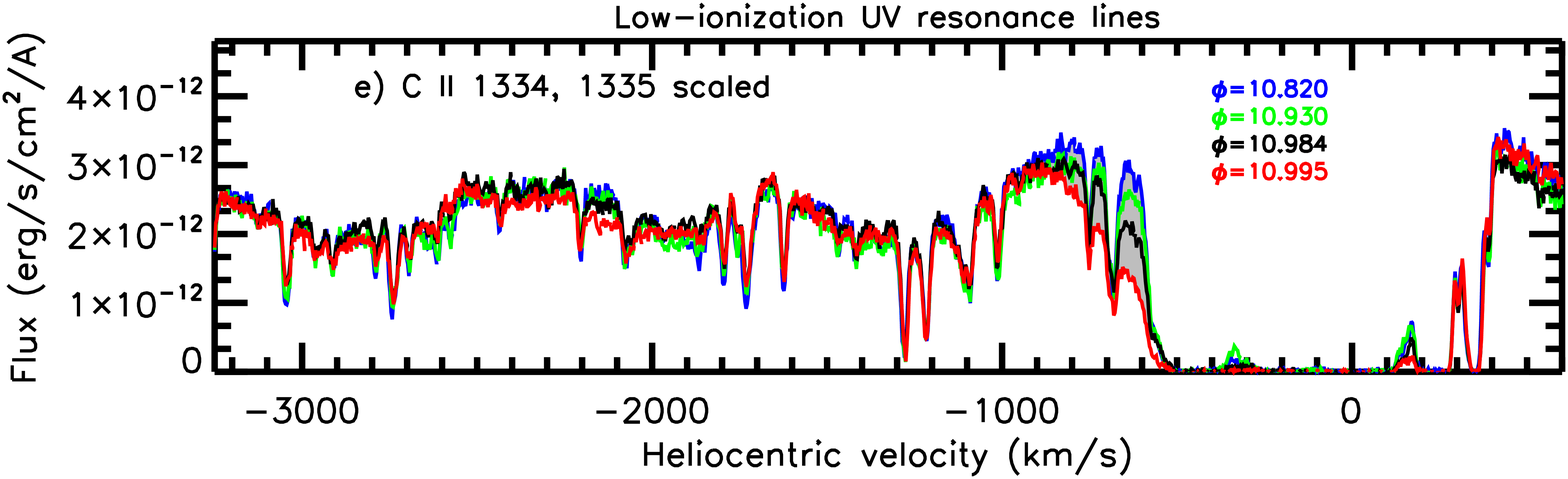}}\\
\resizebox{1.00\hsize}{!}{\includegraphics{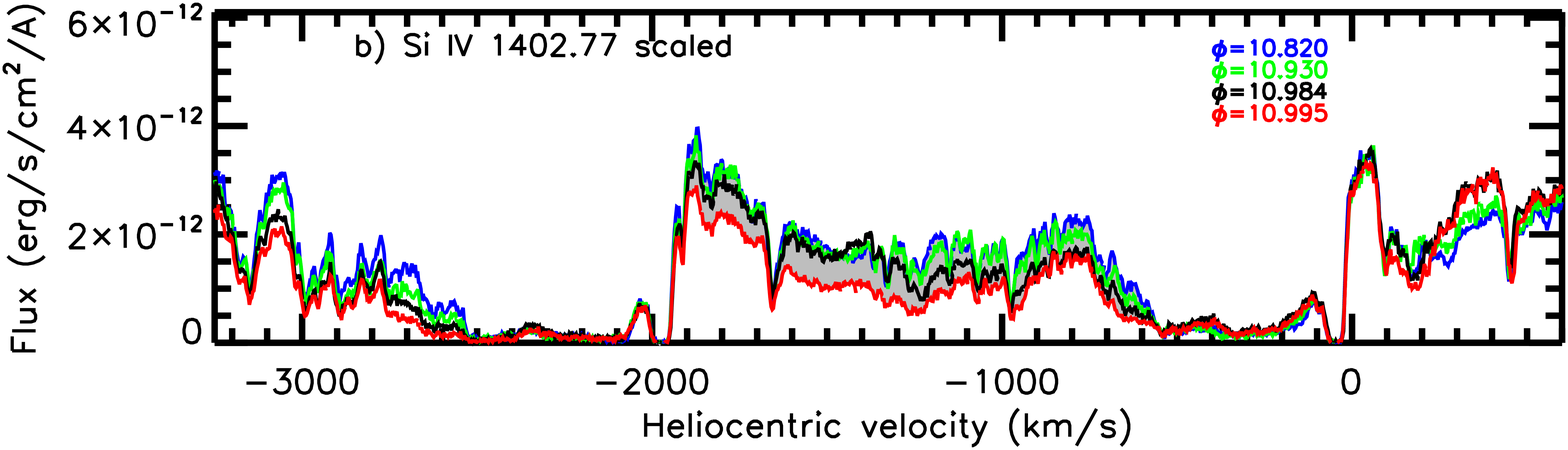} \includegraphics{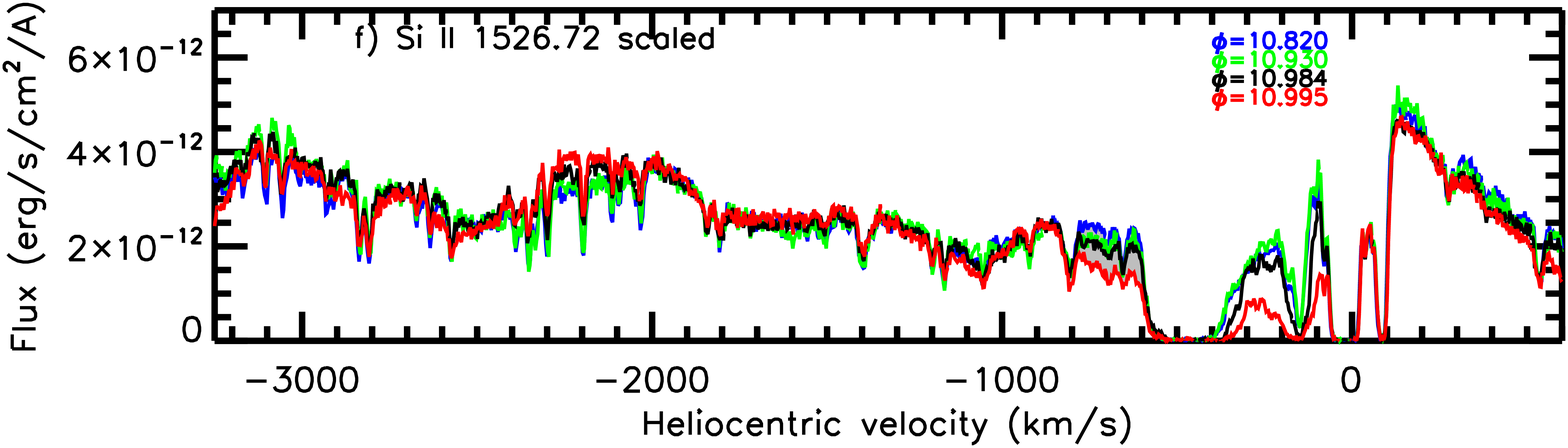}}\\
\resizebox{1.00\hsize}{!}{\includegraphics{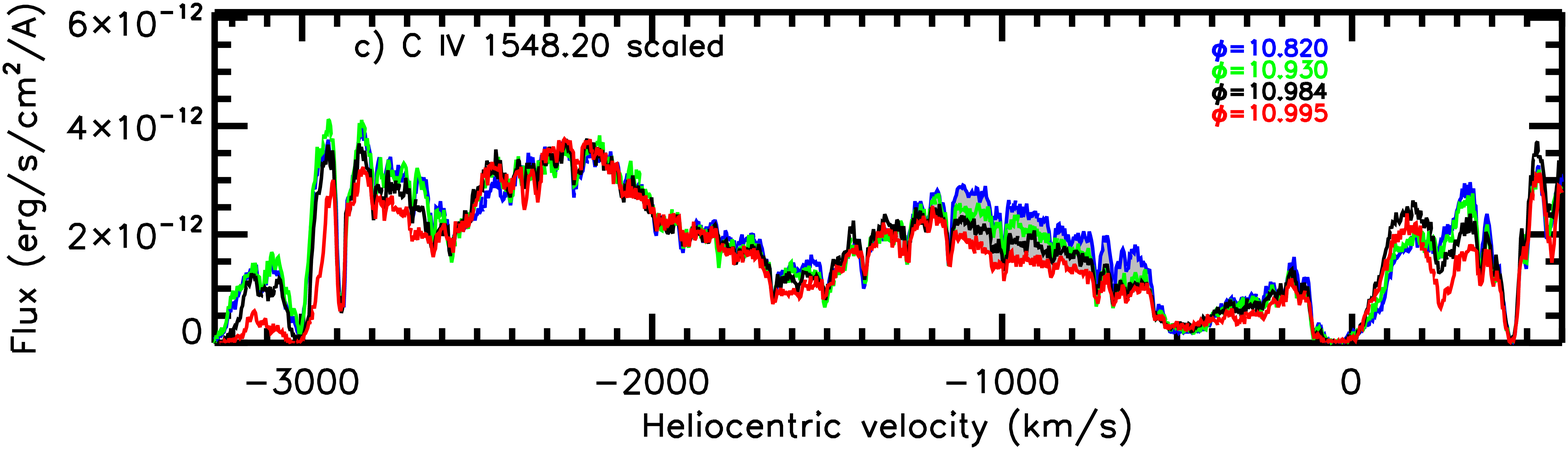}  \includegraphics{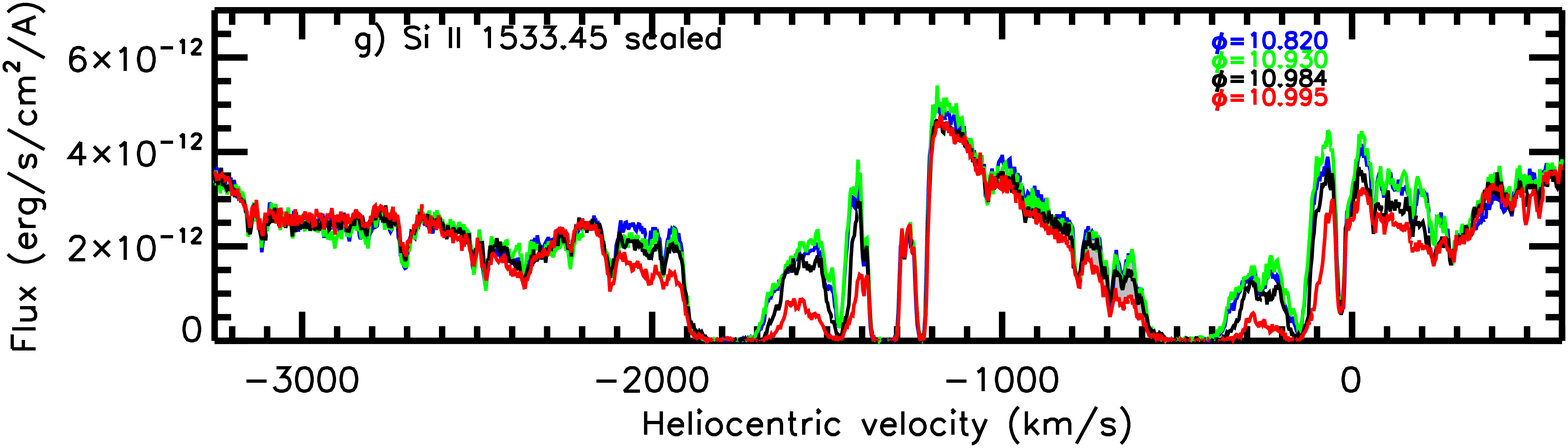}}\\
\resizebox{1.00\hsize}{!}{\includegraphics{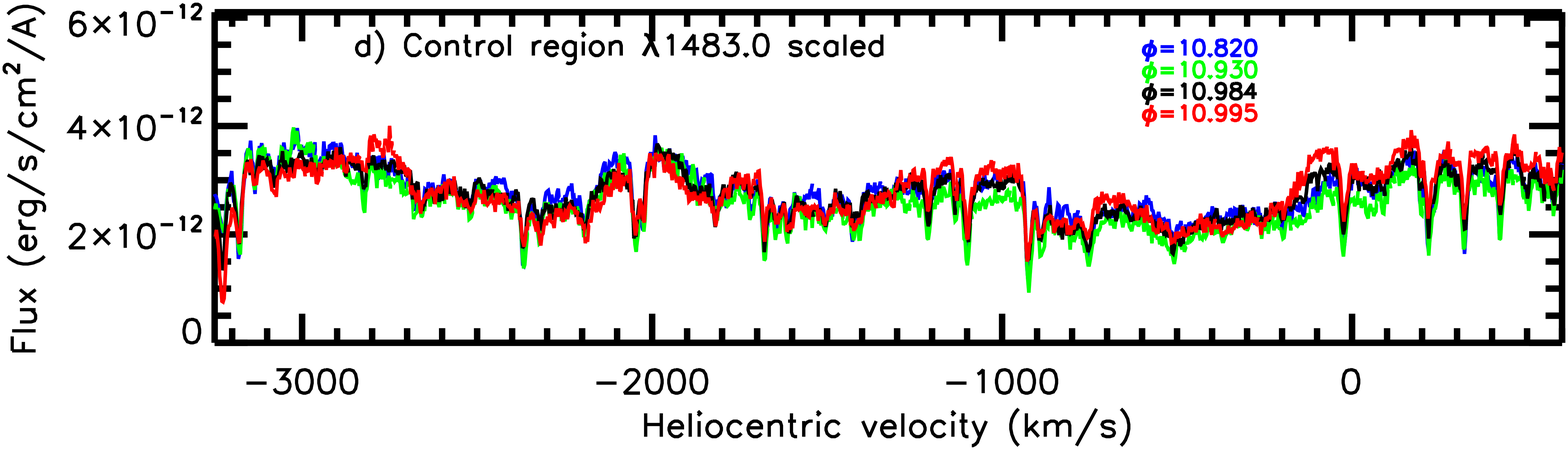}  \includegraphics{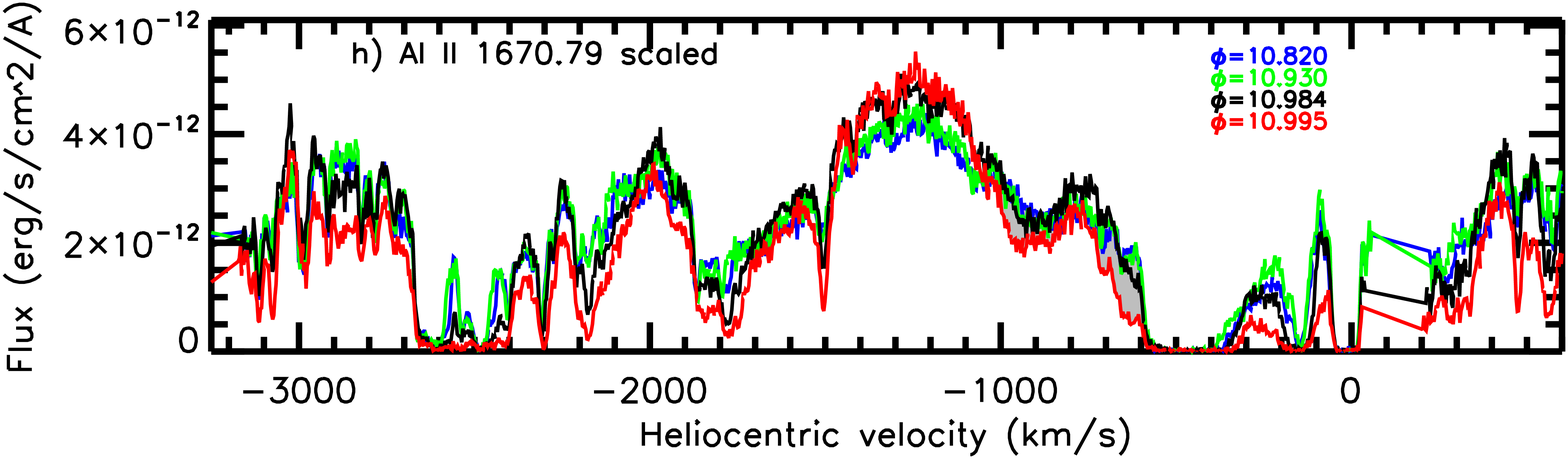}}\\
\caption{\label{highveluv1}Montage of profiles of resonance lines seen in ultraviolet spectra of Eta Car obtained with {\it HST}/STIS across the 2003.5 event at $\phi=10.820$ (blue line), $\phi=10.930$ (green), $\phi=10.984$ (black), and $\phi=10.995$ (red). The continuum level of the spectra taken at $\phi=10.820$, $\phi=10.930$, and $\phi=10.984$ were scaled to approximately match the continuum level of the spectrum taken at $\phi=10.995$. The grey region shows the difference between the spectrum taken at $\phi=10.820$ and at $\phi=10.995$, corresponding to the excess absorption due to the high-velocity material in Eta Car. {\it Left panel:} High-ionization lines. From top to bottom, resonance lines of \ion{Si}{iv} $\lambda$1394,  \ion{Si}{iv} $\lambda$1403, \ion{C}{iv} $\lambda$1548, and a ``control" region around 1483 {\AA} are shown. Notice that little changes are seen in the ``control" region as a function of phase, indicating that the relative variability seen in the UV resonance lines are intrinsic to these lines, and not due to blending. {\it Right panel:} Low-ionization lines. From top to bottom, \ion{C}{ii} $\lambda\lambda$1334, 1335, \ion{Si}{ii} $\lambda$1526, \ion{Si}{ii} $\lambda$1533, and \ion{Al}{ii} $\lambda$1671 are displayed. Note that part of the \ion{Si}{ii} $\lambda$1533 line profile, from $-1200$ to $-2100~\kms$, is contaminated by \ion{Si}{ii} $\lambda$1526. }
\end{figure*}

The scaled line profiles of the strongest, more isolated UV resonance lines in Eta Car are presented in Figure \ref{highveluv1}, where the grey region shows the difference between the spectrum taken at $\phi=10.820$ and at $\phi=10.995$, corresponding to the excess absorption occurring across the spectroscopic event due to the high-velocity material in Eta Car. The absorption components of the low-ionization and the high-ionization resonance lines behave quite differently across the 2003.5 event. Low-ionization resonance lines, such as \ion{C}{ii} $\lambda\lambda$1334, 1335, \ion{Si}{ii} $\lambda\lambda$1526, 1533, and \ion{Al}{ii} $\lambda$1671, show a gradual development of an absorption wing increasing from $-500$ to $-900~\kms$  between $\phi=10.820$ and $\phi=10.984$. At $\phi=10.995$, the spectrum shows a significant increase in the strength of the absorption from $-500$ to $-900~\kms$ (Fig. \ref{highveluv1}, right panel), but the low-ionization UV resonance lines do not show evidence for high-velocity absorption from $-1000$ to $-2000~\kms$ before $\phi=11.0$, as \ion{He}{i} $\lambda$10833 did just before $\phi=12.0$.

The high-ionization ultraviolet resonance lines, \ion{Si}{iv} and \ion{C}{iv}, also show an increase in absorption from $-500$ to $-900~\kms$ before $\phi=11.0$, although, differently from the low-ionization lines, with most of the changes occurring between $\phi=10.820$ and $\phi=10.984$. More noticeable changes occurred after $\phi=10.984$. High-velocity absorption from $-1200$ up to $-2100~\kms$ is seen in the high-ionization \ion{Si}{iv} $\lambda\lambda$1394, 1403 doublet and possibly in the \ion{C}{iv} $\lambda$1548 line.  This high-velocity component becomes noticeably stronger between $\phi=10.984$ and $\phi=10.995$ (Fig. \ref{highveluv1}, left panel). The reality of the high-velocity absorption is confirmed by its presence in both of the \ion{Si}{iv} doublet lines. The weaker \ion{C}{iv} $\lambda$1550 line is severely blended with the stronger \ion{C}{iv} $\lambda$1548 line, and it is impossible to unambiguously judge whether the high-velocity absorption is present in \ion{C}{iv} $\lambda$1550 or not. 

Therefore, ultraviolet resonance lines from low-ionization species, such as \ion{Si}{ii} $\lambda\lambda$1527, 1533, \ion{C}{ii} $\lambda\lambda$1334, 1335, and \ion{Al}{ii} $\lambda$1671, show absorption up to $-800~\kms$, possibly $-1200~\kms$, but the UV resonance lines from high-ionization species, specifically \ion{Si}{iv}, show absorption from $-1200$ to $-2100~\kms$. Thus, the high-velocity absorption originates from a region that is markedly more ionized than the wind of Eta Car~A.

\subsubsection{Optical \ion{He}{i} lines}

\begin{figure}
\resizebox{1.04\hsize}{!}{\includegraphics{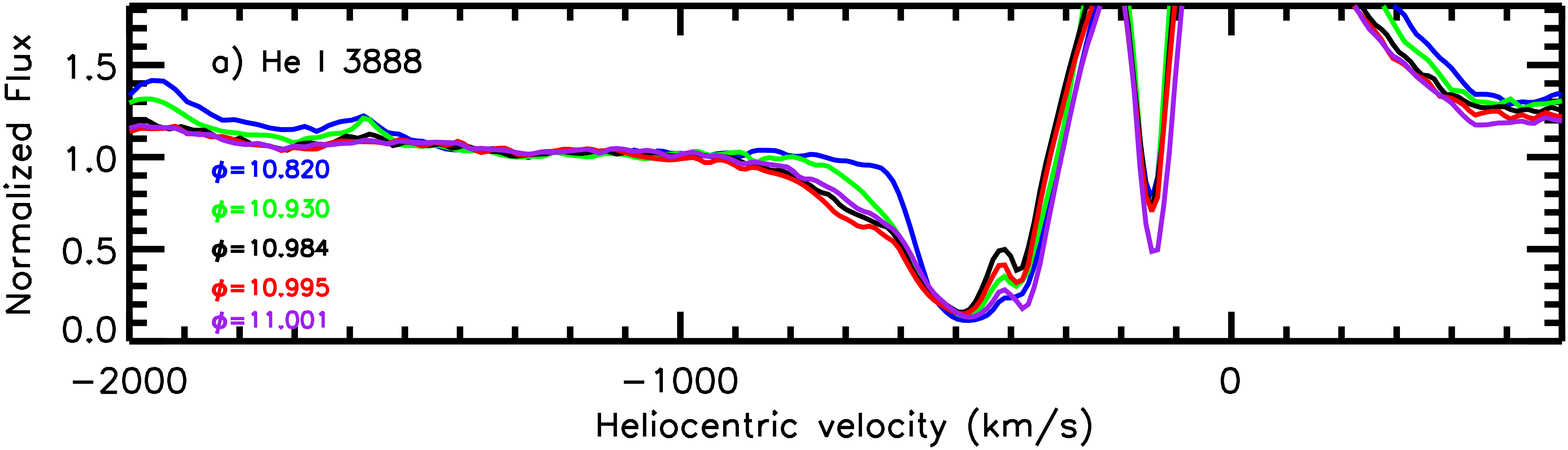}}\\
\resizebox{1.04\hsize}{!}{\includegraphics{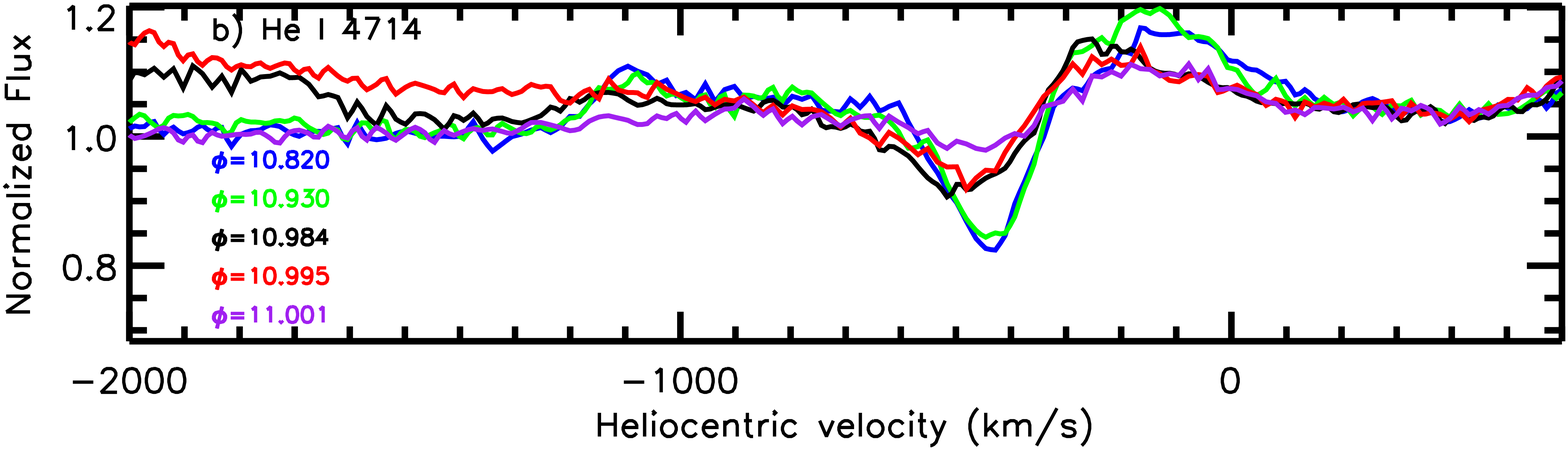}}\\
\resizebox{1.04\hsize}{!}{\includegraphics{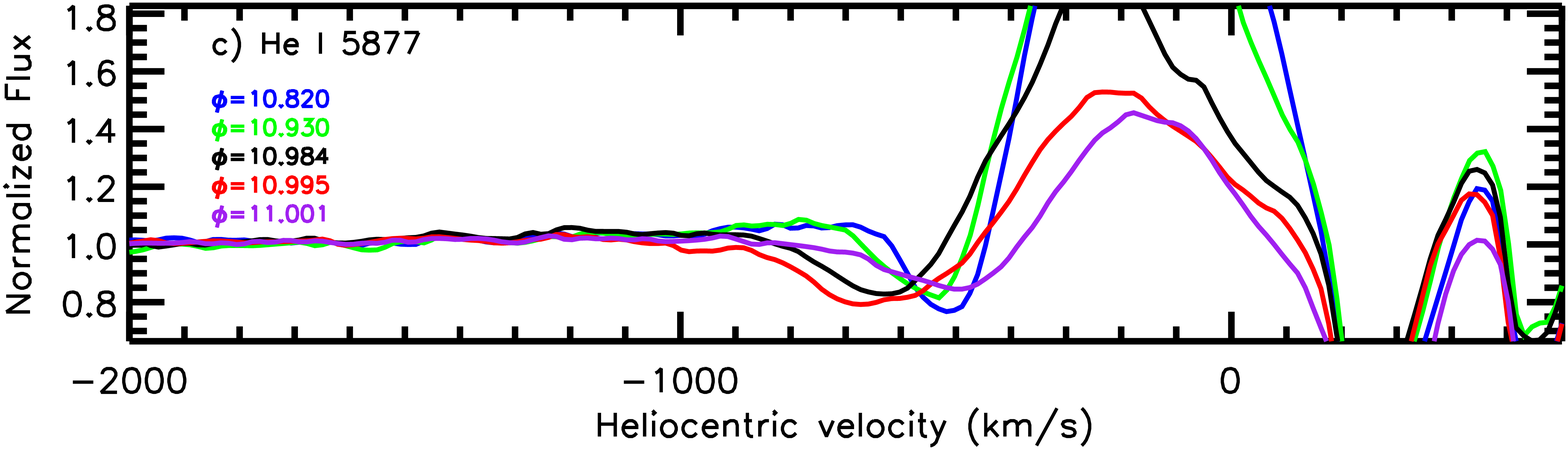}}\\
\resizebox{1.04\hsize}{!}{\includegraphics{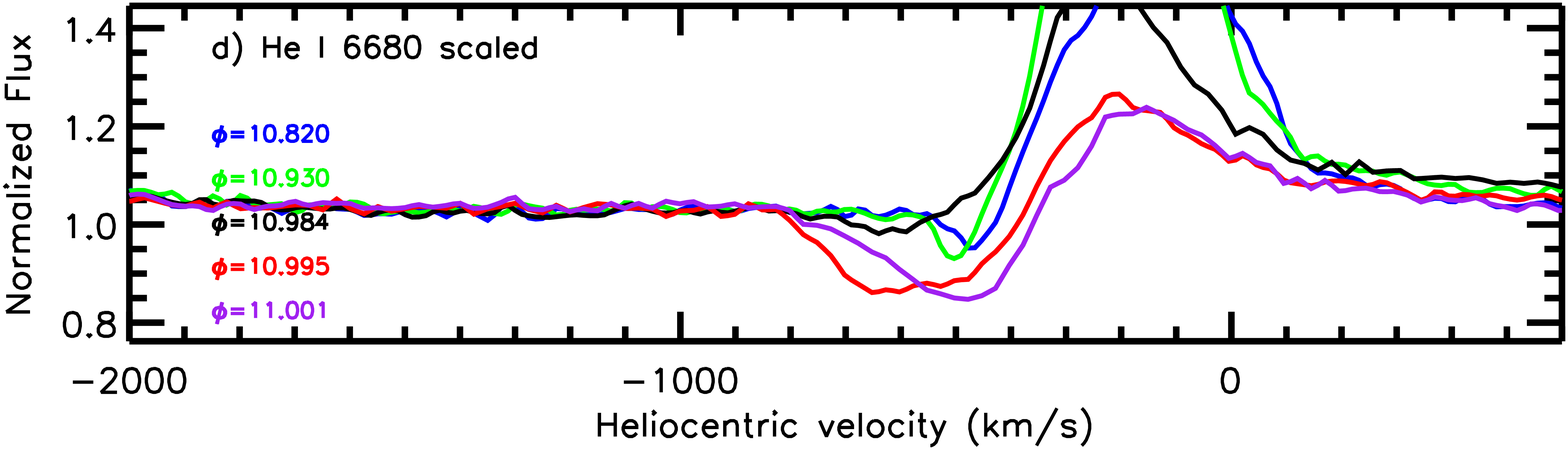}}\\
\resizebox{1.04\hsize}{!}{\includegraphics{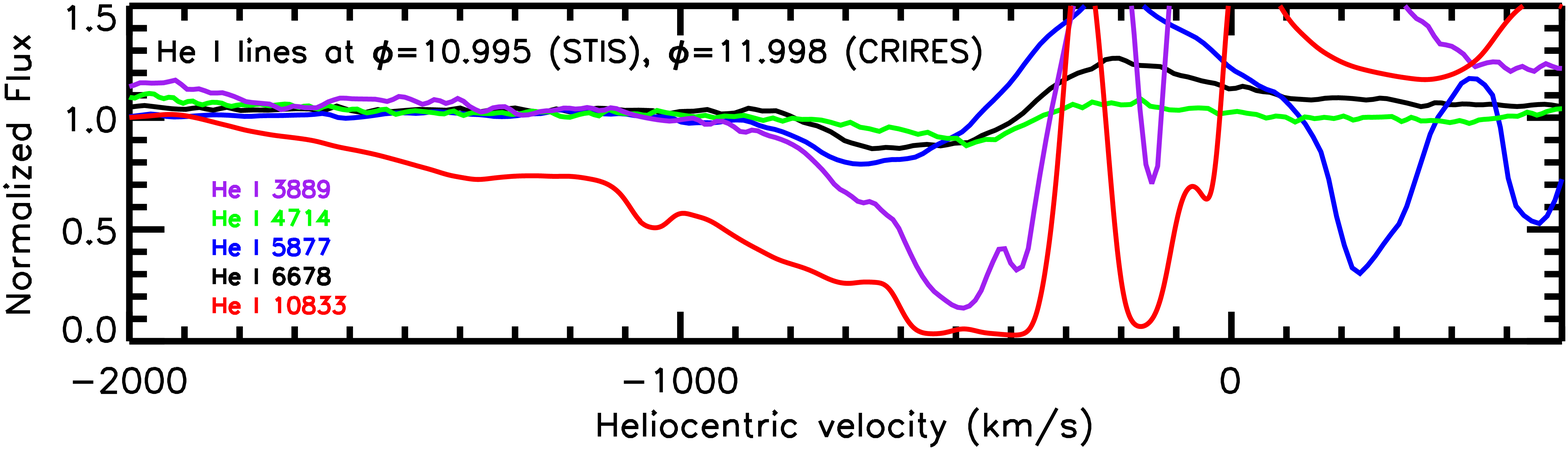}}\\
\caption{\label{highvel_stis_hei}Montage of continuum-normalized \ion{He}{i} line profiles obtained with {\it HST}/STIS across the 2003.5 event. From top to bottom, we present \ion{He}{i} $\lambda$3889 (highly-contaminated by \ion{H}{i} $\lambda$3890, in particular at velocities $v> -800~\kms$), \ion{He}{i} $\lambda$4714 (contaminated by [\ion{Fe}{iii}] $\lambda$4702 emission), \ion{He}{i} $\lambda$5877, and \ion{He}{i} $\lambda$6680 line profiles. The bottom panel displays optical \ion{He}{i} line profiles observed with {\it HST}/STIS at $\phi=10.995$ (2003 Jun 22) with the near-infrared \ion{He}{i} $\lambda$10833 line profile observed with VLT/CRIRES at $\phi=11.998$ (2009 Jan 08). }
\end{figure}

Analysis of optical \ion{He}{i} singlet and triplet line profiles was done by \citet{nielsen07}. Profiles from the same observations
are reproduced in Figure \ref{highvel_stis_hei}. A noticeable high-velocity absorption is apparent in the two triplet line profiles, \ion{He}{i} $\lambda$3888 and $\lambda$5876, extending to $-900$, possibly $-1000~\kms$, at $\phi=10.995$ (2003 Jul 05).
The profiles of singlet lines, such as \ion{He}{i} $\lambda$6680, show less pronounced absorptions up to $-800~\kms$ (note that the \ion{He}{i} $\lambda$4714 line is much weaker and contaminated by [\ion{Fe}{iii}] $\lambda$4702 emission), as noted by \citet{nielsen05}. \citealt{stahl05} also detected similar absorption up to $-750~\kms$ for \ion{He}{i} $\lambda$6680 from the polar spectrum reflected in the Homunculus.
The optical depths of all of these optical \ion{He}{i} lines are substantially less than even that of the \ion{He}{i} $\lambda$20587 line. Hence, the likelihood of seeing absorptions up to $-2000~\kms$ in the \ion{He}{i} optical lines is low.

\section{Duration of the high-velocity absorption component \label{timescale}}

\begin{figure}
\resizebox{\hsize}{!}{\includegraphics{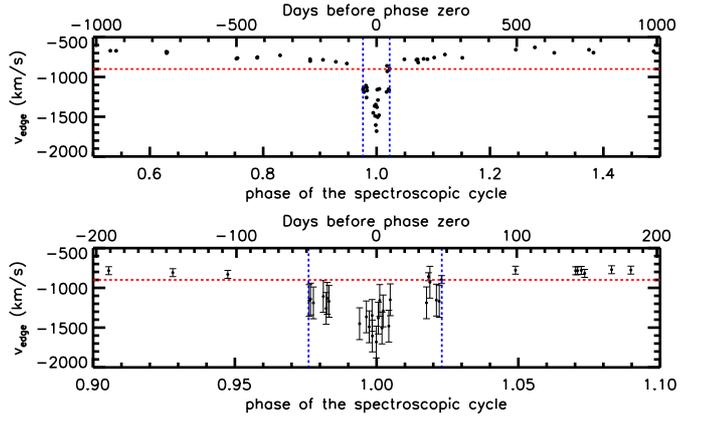}}
\caption{\label{hei10830_lna_maxvel}{\it Top:} Maximum absorption velocity ($v_{\mathrm{edge}}$) observed in the \ion{He}{i} $\lambda$10833 line profile as a function of phase of the spectroscopic cycle for the OPD/LNA dataset from 1992 to 2009, encompassing four cycles, with the data folded around $\phi=1.0$. For clarity, errorbars are presented only in the bottom panel. {\it Bottom:} Zoom-in around the spectroscopic event. To help the reader, in both panels, the horizontal red dashed line denotes the $v_{\mathrm{edge}}=-900~\kms$, while the vertical blue dashed lines represent the range where the high-velocity absorption is detected in \ion{He}{i} $\lambda$10833.
}
\end{figure}

We use the ground-based OPD/LNA data from 1992 to 2009, which are a homogeneous dataset and have a fine time-sampling, to estimate the timescale for presence of the high-velocity absorption in \ion{He}{i} $\lambda$10833. Figure \ref{hei10830_lna_maxvel} presents the maximum velocity of the \ion{He}{i} $\lambda$10833 absorption component as a function of phase, combining data from all cycles available folded around $\phi=1.0$. The timescale depends on the velocity of the material, with the highest velocities likely appearing for the briefest time intervals. However, due to the S/N of the observations and normalization errors, it is not possible to derive quantitatively the variation of the timescale as a function of velocity for the OPD dataset. For that purpose, one needs a much larger amount of high spatial and spectral resolution VLT/CRIRES data during periastron than what is presented in Section \ref{highvelcrires}. Thus, we are unable to compare the duration of the absorption at $-2000~\kms$ relatively to the absorption at $-900~\kms$, for instance. Henceforth, we opted for determining the timescale when gas with velocities more negative than $-900~\kms$ is present, since such velocity is well above the terminal speed of the wind of Eta Car~A and should give a characteristic value for the timescale of the high-velocity absorption component.  

Absorptions with velocities bluer than $-900~\kms$ are detected across $-47~\mathrm{d} \leq \Delta t \leq +46~\mathrm{d}$
($0.976 \leq \phi \leq 1.023$) while the high-velocity absorption is absent in spectra taken at $\Delta t \leq -106~\mathrm{d}$ ($\phi \leq 0.947$) and $\Delta t \geq +100~\mathrm{d}$ ($\phi \geq 1.049$). Hence, based upon the large OPD/LNA dataset, we constrain the duration of the high-velocity absorption component to be 95 to $206~\mathrm{d}$. During most of the spectroscopic cycle, the maximum absorption velocity is $\sim-650~\kms$ (Fig. \ref{hei10830_lna_maxvel}). 

Since a very limited amount of high spatial resolution observations with VLT/CRIRES and {\it HST}/STIS are available, only a lower limit on the timescale of the high-velocity absorption can be obtained, but this estimate agrees well with the value obtained above from the OPD/LNA dataset. The VLT/CRIRES data (Fig. \ref{fig1}) from the 2009.0 event is consistent with the \ion{He}{i} $\lambda$10833 high-velocity absorption ($-1000$ to $-2000~\kms$) appearing between $\phi=11.875$ (2008 May 05) and $\phi=11.991$ (2008 Dec 26), and disappearing before $\phi=12.041$ (2009 Apr 03). The UV resonance line of \ion{Si}{iv} $\lambda$1394 present in the {\it HST}/STIS data indicates that the high-velocity absorption appears between $\phi=10.984$ (2003 Jun 01) and $\phi=10.995$ (2003 Jun 22). The analysis of the ultraviolet absorption is hampered by severe blending with \ion{Fe}{ii} lines after $\phi=11.0$; consequently, the presence of the high-velocity absorption and the timescale are less accurately determined.

\section{Discussion: origin of the high-velocity material in Eta Car \label{disc}}

In the following subsections, we discuss three distinct possibilities that might explain the origin of the high-velocity 
absorption in the spectrum of Eta Car. Although many exotic scenarios could be envisioned, the high-velocity material is most likely due to either a transient episode of high-velocity material ejected by Eta Car~A (Sect. \ref{transient}), directly from the wind of Eta Car~B crossing the line-of-sight (Sect. \ref{etacarb_los}), or formed in a dense, high-velocity part of the wind-wind collision zone (Sect. \ref{wwc_los}). Since we are analyzing absorption lines, the high-velocity 
absorption region must be between the continuum source and the observer. Therefore, the source of the continuum emission is crucial for the interpretation of the origin of the high-velocity gas, and is briefly discussed below.

\subsection{Source of the continuum emission at $1.0~\mu$m \label{cont}} 

The radiative transfer models of \citet{hillier01} have shown that Eta Car~A has an extended photosphere in the near-infrared due to the presence of its dense wind (see their Fig. 8). This causes a huge amount of extended free-free and bound-free emission that is able to explain the quiescent continuum emission at $1.0~\mu$m from the inner regions, as measured by {\it HST}/STIS (see Fig. 4 of \citealt{hillier01}). The extended near-infrared continuum emitting region at $2~\mu$m has been directly resolved by interferometric measurements \citep{vb03,weigelt07}, confirming that the observed size of the $2~\mu$m continuum emission (50\% encircled-energy radius of 4.8 AU) is well reproduced by the \citet{hillier01} wind model of Eta Car~A. This implies that most, if not all, of the quiescent K-band emission is indeed due to free-free and bound-free emission from the wind of Eta Car~A, and that the contribution from hot dust to the K-band emission is negligible within 70 milli-arcseconds of Eta Car~A.  Note that the amount of emission from hot dust would be even smaller at 1.08~$\mu$m than in the K-band. Of course, hot dust is well-known to be present in Eta Car on spatial scales larger than 70 milli-arcseconds (see, e.g., \citealt{chesneau05}), and will certainly contaminate measurements using larger apertures.

Several studies have proposed that dust forms in Eta Car during periastron \citep{diego05,kashi08b}, although \citet{smith10} showed that the dust formation is cycle-dependent, occurring preferentially in the earlier documented spectroscopic events of 1981.4 and 1992.5. Significant dust formation is uncertain during the 2003.5 event \citet{smith10}, and no near-infrared photometry has been reported for the 2009.0 event. The $J$-band flux increased by ca. 25\% just before the 2003.5 event \citep{whitelock04}, which has been interpreted as due to free-free \citep{whitelock04} or hot dust emission \citep{kashi08b}. More importantly, interferometric observations in the K-band during the 2009.0 spectroscopic event, obtained simultaneously to our VLT/CRIRES measurements, do not show a significant change in the size of the K-band emitting region (Weigelt et al. 2010, in preparation), arguing against significant emission from hot dust in the inner 70 milli-arcseconds of Eta Car.

Therefore, a photospheric radius of Eta Car A at 1.08~$\mu$m of 2.2~AU is hereafter assumed as the size of the continuum emission, based on the direct interferometric measurements in the $K$-band (\citealt{vb03,weigelt07}) scaled to 1.08~$\mu$m and on the value that we computed using the CMFGEN radiative transfer model of Eta Car~A \citep{hillier01}.

\subsection{Transient fast material in the wind of Eta Car~A? \label{transient}}

If a binary companion is evoked, the periodicity might be explained as due to brief ejections of high-velocity material
by Eta Car~A triggered during each periastron passage. However, this scenario presents several difficulties, given that previous
spectroscopic observations suggested that the wind of Eta Car~A becomes roughly spherical during periastron \citep{smith03}. 
It would also imply that, during periastron, material from Eta Car~A at $\sim2000~\kms$ (instead of the usual 500--600~\kms) collides with the shock front. This increased velocity from Eta Car~A would produce a much higher
X-ray luminosity than what is currently observed. Both issues could be circumvented if the density and volume-filling factor of the $\sim2000~\kms$ transient wind are sufficiently low so as not to affect the X-ray hardness luminosity and the H$\alpha$ absorption profiles measured by \citet{smith03}. However, it is unlikely that such a thin wind would produce detectable absorption in \ion{He}{i} $\lambda$10833.

The existence of a brief high-velocity wind from Eta Car~A would be very unlikely in a single star scenario, although we cannot rule out that possibility based on our present data. In particular, a single-star scenario would have to invoke a yet unknown mechanism that would produce a periodic episode of high-velocity wind like clockwork every $2022.7 \pm 1.3$ days, as measured by \citet{damineli08_period}.

\subsection{Direct observation of the wind of Eta Car~B? \label{etacarb_los}}

The edge velocity of the high-velocity absorption component seen in \ion{He}{i} $\lambda$10833 and \ion{Si}{iv} $\lambda$$\lambda$1394, 1403 appears to approach the velocity expected of the wind of Eta Car~B, $3000~\kms$,  based upon X-ray spectroscopic modeling by \citet{pc02}. To date, Eta Car~B has not been observed directly. Could the high-velocity absorption component form directly in the wind of 
Eta Car~B? In the next two subsections, we investigate that possibility.

\subsubsection{The wind of Eta Car~B absorbs its own continuum radiation}

Such a hypothesis would correspond to the classical detection of a companion in normal massive binary systems, such as in WR+OB binaries. However, the flux of Eta Car~B is several orders of magnitude lower than that of Eta Car~A in the near-infrared continuum around the \ion{He}{i} $\lambda$10833 \citep{hillier06}. Therefore, even if the wind of Eta Car~B could produce a saturated \ion{He}{i} $\lambda$10833 absorption profile when observed in isolation, an undetectable amount of absorption ($\sim0.5-1\%$) would be seen in the combined spectrum of Eta Car~A and B.

\subsubsection{The wind of Eta Car~B absorbs the continuum radiation from Eta Car~A \label{windeclipse}}

One possible way to observe the wind of Eta Car~B, should it contain significant amounts of neutral He, would be if its \ion{He}{i} $\lambda$10833 absorption zone is extended and dense enough to absorb continuum radiation from Eta Car~A, in a ``wind-eclipse'' scenario. In order to detect the wind of Eta Car~B only during a brief period, at phases $0.976 \leq \phi \leq 1.023$ (i.e., 95\,d), the binary system would again have to be oriented in the sky with a longitude of periastron of $\omega \sim 90\degr$, since Eta Car B would need to be located between the observer and the continuum source for only a brief period around periastron passage. In addition, the material in the wind of Eta Car~B must have a sufficiently high column density of neutral He to absorb enough \ion{He}{i} $\lambda$10833 photons, but this is not predicted by the \citet{hillier06} radiative transfer model either. A much larger $\mdot$ and/or lower $\teff$ would again be needed to produce enough optical depth in the wind of Eta Car~B for \ion{He}{i} $\lambda$10833. In principle, this would argue for the presence of a Wolf-Rayet (WR) instead of an O-type star companion, since WRs have a higher wind density than O-type stars. 

\begin{figure}
\resizebox{\hsize}{!}{\includegraphics{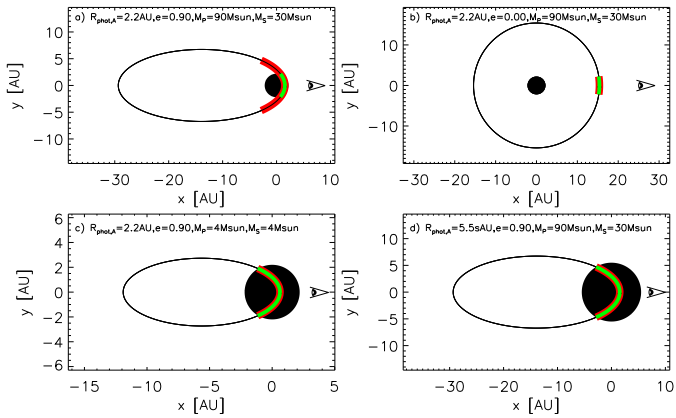}}
\caption{\label{wind_eclipse} Illustration of the pole-on orbital geometry of the Eta Car binary system for different orbital parameters, assuming $P=2022.7$~d, $\omega=90\degr$ and $i=90\degr$ (the observer is located to the right, along the x axis). The part of the orbit highlighted in green corresponds to phases when the wind of Eta Car~B could be able to absorb continuum photons from Eta Car~A (the ``wind-eclipse'' scenario). The red highlighted part of the orbit represents the phases when the high-velocity absorption component is observed in the \ion{He}{i} $\lambda$10833 line assuming, for simplification, that the phase zero of the spectroscopic cycle coincides with the periastron passage. In {\it a)} we assume masses of 90 and $30~\msun$ for Eta Car~A and B, respectively, a photospheric radius of Eta Car~A at 1.08~$\mu$m of 2.2~AU, and $e=0.9$. In {\it b)} we vary the eccentricity to $e=0$; {\it c)} assumes unrealistic masses of $4~\msun$ for both Eta Car~A and B; and {\it d)} assumes a photospheric radius of Eta Car~A at 1.08~$\mu$m of 5.5~AU. }
\end{figure}

Even in the unlikely possibility that the wind of Eta Car B could significantly absorb \ion{He}{i} $\lambda$10833 photons, the ``wind-eclipse'' scenario also fails to reproduce the observed duration of the high-velocity absorption component for the assumed orbital parameters, even allowing for significant uncertainties in these parameters.  We show in Figure \ref{wind_eclipse}a  a pole-on view of the geometry of the orbit for the ``wind-eclipse'' scenario, assuming masses of 90 and $30~\msun$ for Eta Car~A and B, respectively, orbital period of $P=2022.7$~d, semi-major axis of $a=15.4$~AU, eccentricity of $e=0.9$, and  $\omega=90\degr$. A photospheric radius at 1.08~$\mu$m  of 2.2~AU is assumed for Eta Car~A, as discussed in Section \ref{cont}. In this subsection, we assume an inclination angle of $i=90\degr$ to derive an upper limit for the timescale of the ``wind-eclipse''. For (more realistic) lower inclination angles, an even shorter timescale will be obtained.

From Figure \ref{wind_eclipse}a it is apparent that, with the parameters described above, the wind of Eta Car~B is in front of the continuum source due to Eta Car~A during a much shorter (by a factor of $\sim4$) time interval ($0.994 \leq \phi \leq 1.006$, green line) than observed (at least $0.976 \leq \phi \leq 1.023$, red line). Unrealistic values for the eccentricity ($e=0$, Fig. \ref{wind_eclipse}b), combined mass of the stars (maximum of $8~\msun$, Fig. \ref{wind_eclipse}c), or a larger photospheric radius of Eta Car A (5.5~AU, Fig. \ref{wind_eclipse}d) would be required in order for this scenario to work.
Alternatively, unrealistically large amounts of hot dust emission located conveniently behind the wind of Eta Car B, which seems unlikely and is not supported by the available observations, would be required to make this scenario to work.

An additional issue regards the amount of absorption observed in \ion{He}{i} $\lambda$10833. Since the observed high-velocity absorption spans velocities from $-800$ up to $-2000~\kms$, the absorption necessarily has to occur in the acceleration zone of Eta Car~B, before the wind reaches the supposed terminal velocity of $3000~\kms$. Based on the CMFGEN radiative transfer model of the wind of Eta Car~B, the acceleration zone of the wind of Eta Car~B is relatively compact compared to the size of the photosphere of Eta Car~A. Consequently, if the wind of Eta Car~B is to absorb continuum photons from Eta Car~A, the coverage of such a continuum source would be very small.

We conclude that it is unlikely that the high-velocity absorption component originates in the wind of Eta Car~B.

\subsection{High-velocity, shocked material from the wind-wind collision zone \label{wwc_los}}

The high-velocity absorption may originate in shocked material from the wind-wind collision zone that crosses our line-of-sight to Eta Car briefly across periastron. Such a hypothesis has been already suggested  by \citet{damineli08_multi} to explain the behavior of \ion{He}{i} $\lambda$10833, assuming $\omega=270\degr$, and by \citet{kashi09b}, who instead derived $\omega=90\degr$ from their analytical modeling. 

Here we use the aid of three-dimensional (3-D) hydrodynamical simulations of the Eta Car binary system to investigate where high-velocity material can be found in the system, and at which epochs. We qualitatively compare our observations with the 3-D hydrodynamical simulations with the goal of constraining which orbital orientation is more consistent with our data. Specifically, we aim at obtaining for which inclination angles and orbital orientations there is high-velocity gas, with velocities between $-800$ and $-2000~\kms$, in our line-of-sight to Eta Car~A across periastron, and at which distances that gas is located. The presence of high-velocity gas in our-line-of-sight is a necessary, but not sufficient condition for the presence of high-velocity absorption in a given spectral line. The amount of absorption of a spectral line will depend on the population of the lower energy level related to that line, which is regulated by the ionization stage of the gas. The 3-D hydrodynamical simulations allow us to analyze the hydrodynamics of the material flowing from the wind-wind collision zone with a much higher precision than in the analytical models of \citet{kashi09b}, in particular for epochs across periastron, when the high-velocity material has been detected. For these epochs, the structure of the wind-wind collision zone is severely distorted, with the arms of the bowshock being wrapped around Eta Car~A \citep{okazaki08,parkin09}.

\begin{figure}
\resizebox{\hsize}{!}{\includegraphics{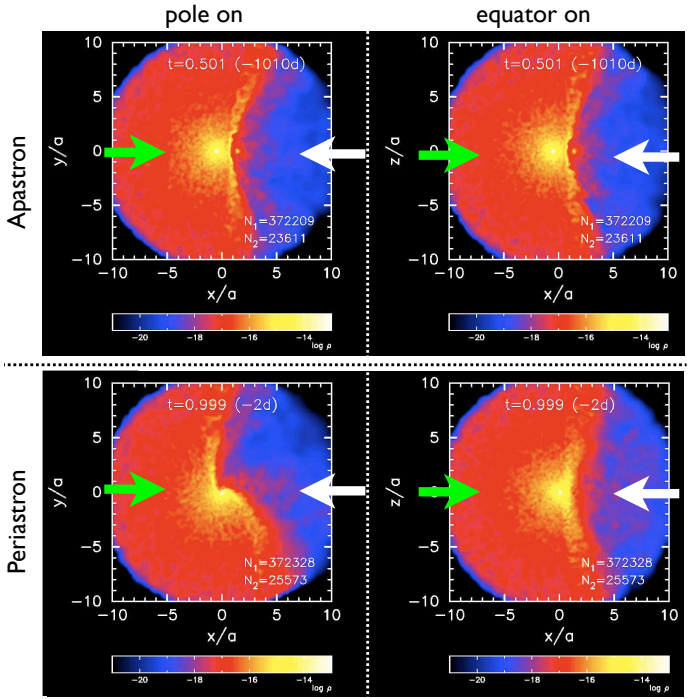}}
\caption{\label{sph1} Illustration of the wind-wind collision and orbital geometry from different vantage points of the Eta Car system, based on snapshots from the 3-D hydrodynamic simulations similar to those of \citet{okazaki08}, but including adiabatic cooling. The color scale refers to the 2-D density structure, and the spatial scales are in units of the semi-major axis $a$ of the orbit ($a=15 AU$). The top row shows the system configuration during apastron, while the bottom row refers to periastron. The wind from Eta Car~A, the wind-wind collision zone, and the wind from Eta Car~B are seen from a pole-on view (left panels) and from the equator (right). The white arrow corresponds to $\omega=270\degr$, while the green arrow corresponds to $\omega=90\degr$. }
\end{figure}

We use 3-D simulations which are similar to and have the same parameters as those presented in \citet{okazaki08}, with the exception that adiabatic cooling has been included\footnote{3-D simulations from \citet{parkin09} show similar hydrodynamics as in the \citet{okazaki08} simulations.}. The simulations assume the following parameters: for Eta Car~A, a mass of $90~\msun$, radius of $90~\rsun$, mass-loss rate of $2.5\times10^{-4}~\msunyr$, and wind terminal velocity of $500~\kms$; for Eta Car~B, a mass of $30~\msun$, radius of $30~\rsun$, mass-loss rate of $10^{-5}~\msunyr$, and wind terminal velocity of $3000~\kms$; orbital period of $P=2024~\mathrm{d}$, eccentricity of $e=0.9$, and semi-major axis of $a=15.4~\mathrm{AU}$. We refer the reader to \citet{okazaki08} for further details. Figure \ref{sph1} presents 2-D slices of the wind-wind collision zone geometry based on the 3-D hydrodynamical simulations of Eta Car to help better visualize the geometry of the binary system.

Time-dependent, multi-dimensional radiative transfer modeling of the outflowing material from the wind-wind collision zone is needed to obtain the physical conditions of the high-velocity gas. That is well beyond the scope of this paper, and we will defer such analysis for future work. Since we did not compute a multi-dimensional radiative transfer model, we are able to obtain only {\it total} column densities from the SPH simulations, but not the column density of the population of the lower energy level of \ion{He}{i} $\lambda$10833 (2s\,\element[][3]{S}). The total column density computed here (hereafter referred to as ``column density") provides an upper limit for the amount of absorption. Thus, a low column density at higher velocities implies that no high-velocity absorption will be present. However, a high column density does not necessarily mean that a strong absorption line will be detected. In particular, the current 3-D simulations that we use do not account for radiative cooling, which makes it difficult to estimate the actual temperature and ionization structure of the high-velocity material in the wind-wind collision zone. In this Section, we assume that this material is able to efficiently cool and to produce the observed high-velocity absorption if the column density and velocity in the line-of-sight to Eta Car~A are high enough. 

Hereafter, for simplicity we assume that the phase zero of the spectroscopic cycle (derived from the disappearance of the narrow emission component of \ion{He}{i} $\lambda$6678) coincides with phase zero of the orbital cycle (periastron passage). Note that in a highly-eccentric binary system like Eta Car the two values are not expected to be shifted by more than a few weeks. Such time shift would only cause a small change of $10\degr-20\degr$ in the best value of $\omega$, which will not affect our conclusions.

As discussed in Section \ref{cont}, the main source of continuum radiation at $1.0~\mu$m is the free-free and bound-free emission from the wind of Eta Car~A, and we analyze the physical conditions of the gas between the observer and Eta Car~A.

\subsubsection{Orbital plane is aligned with the Homunculus equatorial plane}

\begin{figure*}

\resizebox{0.33\hsize}{!}{\includegraphics{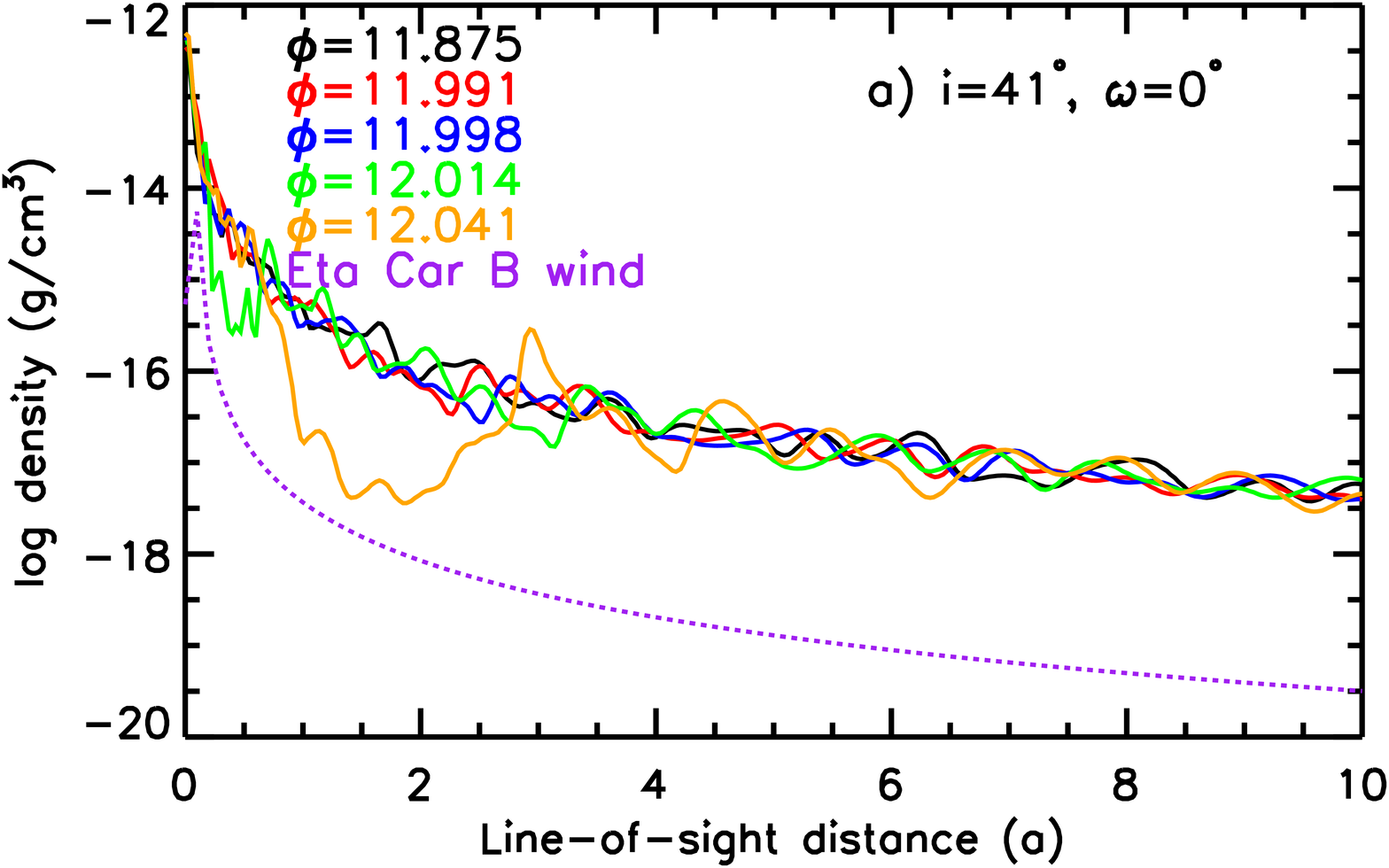}}
\resizebox{0.33\hsize}{!}{\includegraphics{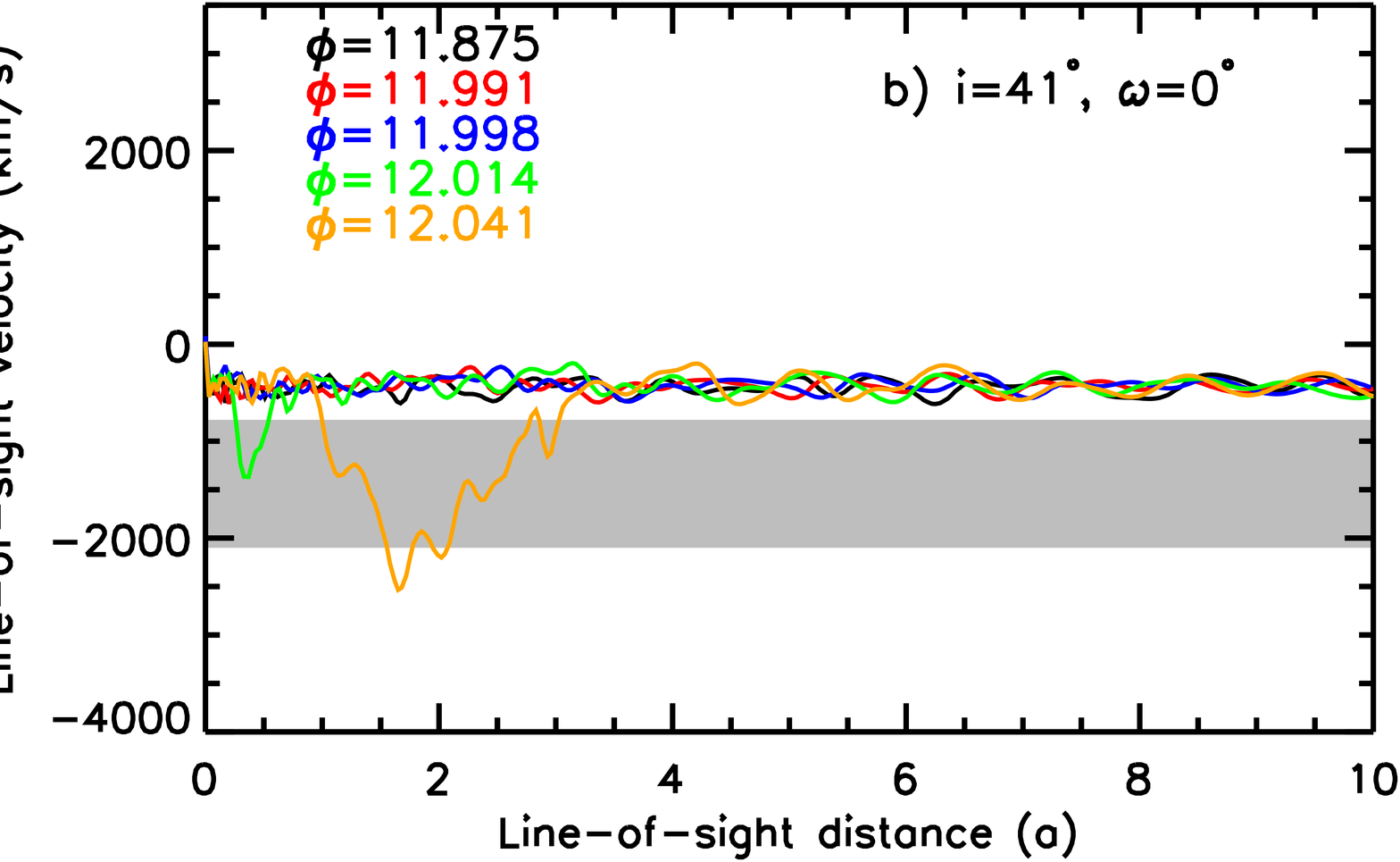}}
\resizebox{0.33\hsize}{!}{\includegraphics{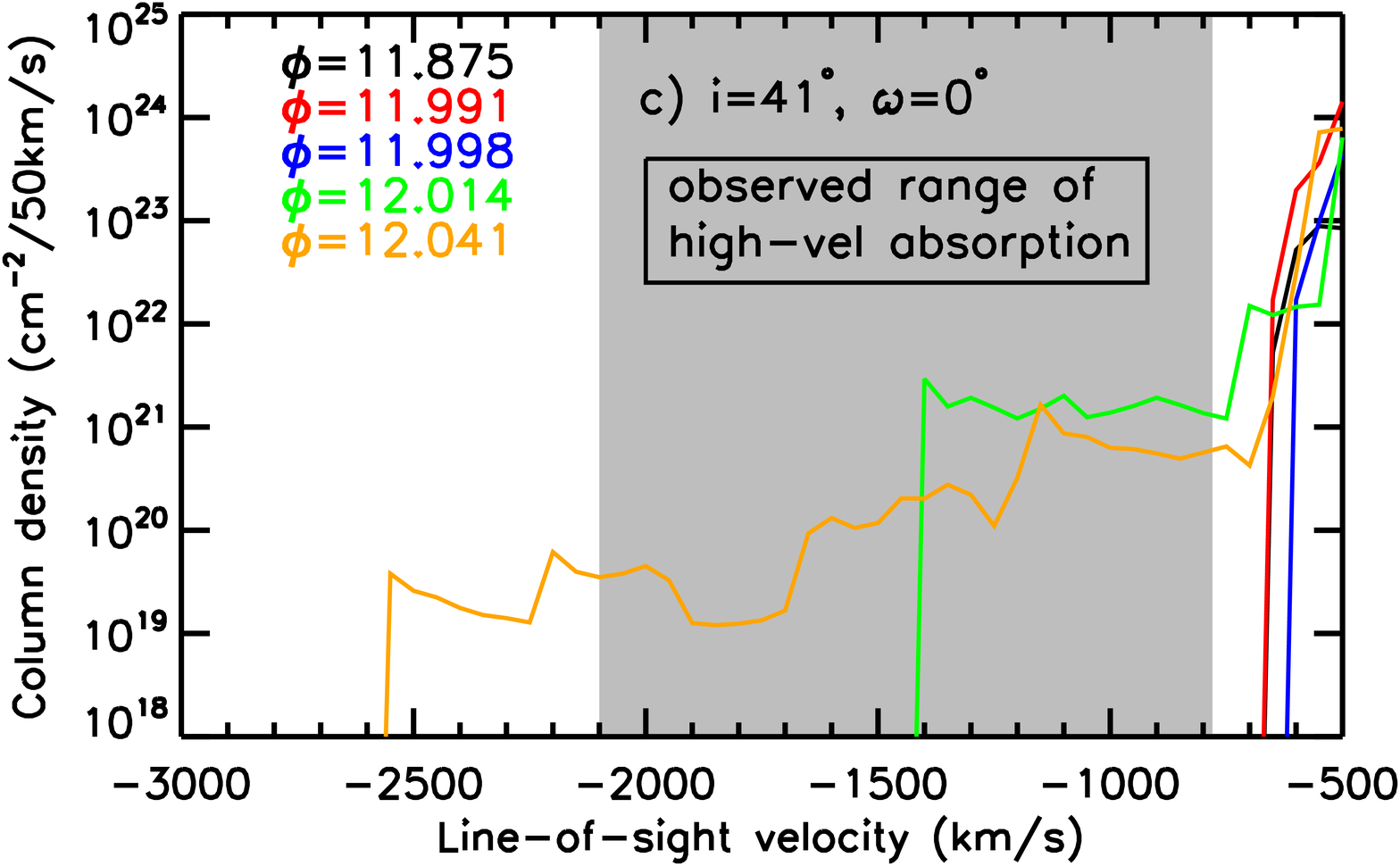}}\\

\resizebox{0.33\hsize}{!}{\includegraphics{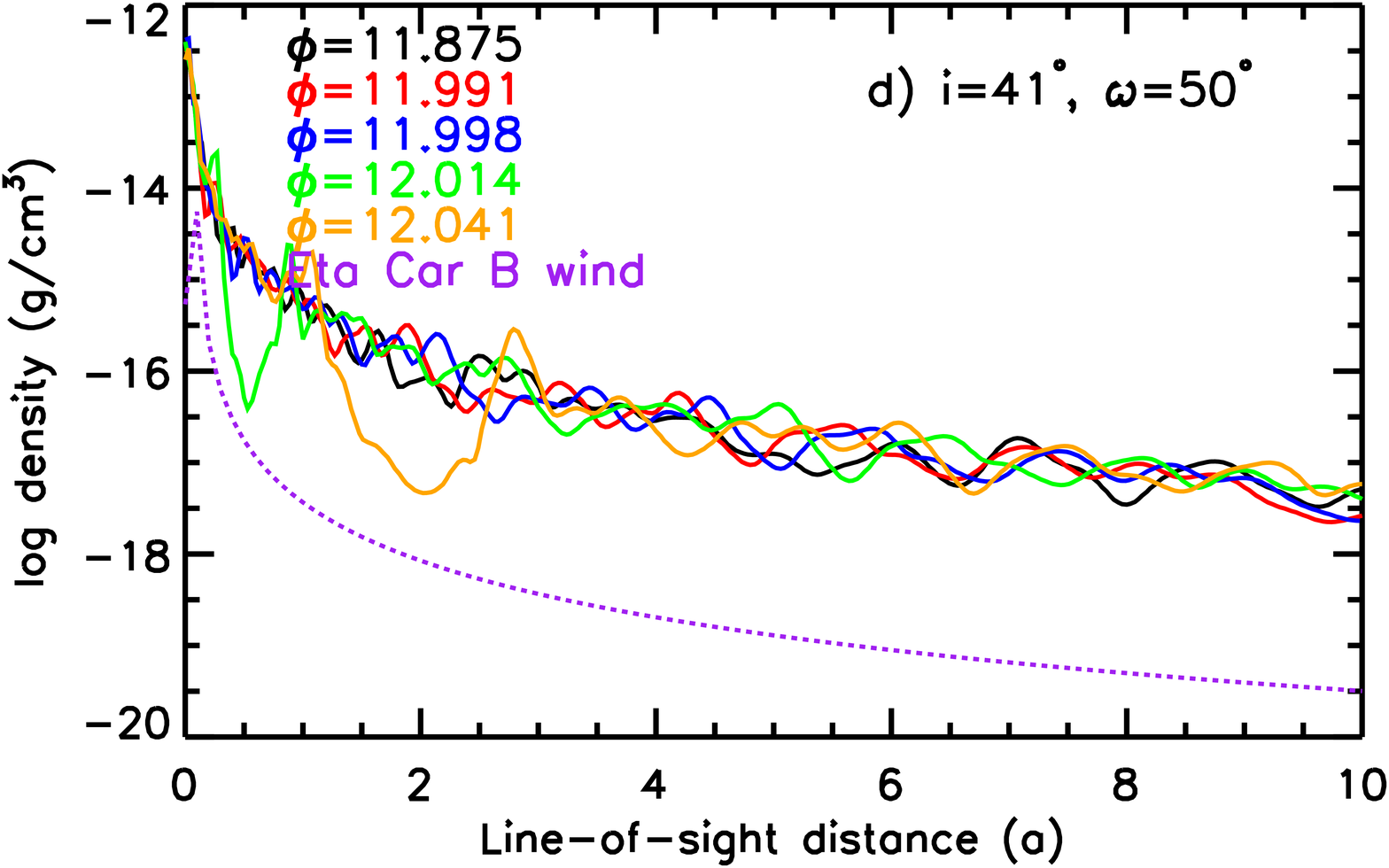}}
\resizebox{0.33\hsize}{!}{\includegraphics{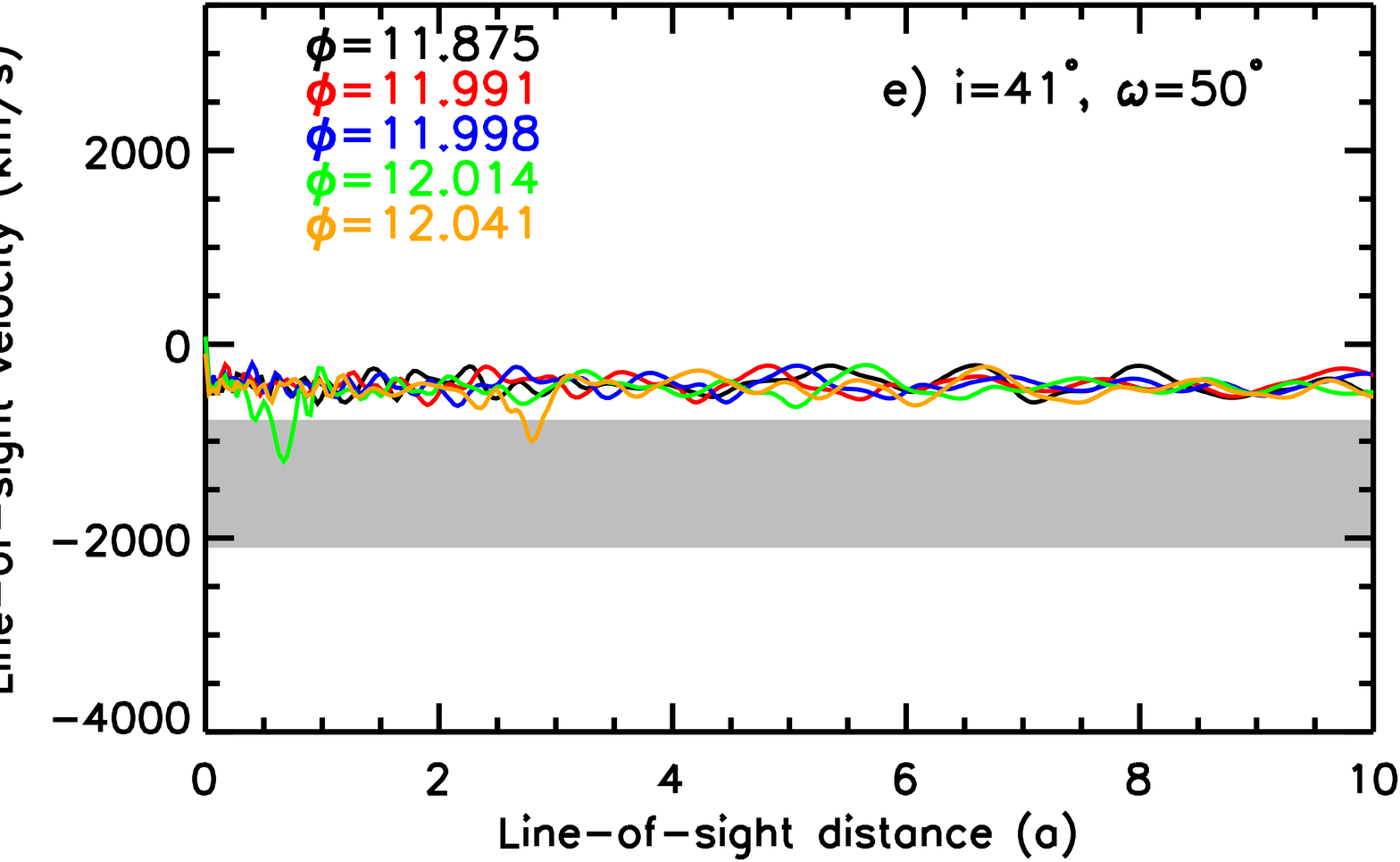}}
\resizebox{0.33\hsize}{!}{\includegraphics{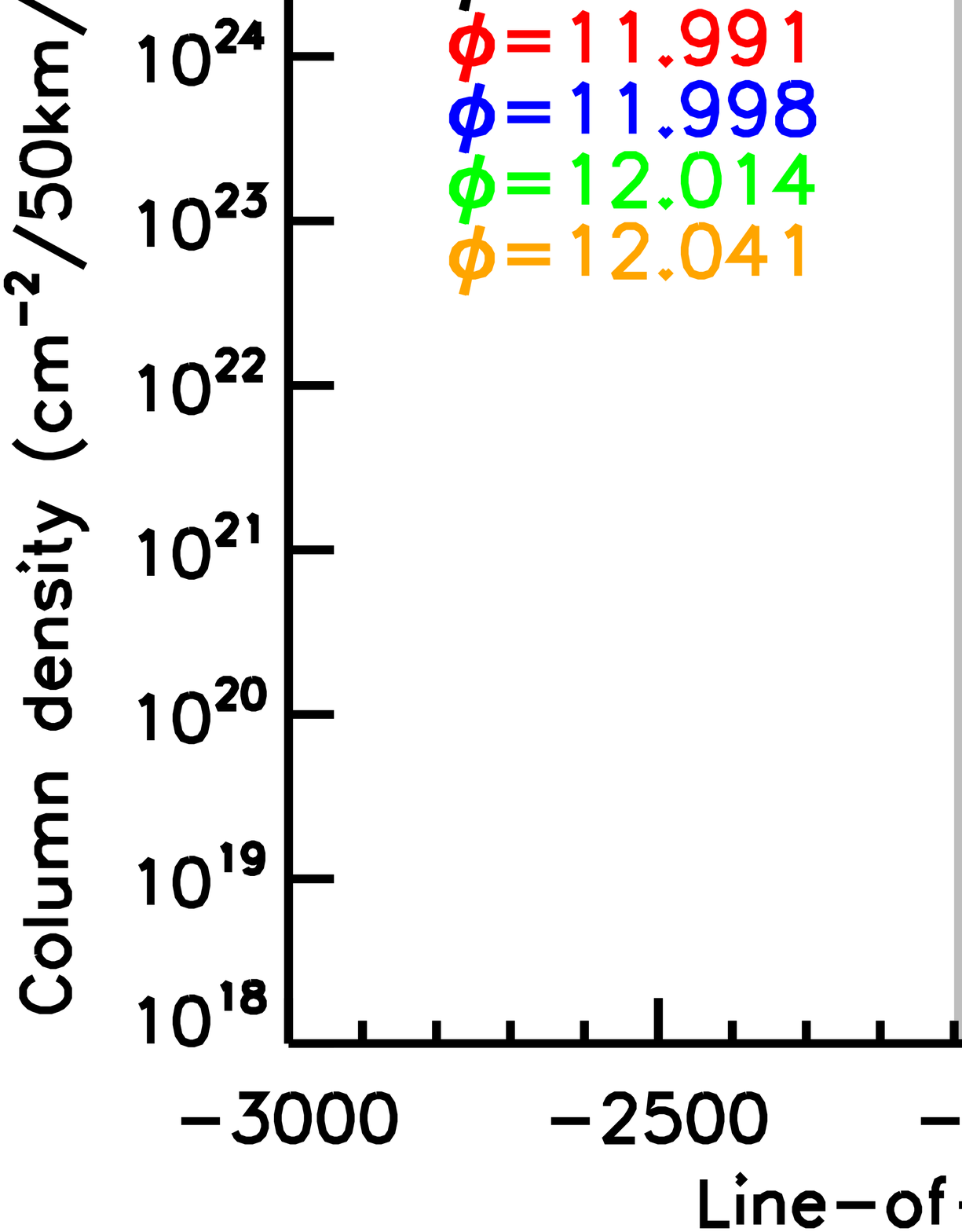}}\\

\resizebox{0.33\hsize}{!}{\includegraphics{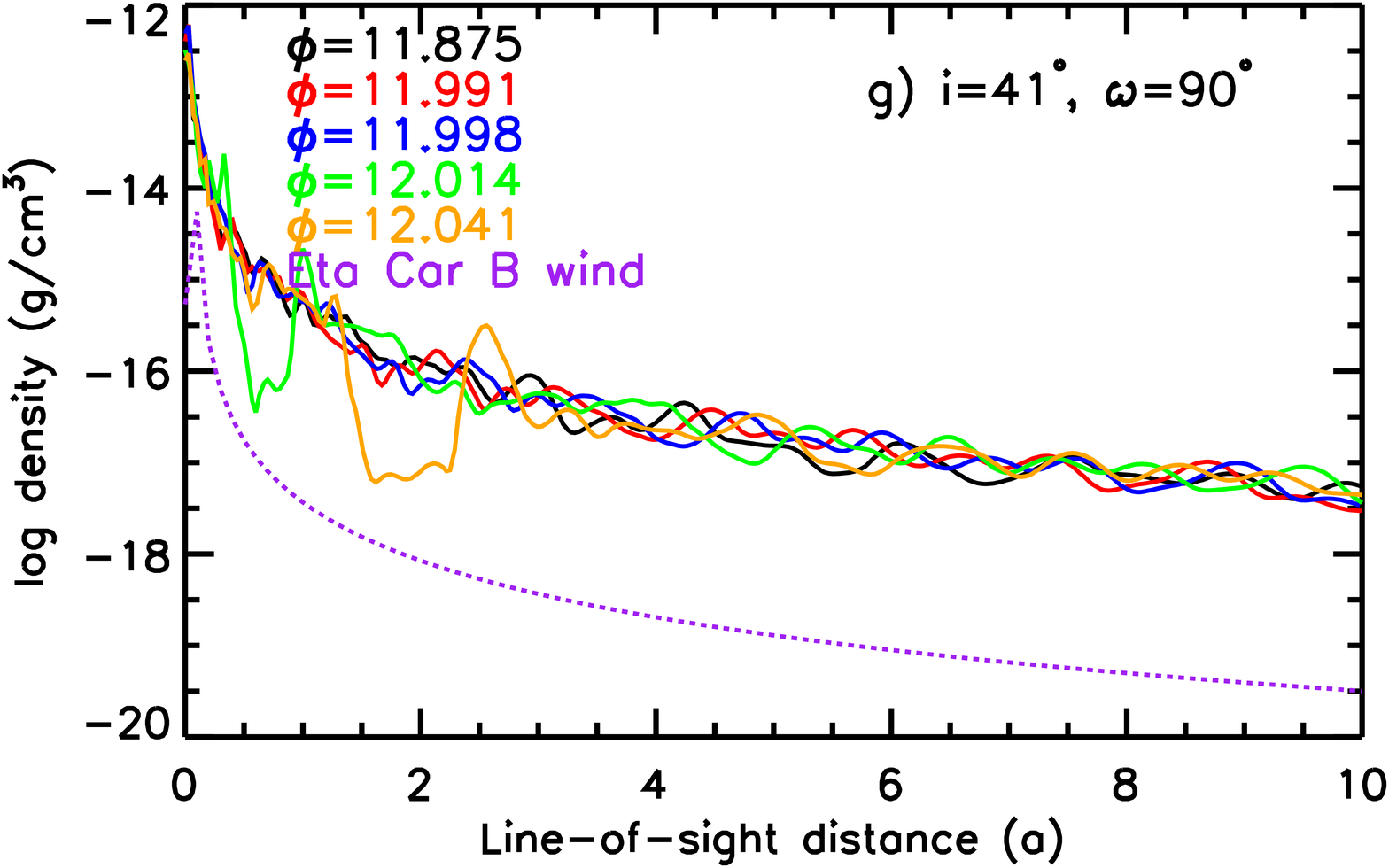}}
\resizebox{0.33\hsize}{!}{\includegraphics{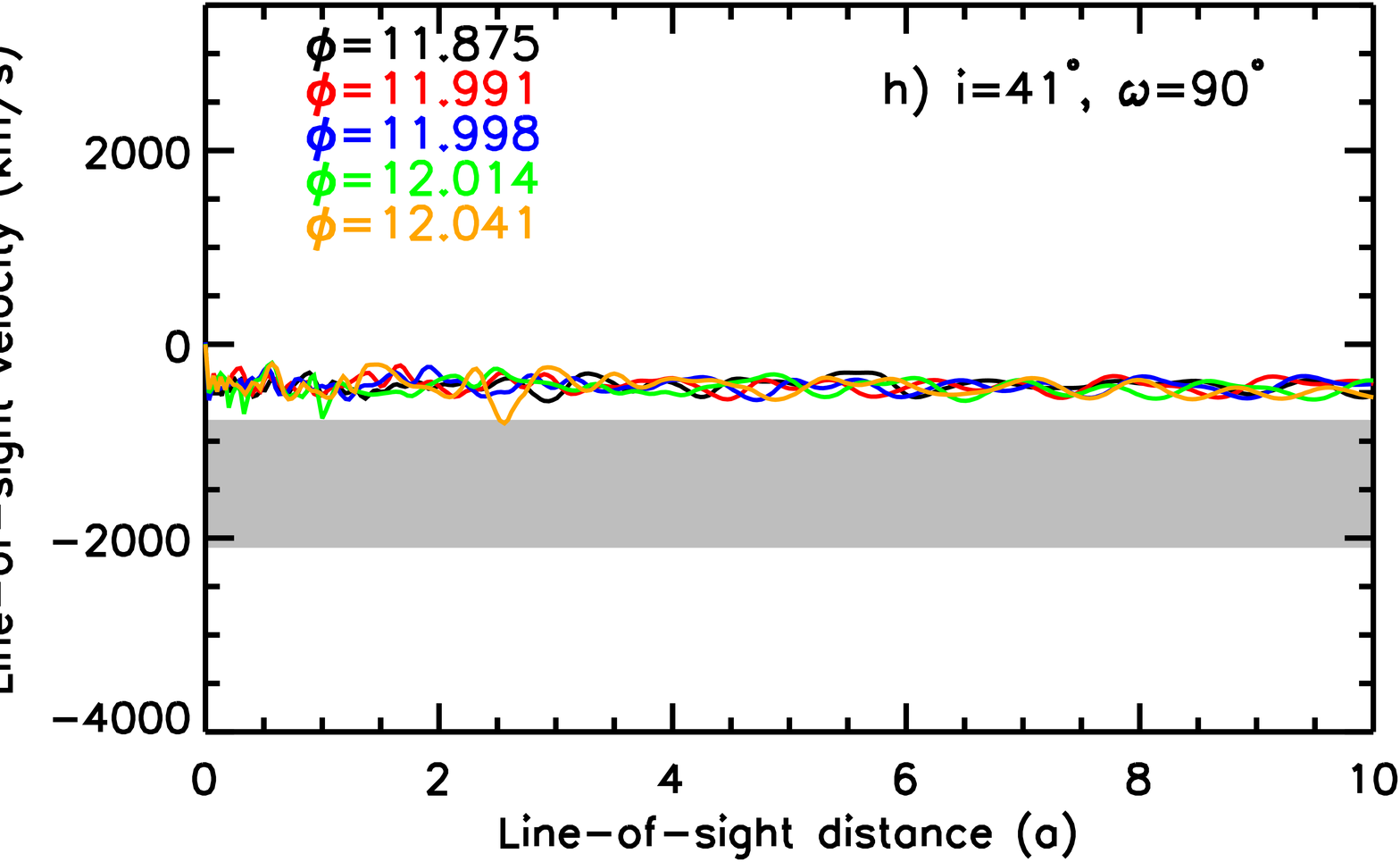}}
\resizebox{0.33\hsize}{!}{\includegraphics{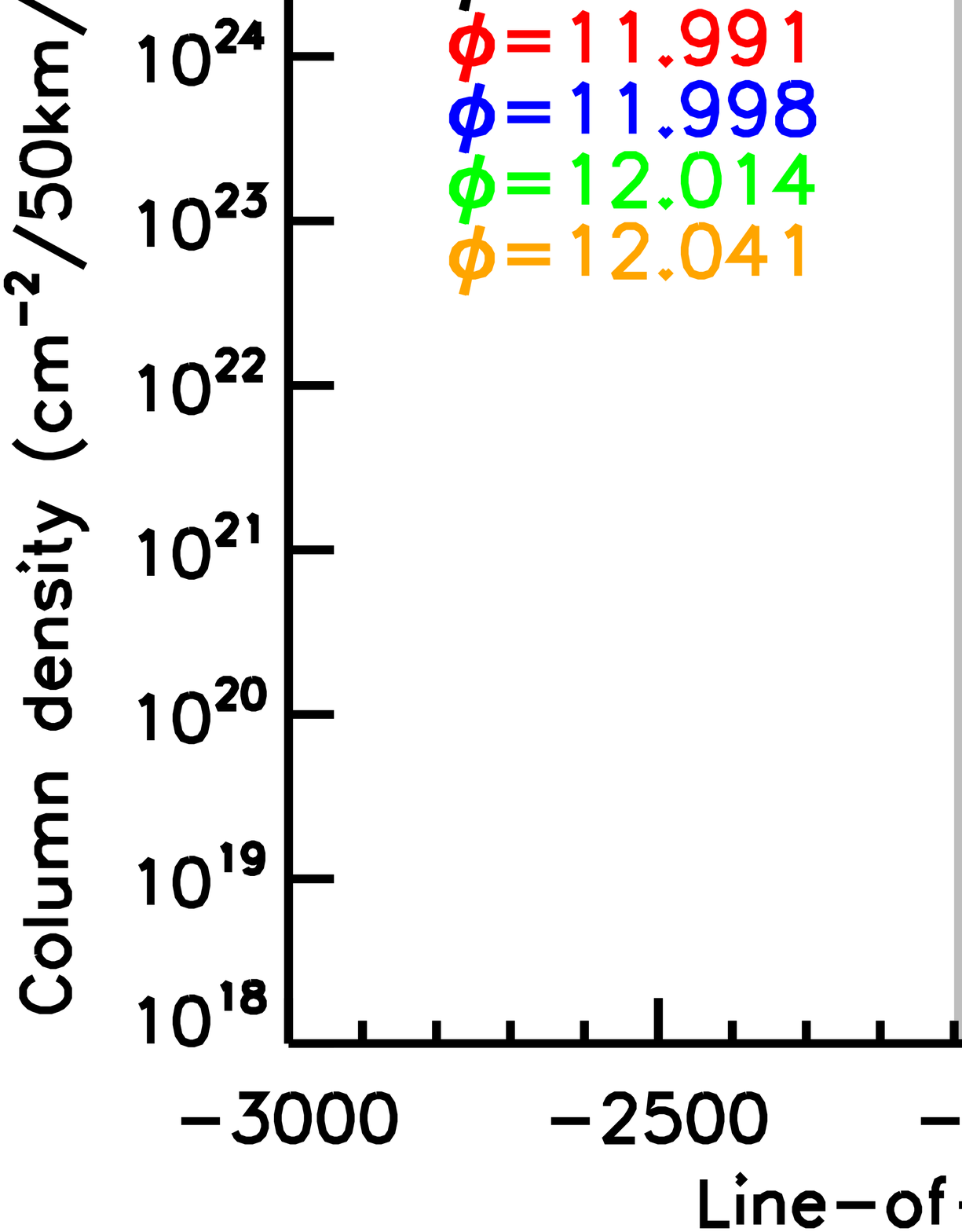}}\\

\resizebox{0.33\hsize}{!}{\includegraphics{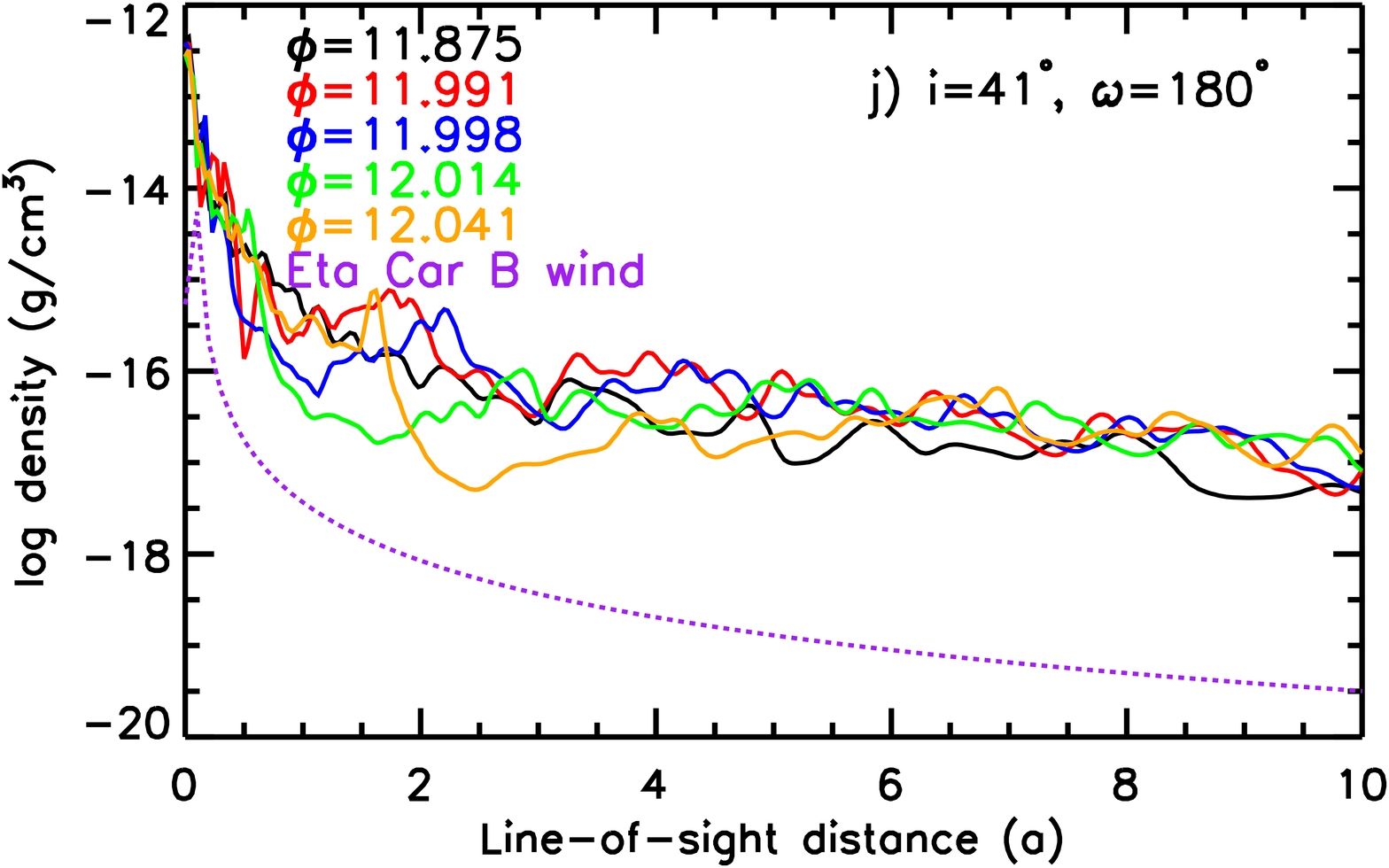}}
\resizebox{0.33\hsize}{!}{\includegraphics{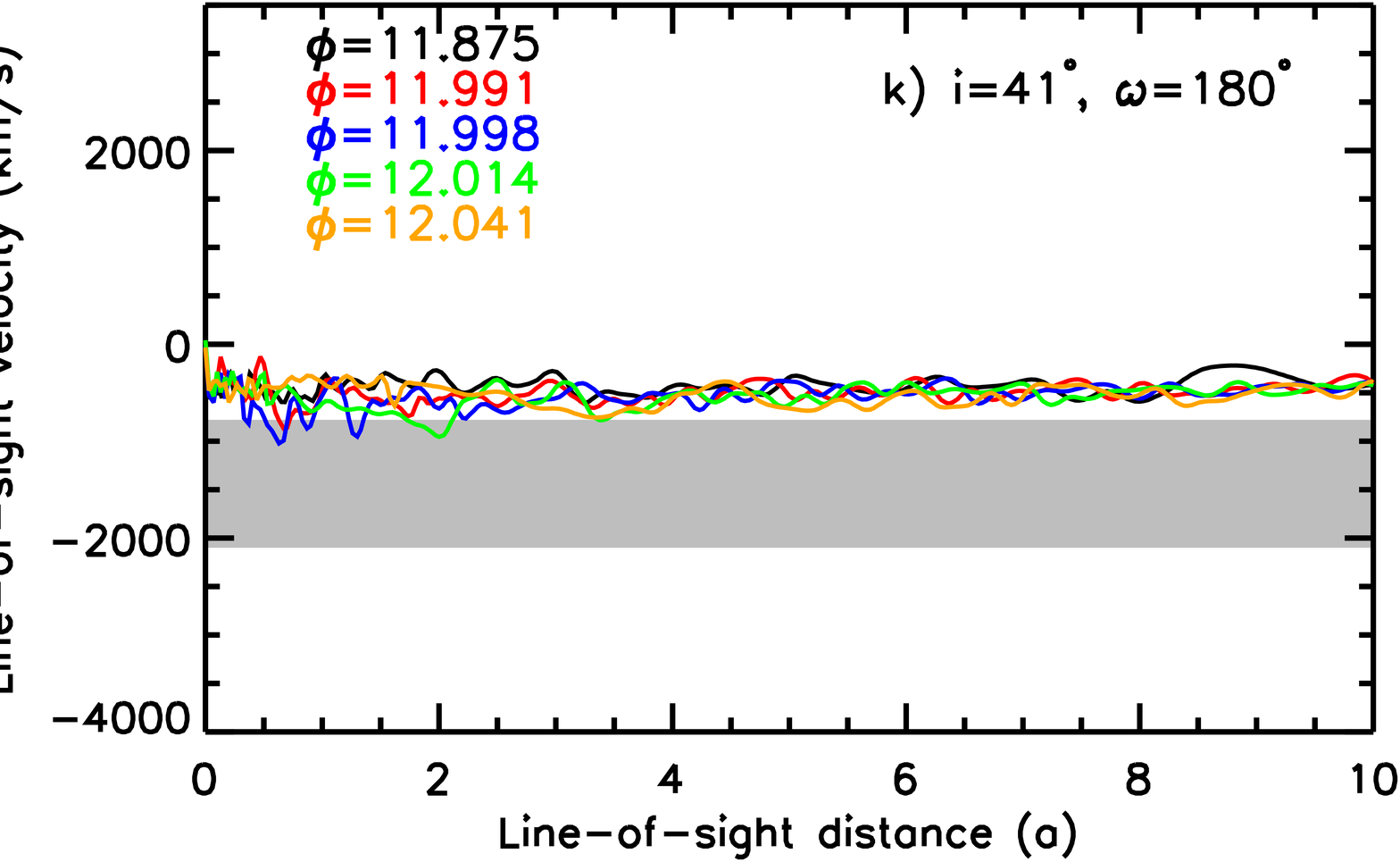}}
\resizebox{0.33\hsize}{!}{\includegraphics{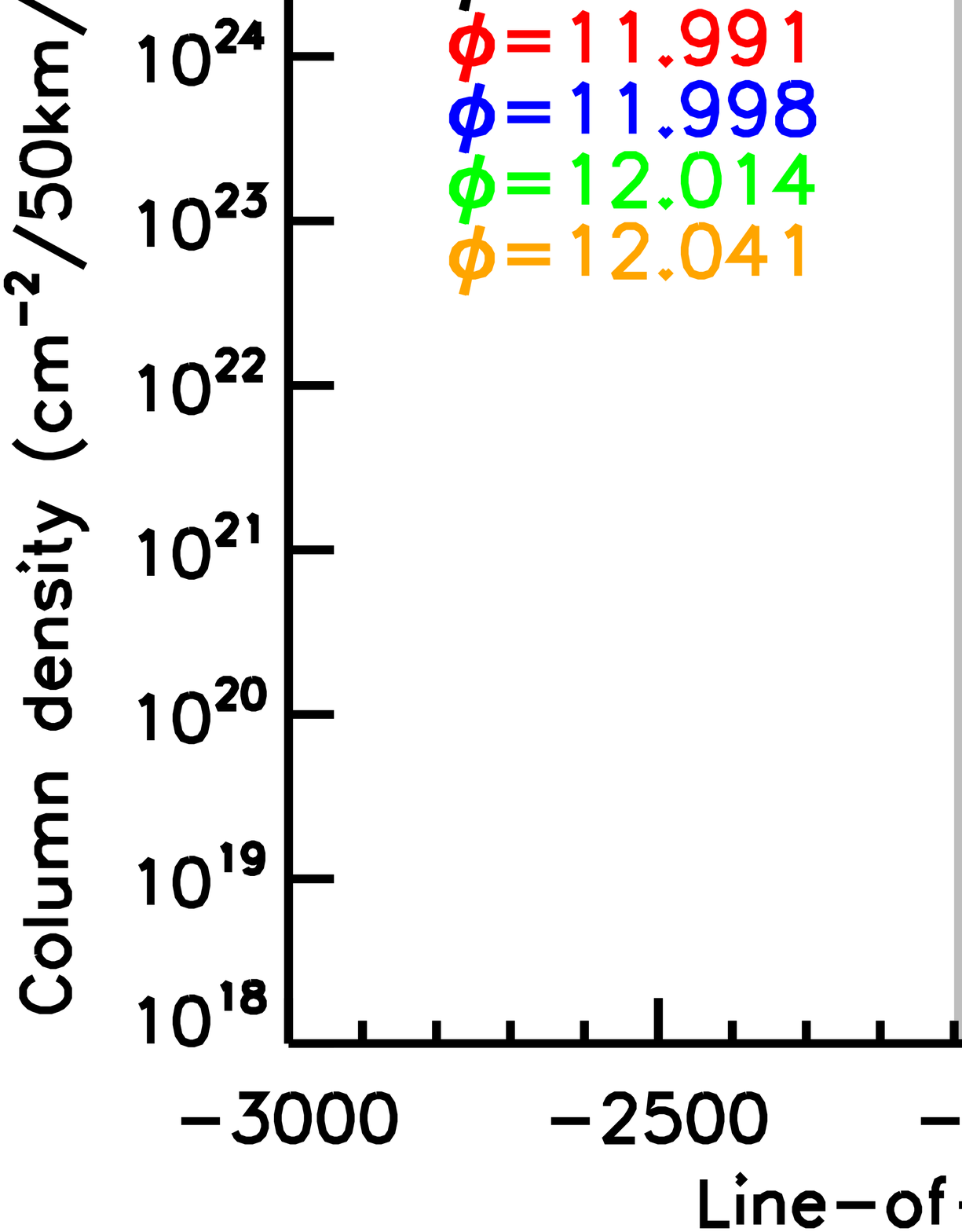}}\\

\resizebox{0.33\hsize}{!}{\includegraphics{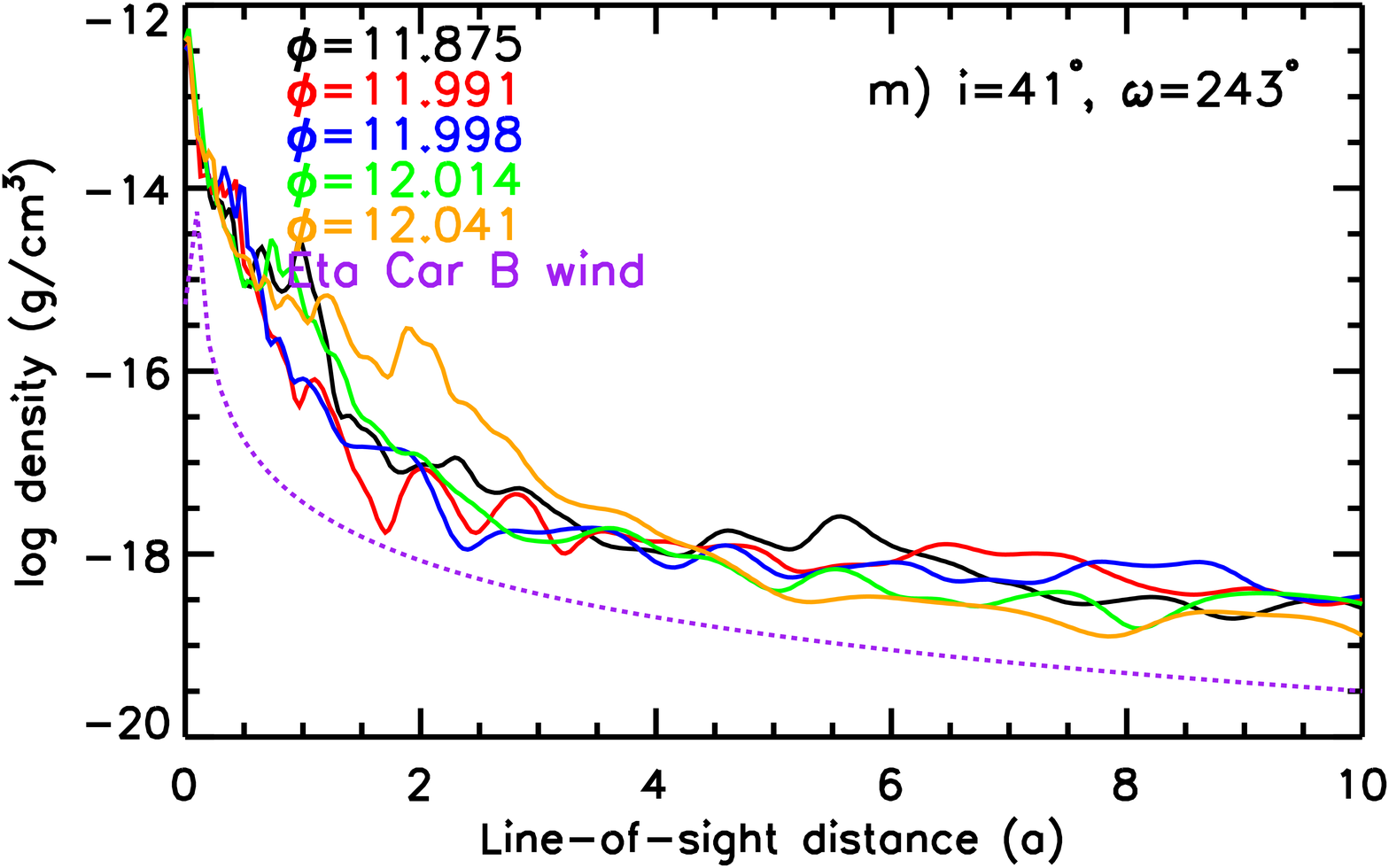}}
\resizebox{0.33\hsize}{!}{\includegraphics{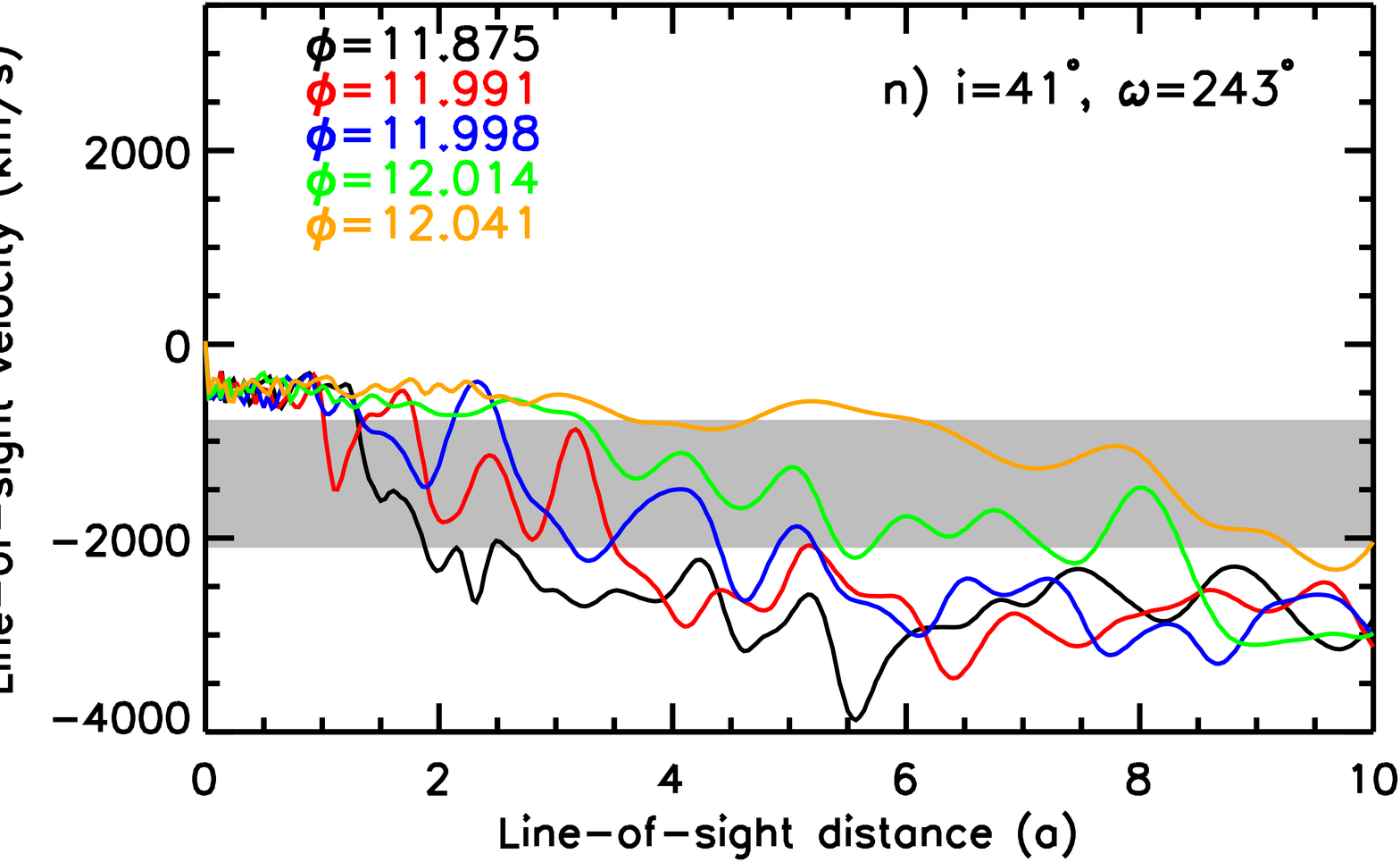}}
\resizebox{0.33\hsize}{!}{\includegraphics{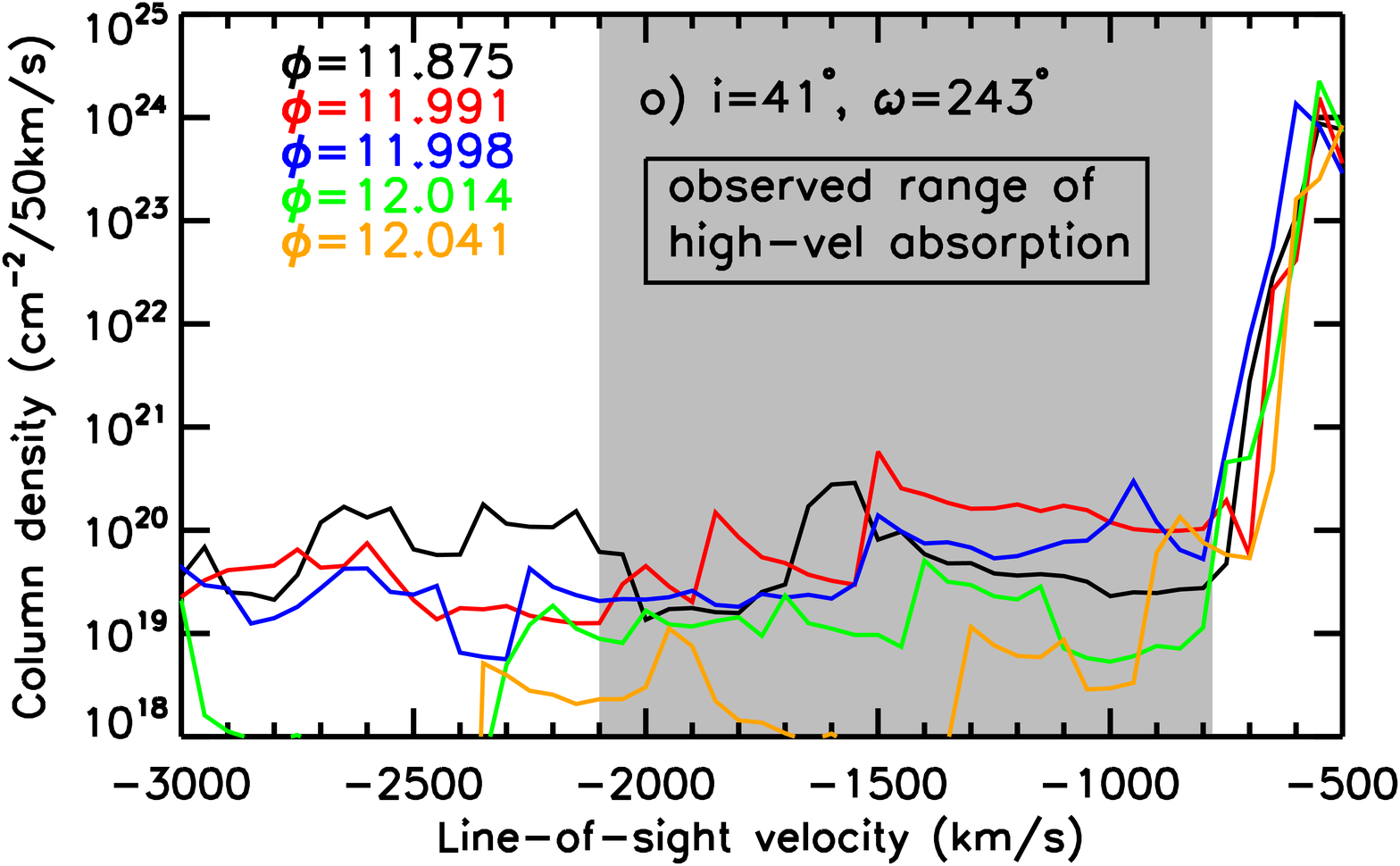}}\\

\resizebox{0.33\hsize}{!}{\includegraphics{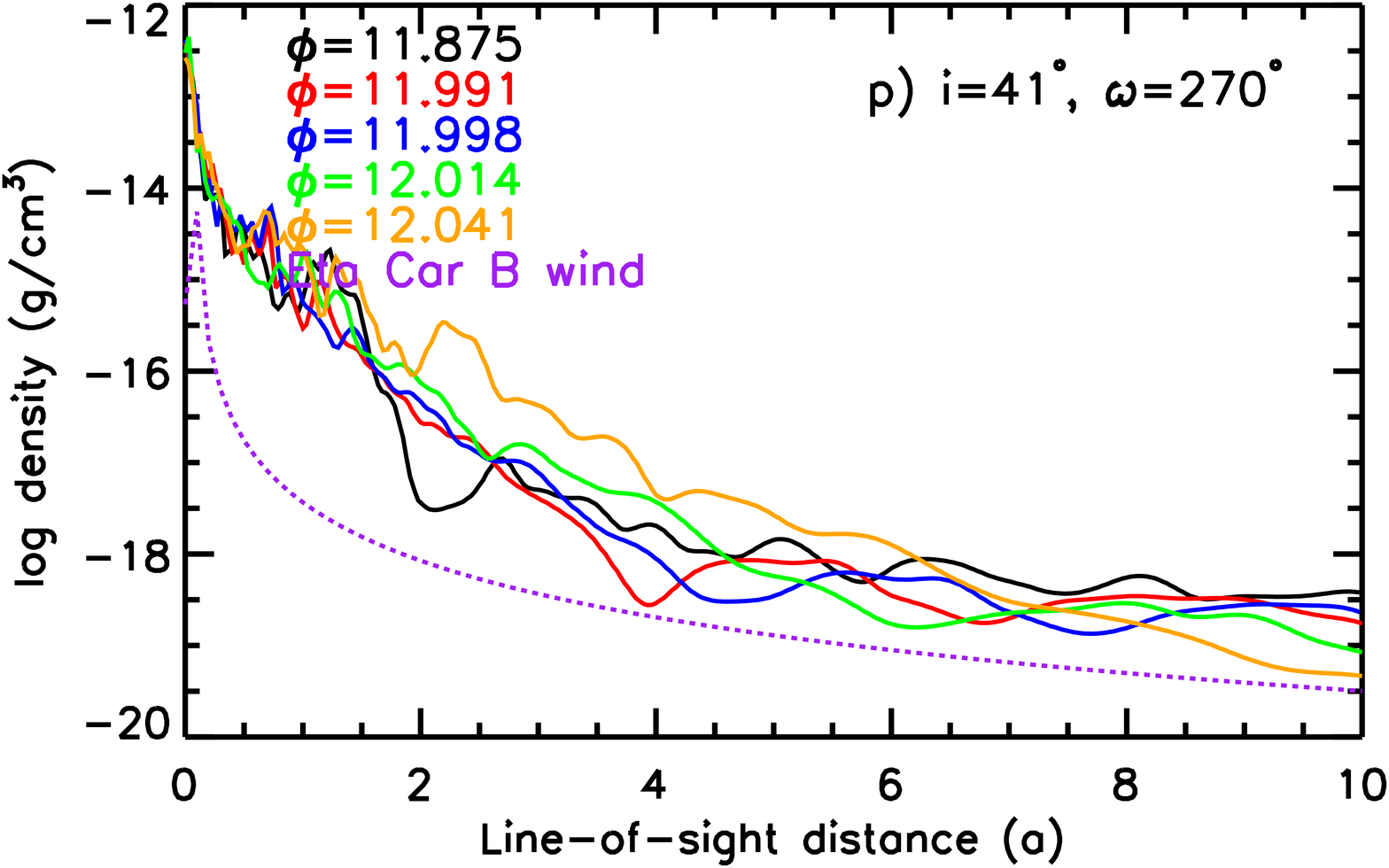}}
\resizebox{0.33\hsize}{!}{\includegraphics{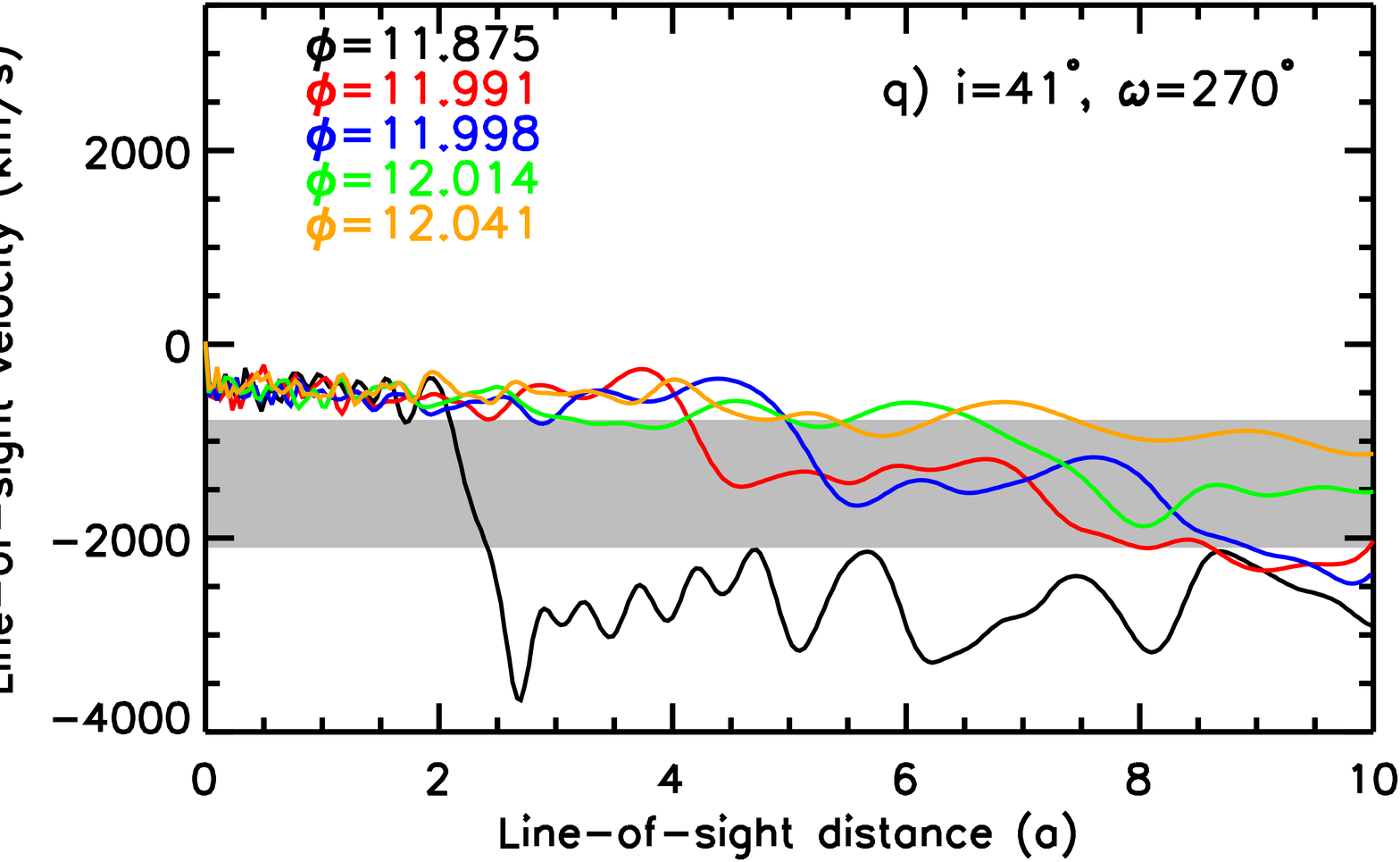}}
\resizebox{0.33\hsize}{!}{\includegraphics{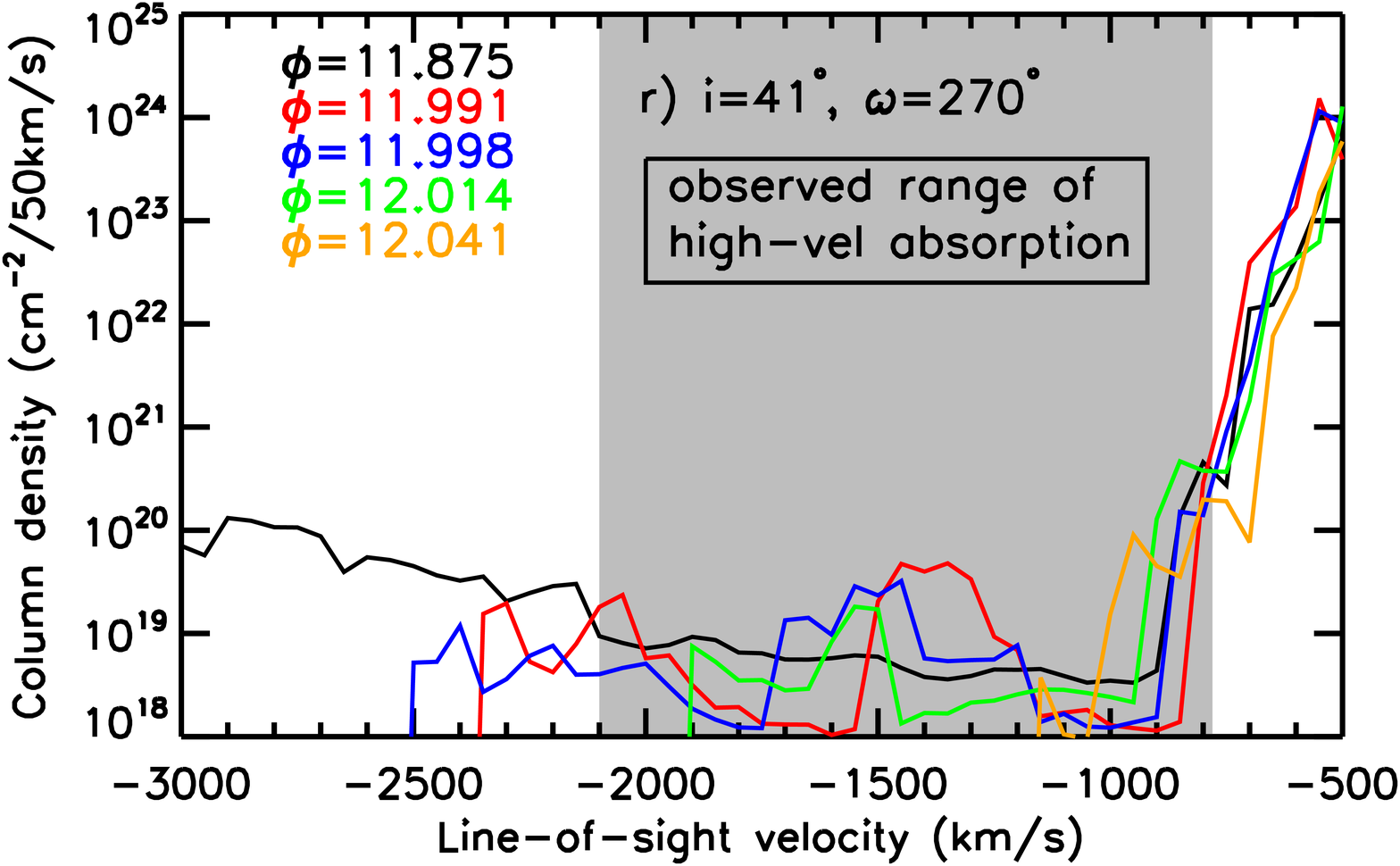}}\\
\vspace{-0.5cm}
\caption{\label{sph_tom1}  {\it Left:} Log density of material as a function of line-of-sight distance to Eta Car~A (in units of the semi-major axis $a=15.4$~AU) for selected orbital phases. {\it Middle:} Line-of-sight velocity of material along the same assumed line-of-sight to Eta Car~A. {\it Right:} Log of the column density (in units of cm$^2$ per 50~\kms bin) along the same assumed line-of-sight to the primary star. All panels assume $i=41\degr$ and, from top to bottom, $\omega=0\degr$, $50\degr$, $90\degr$, $180\degr$, $243\degr$, and $270\degr$, respectively. The grey region corresponds to the observed range of high-velocity absorption.}
\end{figure*}

In this Section we investigate the hydrodynamics of the material in line-of-sight to Eta Car~A viewing the system from different $\omega$, assuming that the orbital plane has is aligned with the Homunculus equatorial plane and, thus, the inclination of the orbit is $i=41\degr$ \citep{smith06}. Since we are analyzing an absorption line, Figure \ref{sph_tom1} presents the one-dimensional density, velocity, and column density structure of the gas for different lines-of-sight to Eta Car~A at orbital phases when VLT/CRIRES observations were available.

For $\omega = 90\degr$, the line-of-sight to Eta Car~A contains only high-density material from the wind of Eta Car~A, with $v\sim -500~\kms$, before periastron (Fig. \ref{sph_tom1}g--h). After periastron, a patch of shocked material crosses the line-of-sight, but it does not possess material faster than $-800~\kms$ and, as a consequence, produces a negligible column density of high-velocity material in the range of $-800$ to $-2000~\kms$ (Fig. \ref{sph_tom1}i). For a lower value of $\omega = 50\degr$, part of the wind of Eta Car~B crosses our line-of-sight to Eta Car~A after periastron, producing a considerable amount of column density of high-velocity material up to $-1300~\kms$ (Fig. \ref{sph_tom1}d--f). This is exactly the ``wind-eclipse'' scenario described above (Sect. \ref{windeclipse}) and, as one can see in Figure \ref{sph_tom1}f, the duration of the high-column density of high-velocity material in our line-of-sight to Eta Car~A is very brief. Therefore, based on the hydrodynamics predicted from detailed 3-D hydrodynamical simulations, we can rule out that the Eta Car system has orbital orientations around $\omega\sim 50\degr-90\degr$ if the high-velocity gas originates in shocked material from the wind-wind collision zone. We also obtained that orbital orientations with $\omega=0\degr$ (Fig. \ref{sph_tom1}a--c) and $\omega=180\degr$ (Fig. \ref{sph_tom1}j--l) do not provide high-velocity gas with sufficient column density in the line-of-sight to Eta Car~A during the observed duration of the high-velocity absorption. 

We find that the 3-D hydrodynamical simulations require an orbital orientation with $\omega$ in the range 240\degr--270$\degr$ to produce high-velocity gas with enough density in the line-of-sight towards Eta Car~A (Figs. \ref{sph_tom1}m--o, \ref{sph_tom1}p--r). For these orbital orientations, the density of the high-velocity gas is approximately an order of magnitude higher than expected from the wind of Eta Car~B, while the velocities are significantly lower compared to the value expected for the wind of Eta Car~B. These physical conditions show that the dense, high-velocity gas which is in our line-of-sight to Eta Car~A originates from shocked material from the wind-wind collision zone. For $\omega = 243\degr$, the high-velocity material is located between 1 and 3 semi-major axis (15 to 45 AU; Figs. \ref{sph_tom1}m--o). The radial dependence of the density follows roughly an $r^{-2}$ law, which closely resembles that of a stellar wind and might explain why the observed high-velocity absorption profiles are rather smooth and broad.

For $\omega = 270\degr$, the increase in column density around $\phi=11.991-11.998$ occurs at a velocity region around $-1200$ to $-1800~\kms$, and the column density of the gas with velocities $-800$ to $-1200~\kms$ actually decreases before periastron and increases after periastron (Fig. \ref{sph_tom1}p--r). Qualitatively, we would expect the opposite behavior to explain the increase in high-velocity absorption. However, ionization and radiative transfer effects might also play a role in determining the amount of absorption.

\begin{figure}
\resizebox{0.93\hsize}{!}{\includegraphics{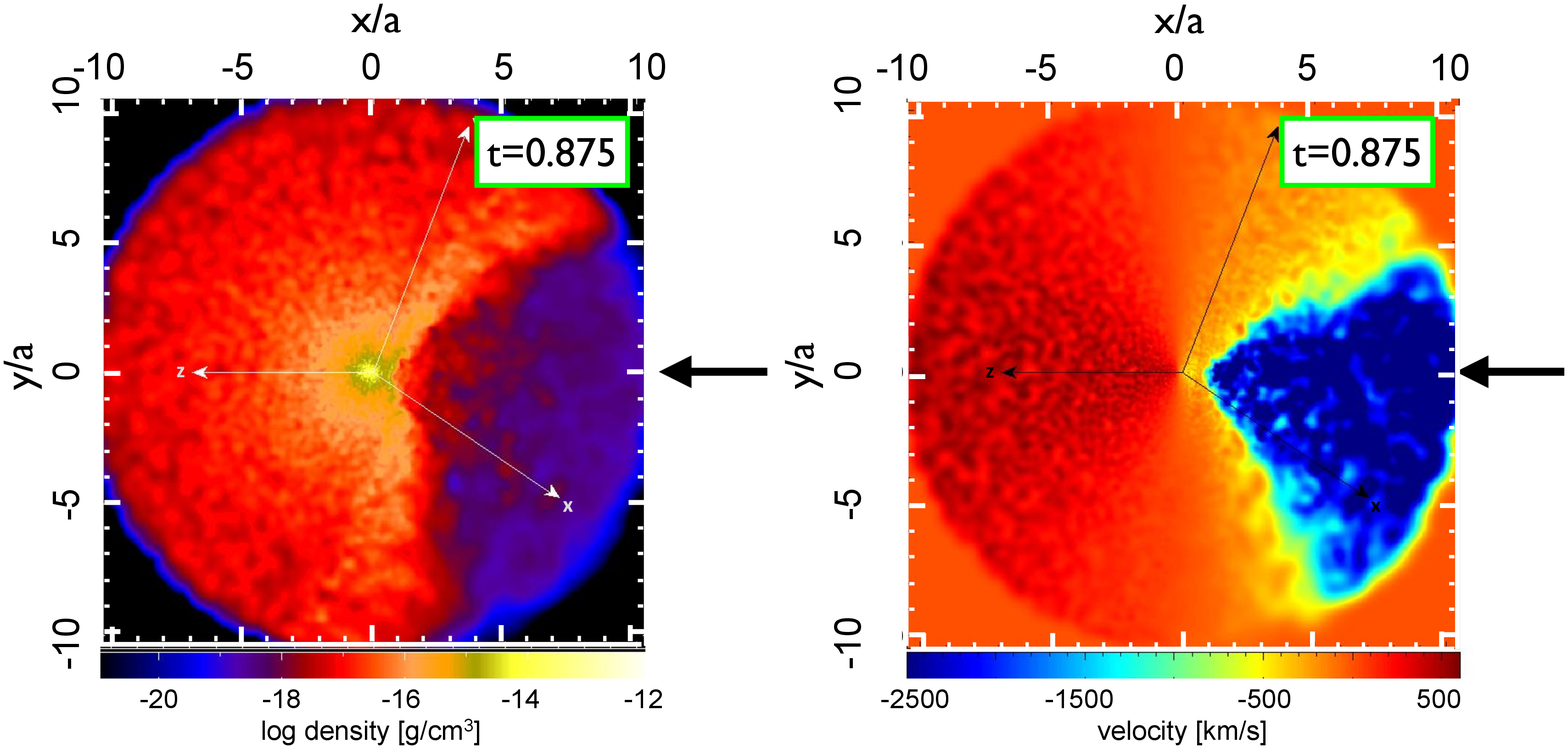}}\\
\resizebox{0.93\hsize}{!}{\includegraphics{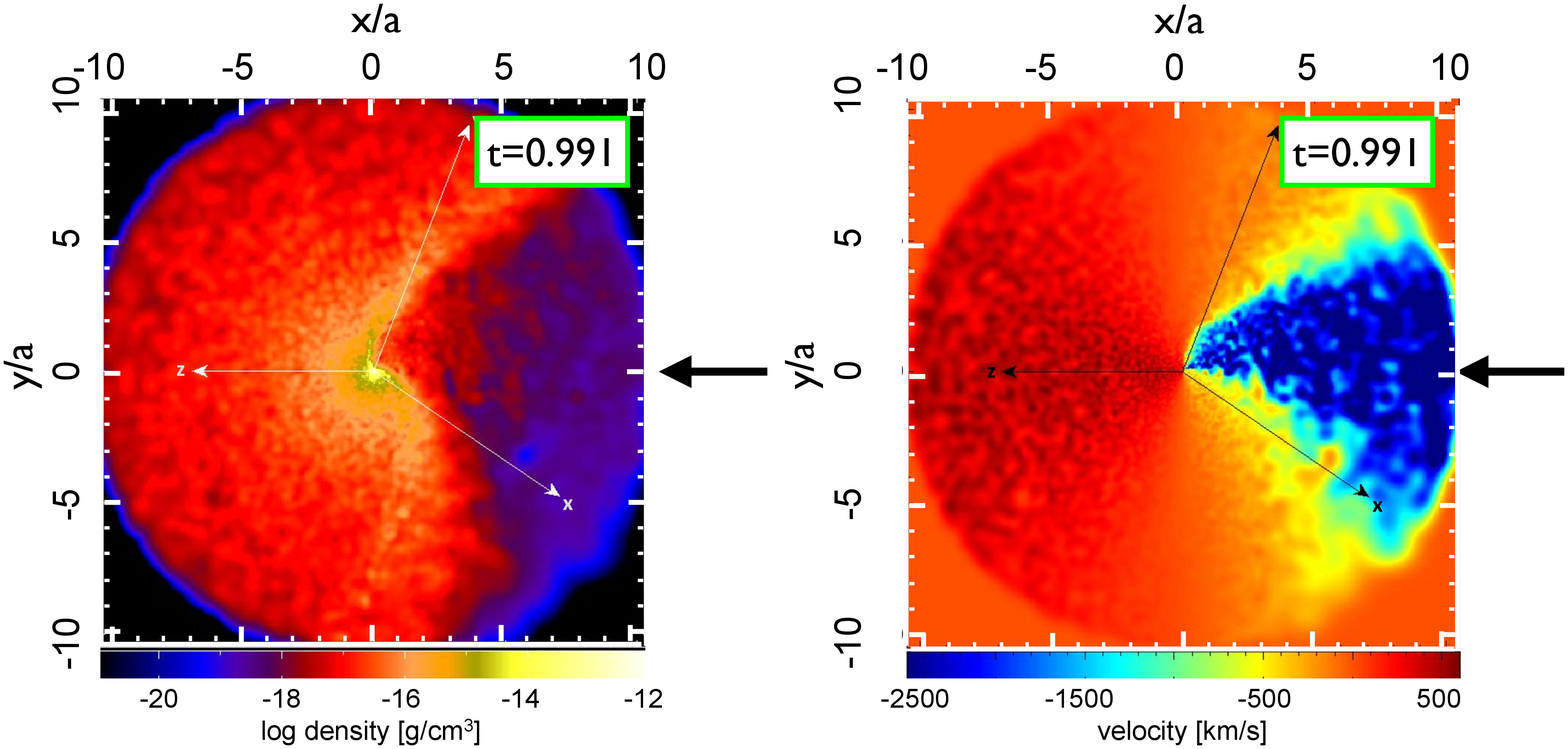}}\\
\resizebox{0.93\hsize}{!}{\includegraphics{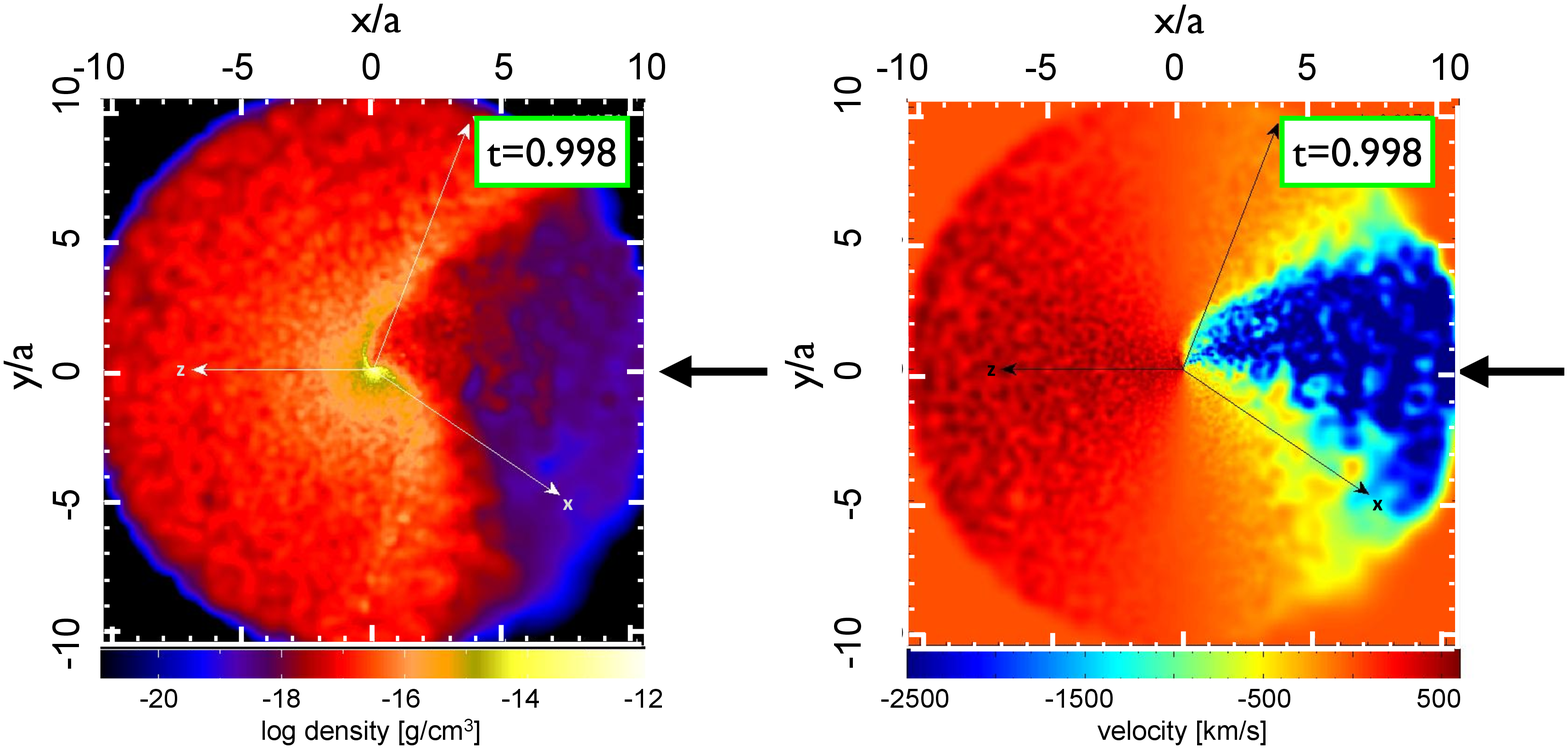}}\\
\resizebox{0.93\hsize}{!}{\includegraphics{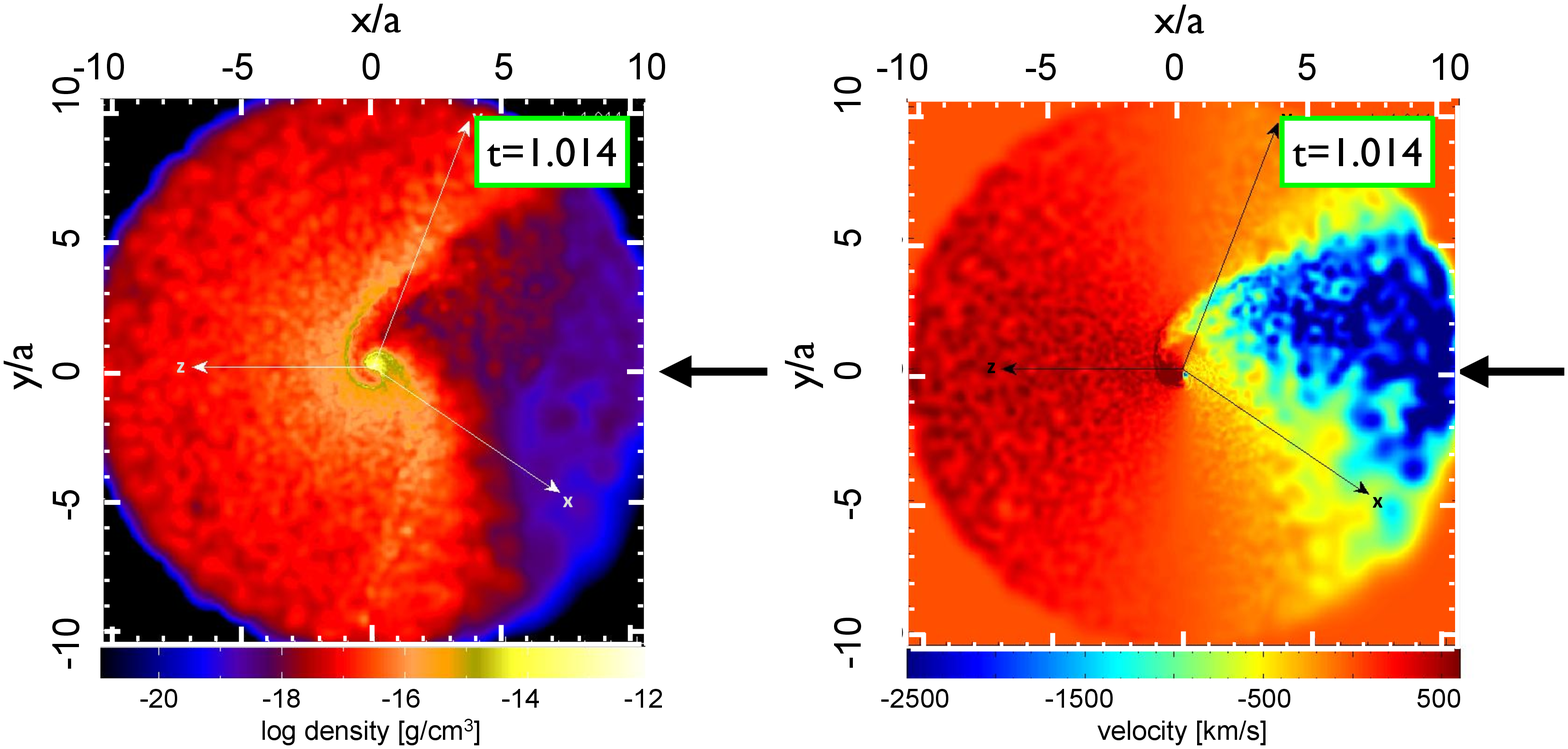}}\\
\resizebox{0.93\hsize}{!}{\includegraphics{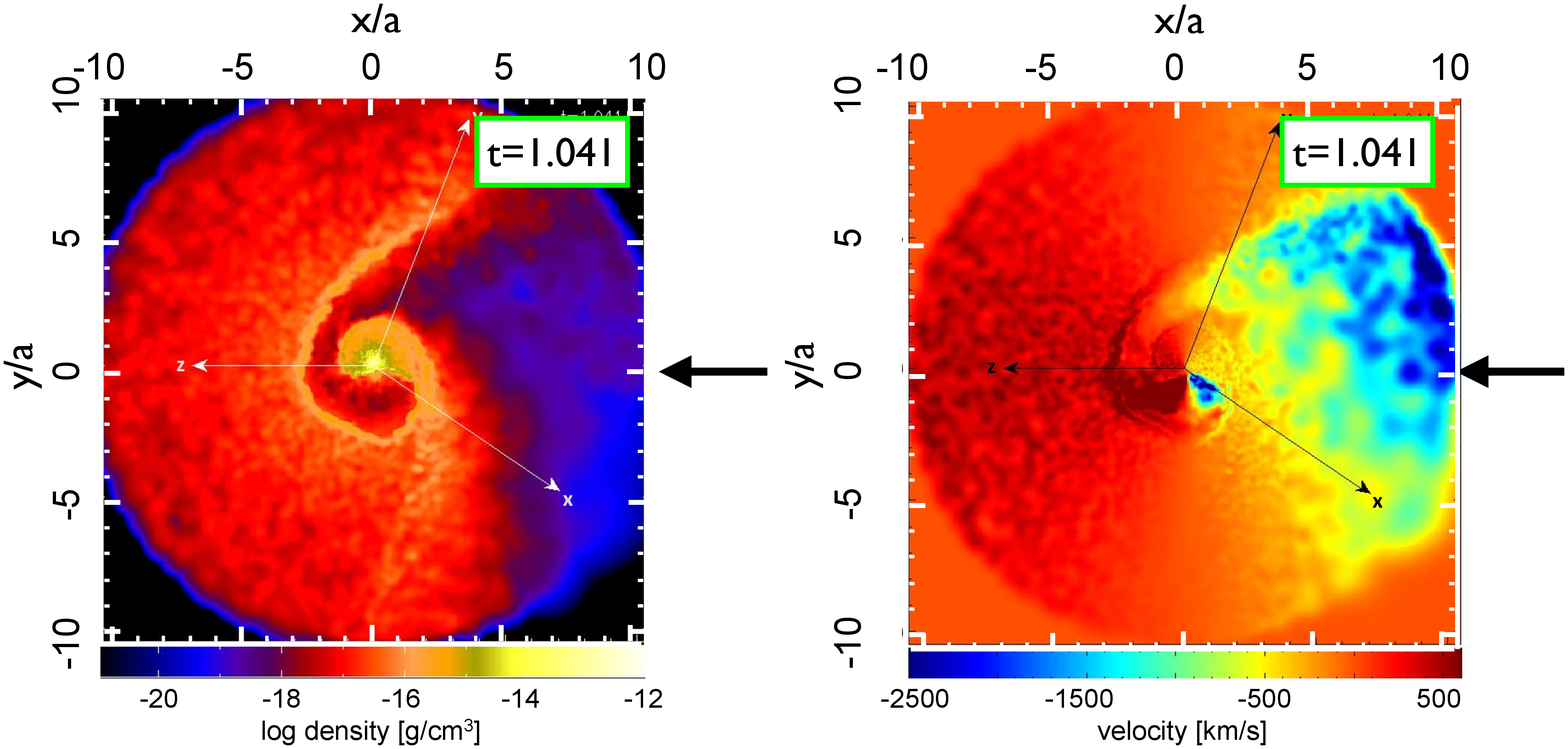}}\\
\vspace{-0.5cm}
\caption{\label{sph_tom3} 2-D slices in the plane containing the observer's line-of-sight and Eta Car~A for our preferred orbital orientation of $i=41\degr$ and a longitude of periastron of $\omega = 243\degr$, based on 3-D hydrodynamic simulations of the Eta Car binary system similar to those presented by \citet{okazaki08}. The observer is located on the right, along the abscissa axis (indicated by the arrows in each panel). The semi-major axis of the orbit is the arrow labeled x, the semi-minor axis the arrow y, and the orbital axis the arrow z. From top to bottom, each row corresponds to orbital phases of $\phi=11.875$, 11.991, 11.998, 12.014, and 12.041, respectively. 
{\it Left}: Density in logarithmic scale as a function of line-of-sight distance from Eta Car~A, in units of the semi=major axis $a=15.4$~AU. {\it Right:} Line-of-sight velocity of material through the same plane. Material is color coded to line-of-sight velocity towards (blue) or away (red) from the observer.}
\end{figure}

A better qualitative agreement between our data and the models is obtained for $i=41\degr$ and $\omega = 243\degr$, which is in line with the values obtained by \citet{okazaki08} and \citep{parkin09} to fit the X-ray lightcurve of Eta Car. For this orientation, there is an overall increase of the column density of the gas with velocities between $-800$ to $-2000~\kms$ from $\phi=11.875$ to $\phi=11.991-11.998$ (Fig. \ref{sph_tom1}m--o), which corresponds to the phase range when the high-velocity component appears in the observations (Section \ref{timescale}). There is still significant column density at $\phi=11.875$ 
at some velocities, in particular around $-1600~\kms$, which is due to a blob expanding at that velocity. A steep overall decrease occurs at phases $\phi=12.014-12.040$, agreeing qualitatively with the disappearance of high-velocity absorption. Therefore, it is very likely that the huge changes in the column density of the high-velocity material from the wind-wind collision zone, which occurs across periastron, are one of the main explanations for the brief appearance of the high-velocity absorption component.  

For some velocity ranges (e.g., $-1100$ to $-1400~\kms$), the column density is higher at $\phi=11.991$ than at $\phi=11.998$, which is opposite to what one would naively expect if the total column densities computed here corresponded directly to a certain amount of line absorption. If ionization effects occur, the behavior of the column density of the population of the lower energy level of \ion{He}{i} $\lambda$10833 (2s\,\element[][3]{S}) as a function of velocity would differ from that of the total column density. Since the distance of Eta Car~B to the high-velocity material and its optical depth change significantly across periastron due to the high orbital eccentricity, ionization effects are indeed expected to happen. These probably play a role in the observed duration of the high-velocity absorption as well as to explain why no high-velocity material is detected with velocities from $-2000$ to $-3000~\kms$, even if a high {\it total} column density is predicted by the SPH models. Note that, with the qualitative comparison done here, we do not aim at explaining the exact behavior of the column density as a function of velocity, nor to claim that the column density derived from the SPH simulations is able to explain the amount of absorption at each velocity. In order to do that, a proper radiative transfer model including the ionizing flux of Eta Car B and of the wind-wind collision-zone would be needed, which is well beyond the scope of this paper.

The hydrodynamical structure of the wind-wind interaction zone is extremely complex, in particular across periastron when the high-velocity absorption is observed. To illustrate the complex geometry and dynamics of the wind-wind interaction in Eta Car, we present in Figure \ref{sph_tom3} 2-D slices of density and velocity in the plane containing the observer's line-of-sight and Eta Car~A for our best orientation of $i=41\degr$ and $\omega = 243\degr$. We notice that the region responsible for the high-velocity absorption, located between 1 and 3 semi-major axis (15 to 45 AU), is clumpy and increases in density across periastron. However, after periastron, there is a major decrease in the wind velocity in the line-of-sight to the primary star, explaining the quick disappearance of the high-velocity absorption. 

\begin{figure}
\resizebox{\hsize}{!}{\includegraphics{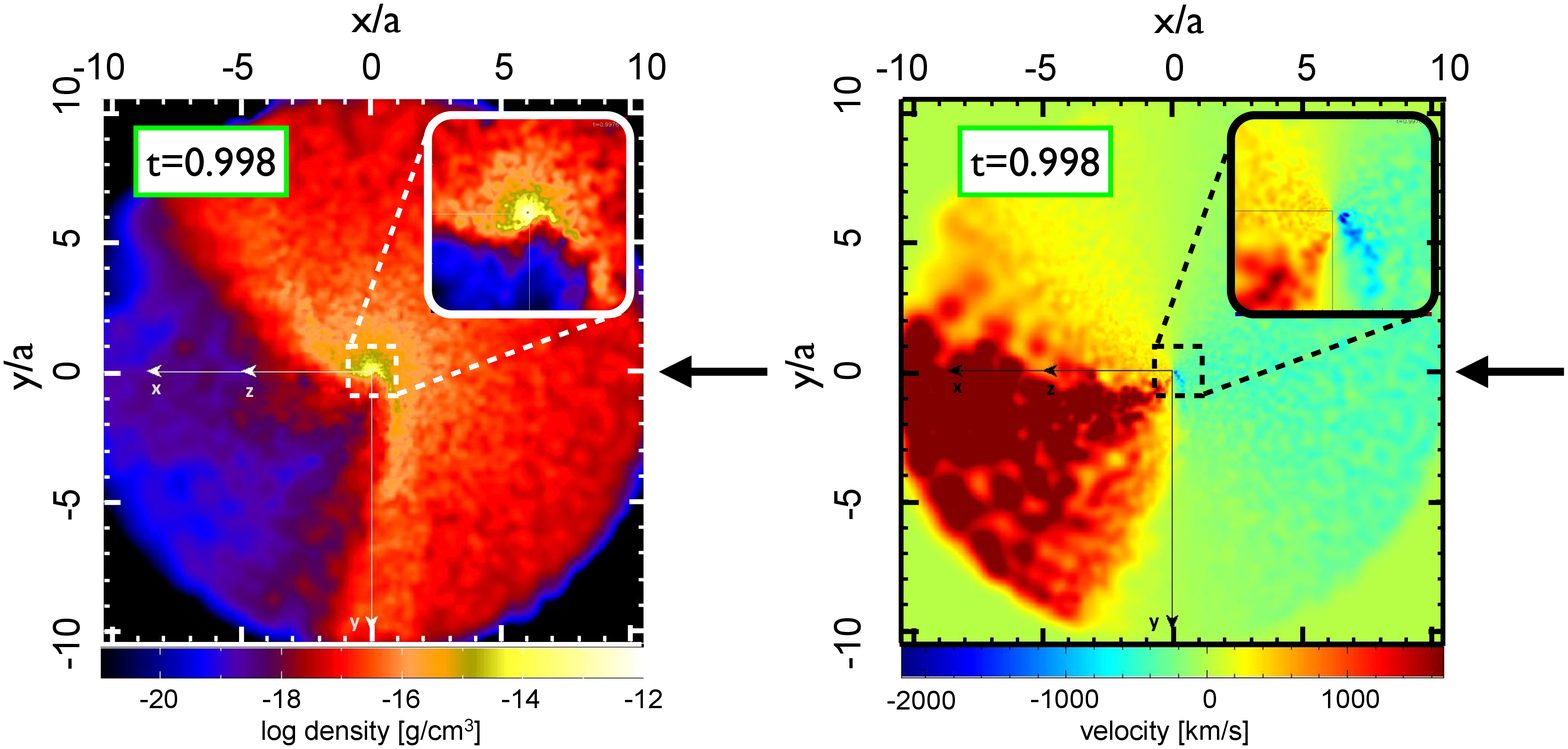}}\\
\caption{\label{omega90} 2-D slices in the plane containing the observer's line-of-sight and Eta Car~A, similar to Fig. \ref{sph_tom3},  but for an orientation of $i=60\degr$, longitude of periastron of $\omega = 90\degr$, and orbital phase $\phi=11.998$. The observer is located on the right, along the abscissa axis (indicated by the arrows in each panel). 
{\it Left}: Density in logarithmic scale as a function of line-of-sight distance from Eta Car~A, in units of the semi=major axis $a=15.4$~AU. {\it Right:} Line-of-sight velocity of material through the same plane. Material is color coded to line-of-sight velocity towards (blue) or away (red) from the observer. The insets show a zoom-in around the inner $\pm 1a$ region. }
\end{figure}

Using relatively simple analytical models for the wind-wind collision zone, \citet{kashi09b} obtained a different value of $\omega=90\degr$ as their best-fit toy model. We suggest that the complex hydrodynamics of the Eta Car system across periastron, which was not included in the \citet{kashi09b} calculations, and the different assumption for the source of the continuum emission (extended hot dust emission on scales of $\sim30$ AU) are the main reasons for the very different value of $\omega$ found by these authors. We show in Figure \ref{omega90} 2-D slices of density and velocity in the plane containing the observer's line-of-sight and Eta Car~A for $i=60\degr$ and $\omega = 90\degr$. As discussed above, there is no high-velocity material from the wind-wind collision zone in line-of-sight to Eta Car~A, and only high-velocity material from the wind of the Eta Car Car~B is in line-of sight (as in the ``wind-eclipse" scenario described in Section \ref{windeclipse}, which does not explain the observations). Since \citet{kashi09b} assumed that the absorption region was compact ($\sim$ a few AU), it is also unclear how such a compact region would cover a significant fraction of their extended continuum source ($\sim30$ AU) in order to reproduce the significant high fraction of continuum coverage (30--50\%) inferred from the amount of high-velocity absorption reported in our present paper.

\subsubsection{A tilted orbital plane relative to the Homunculus equatorial plane?}

\begin{figure*}

\resizebox{0.33\hsize}{!}{\includegraphics{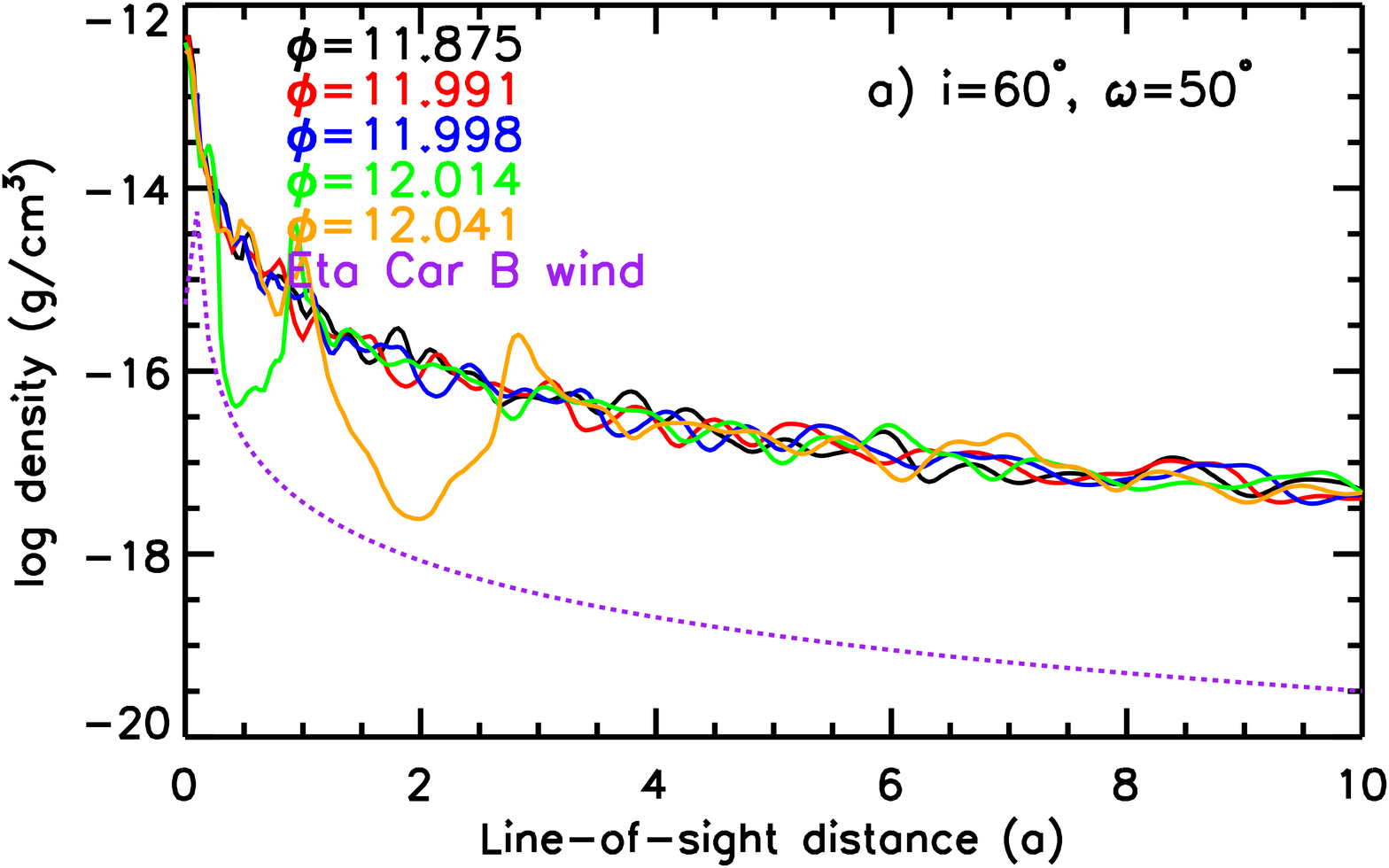}}
\resizebox{0.33\hsize}{!}{\includegraphics{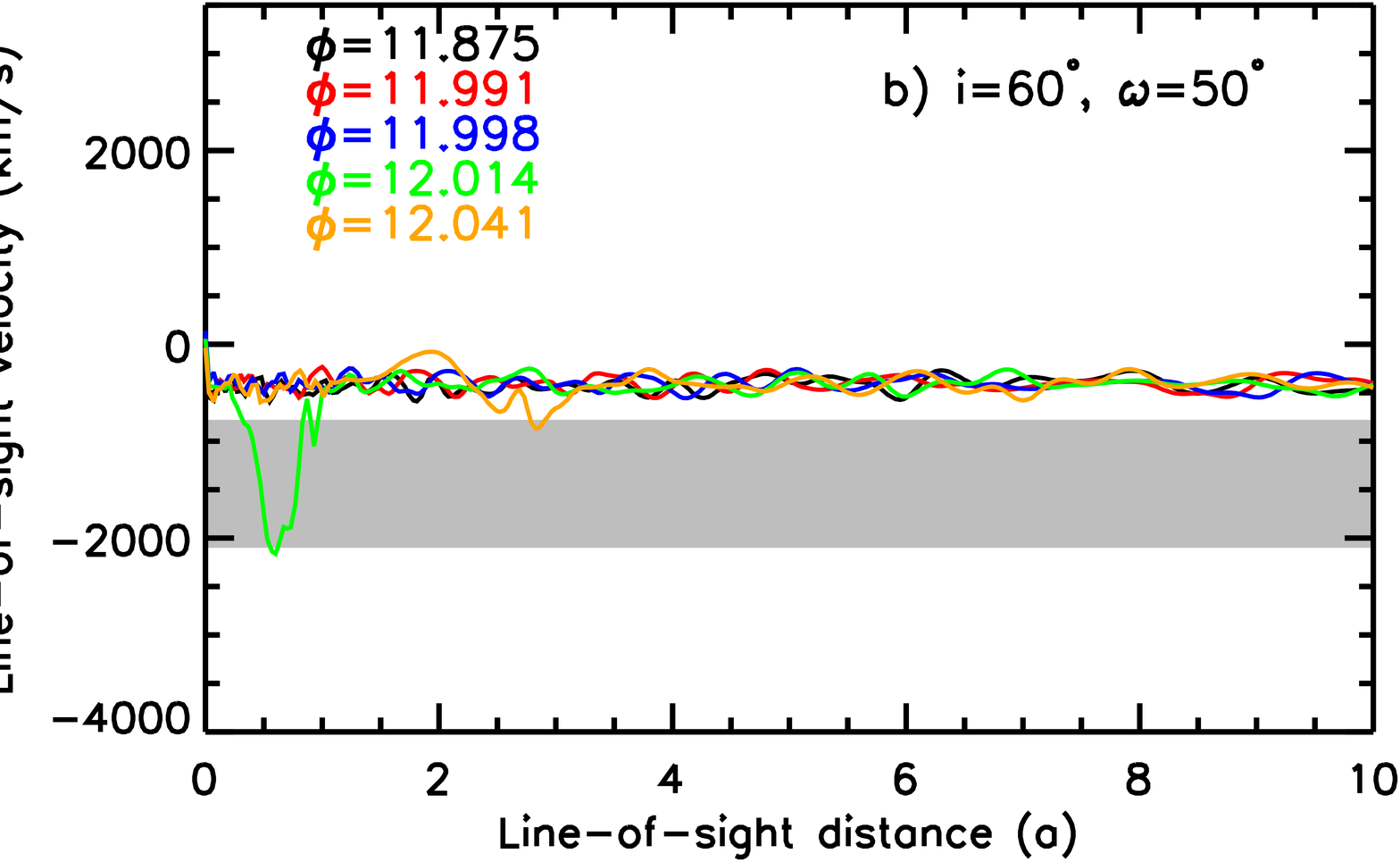}}
\resizebox{0.33\hsize}{!}{\includegraphics{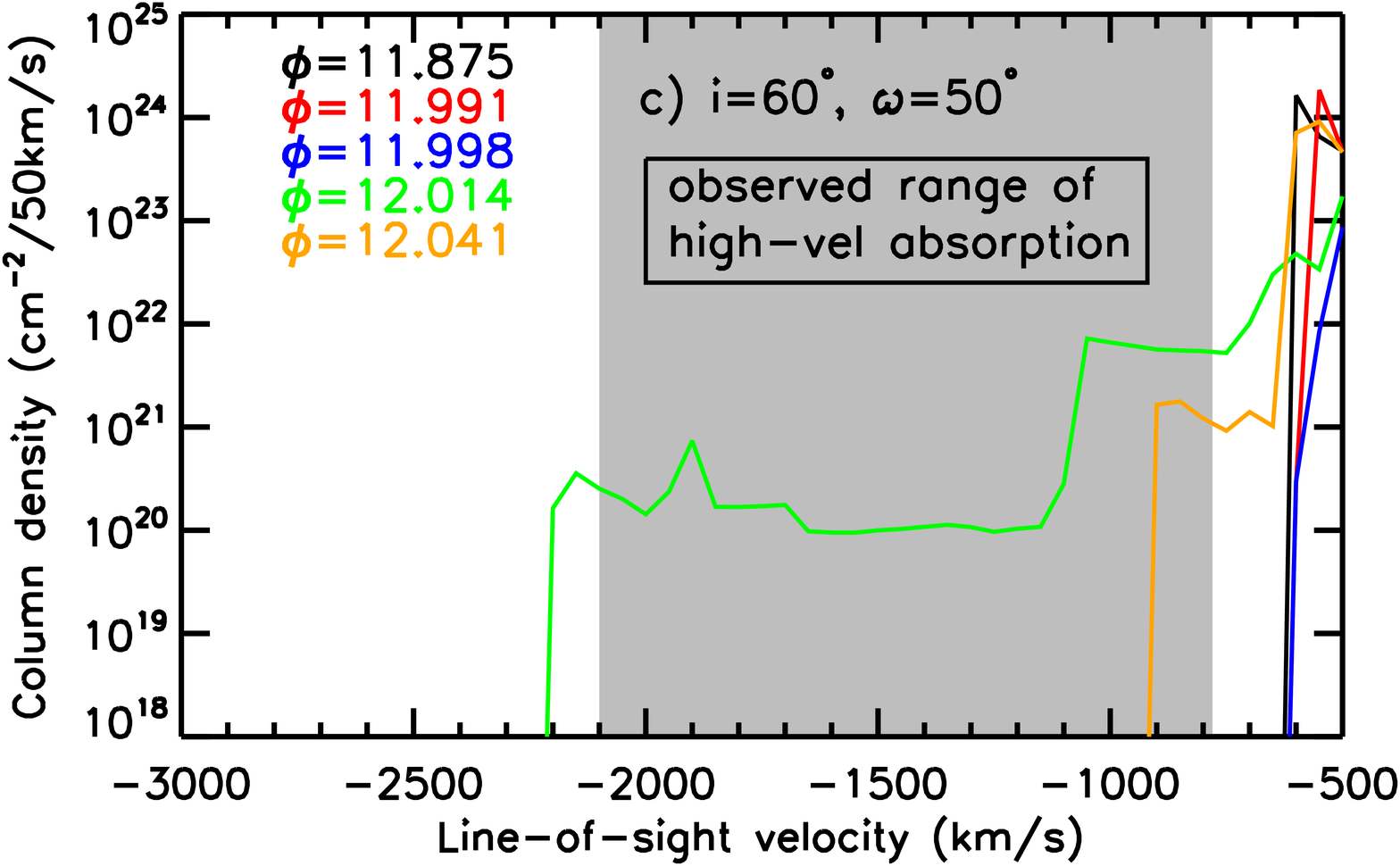}}\\

\resizebox{0.33\hsize}{!}{\includegraphics{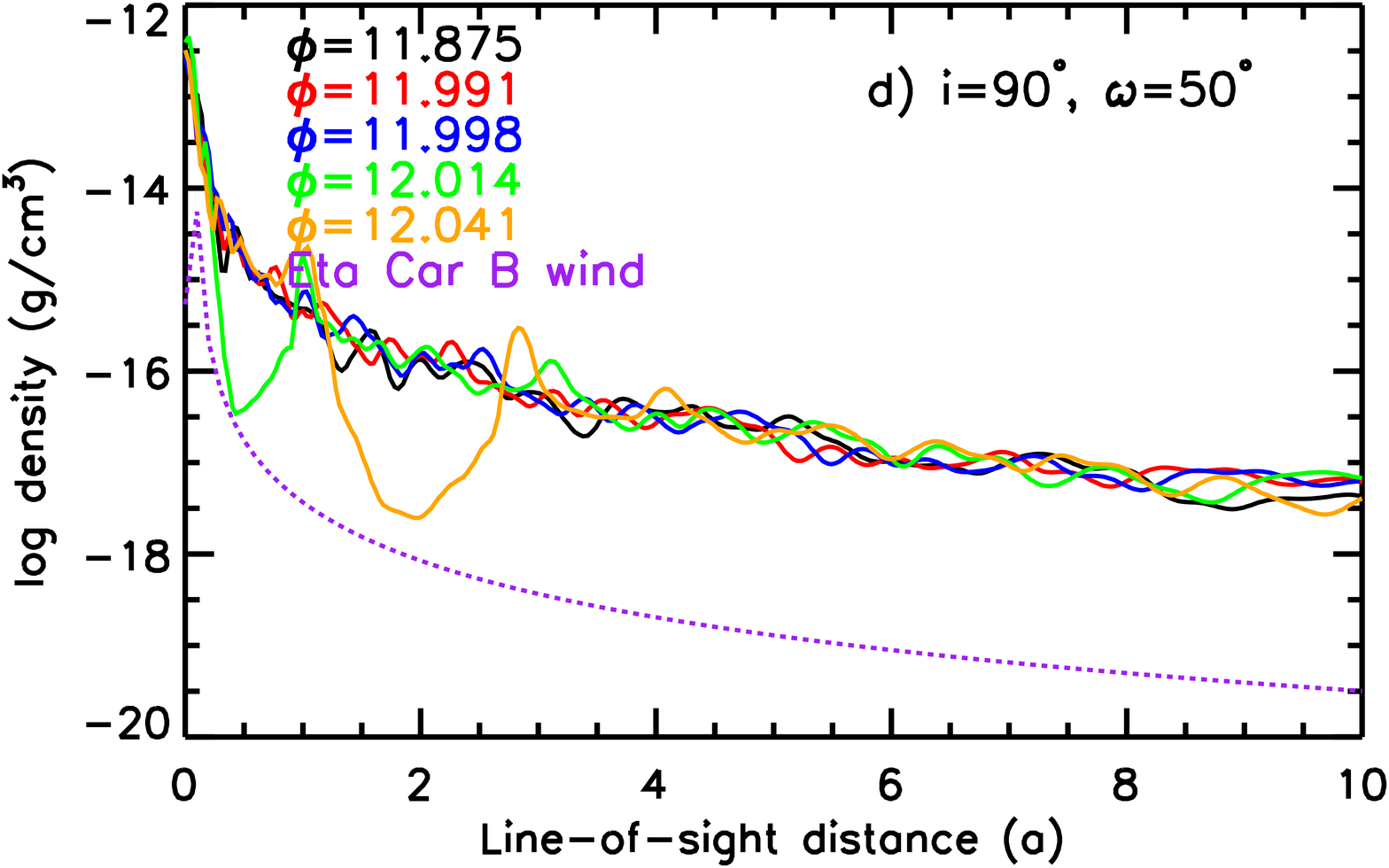}}
\resizebox{0.33\hsize}{!}{\includegraphics{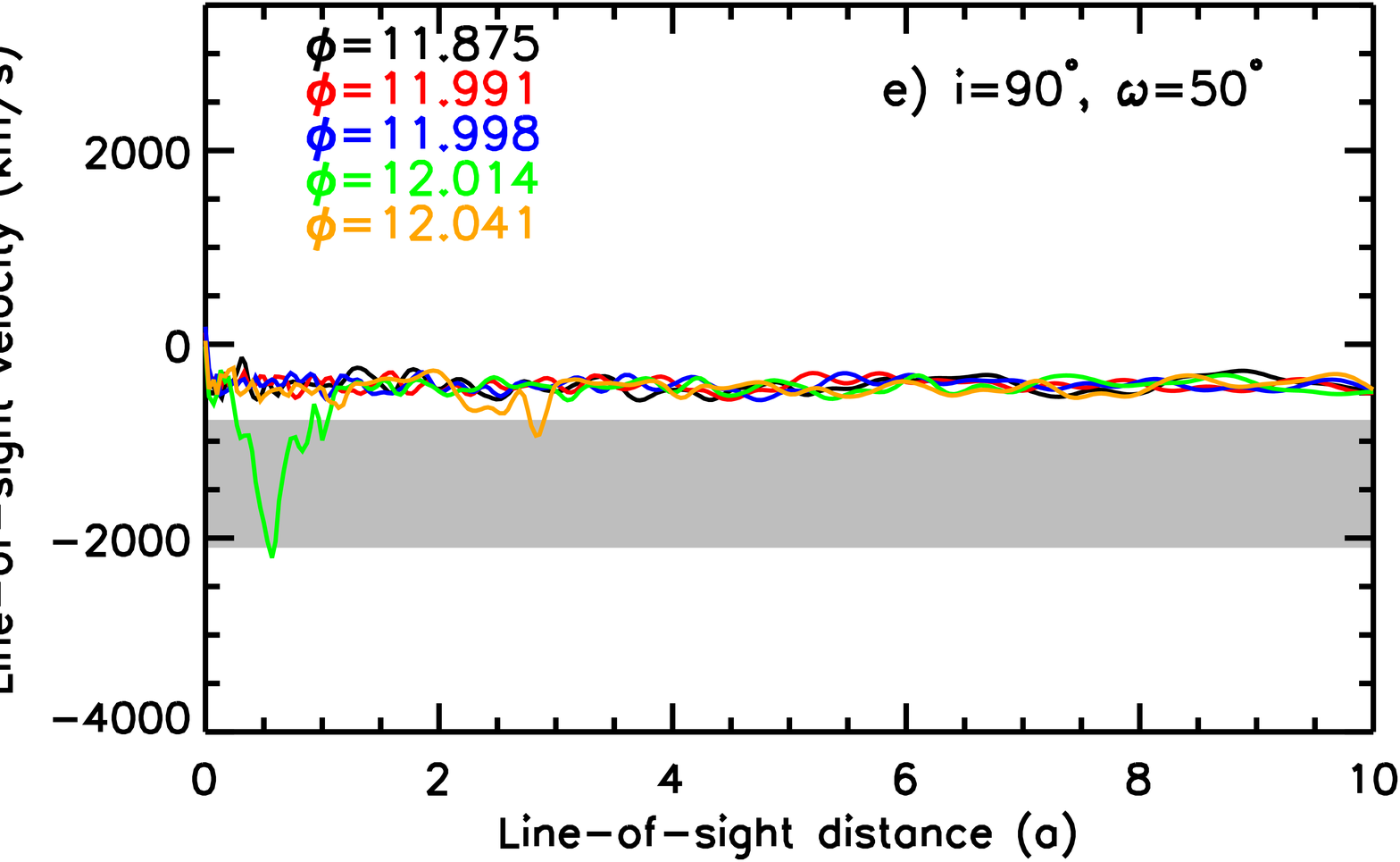}}
\resizebox{0.33\hsize}{!}{\includegraphics{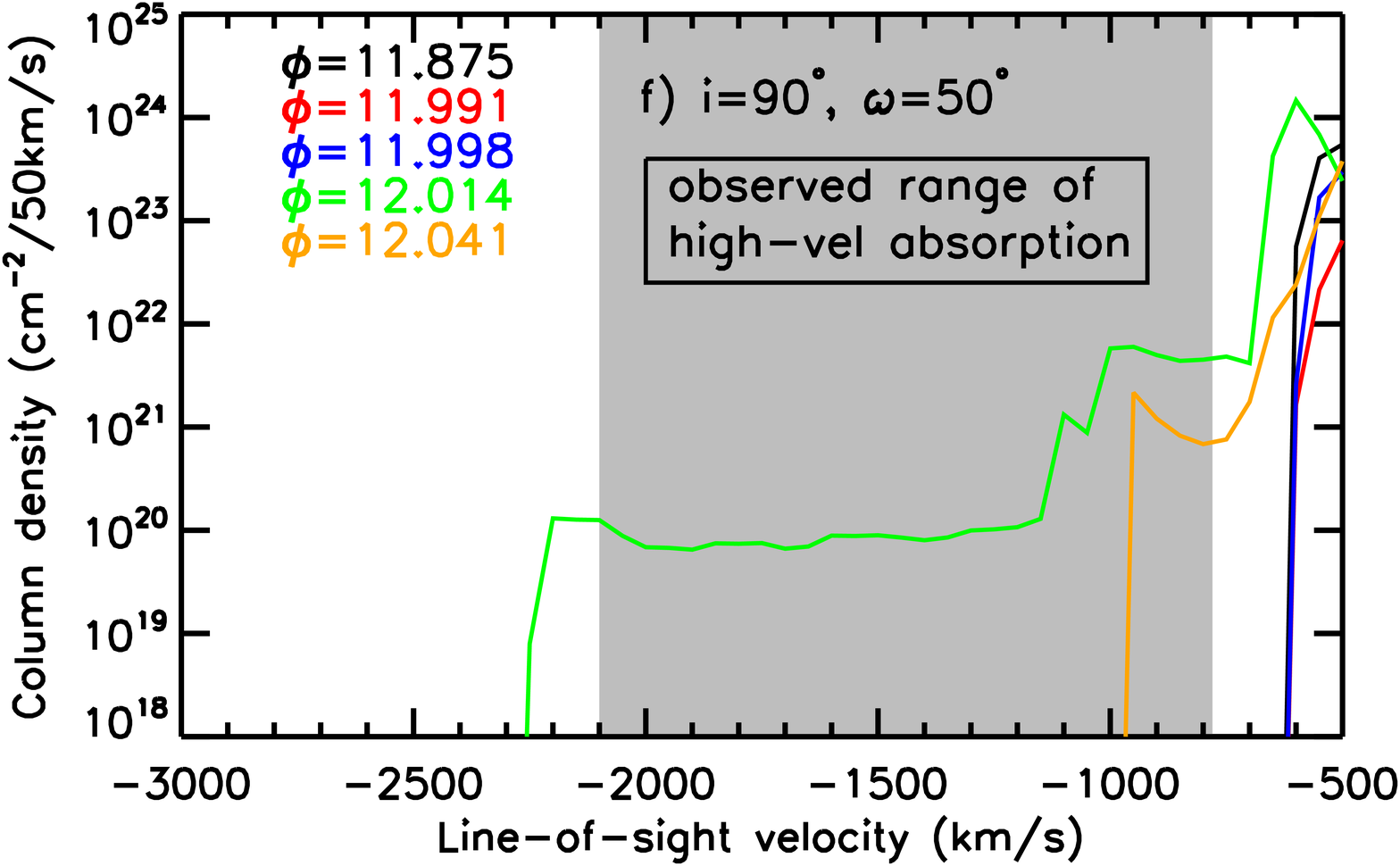}}\\

\resizebox{0.33\hsize}{!}{\includegraphics{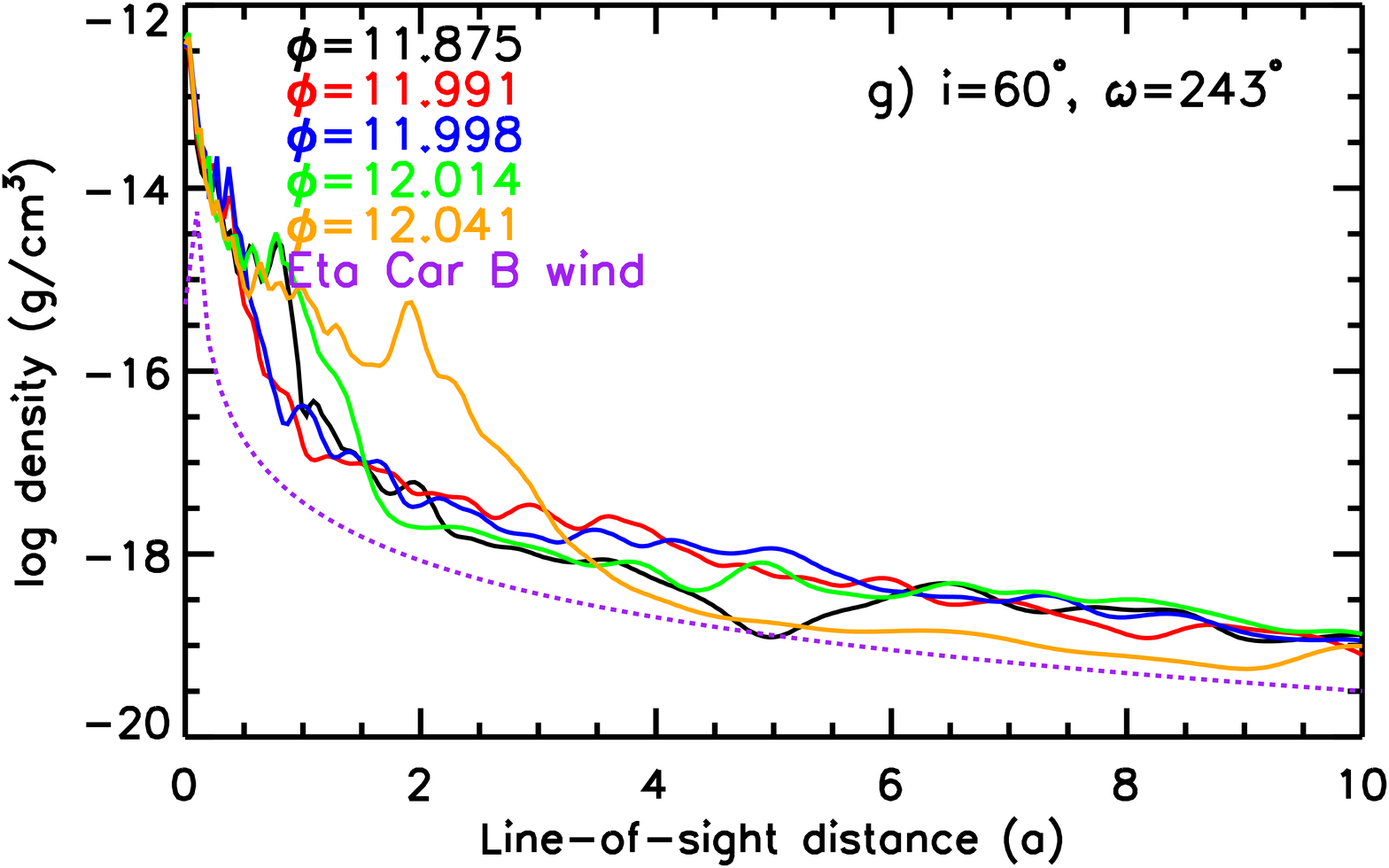}}
\resizebox{0.33\hsize}{!}{\includegraphics{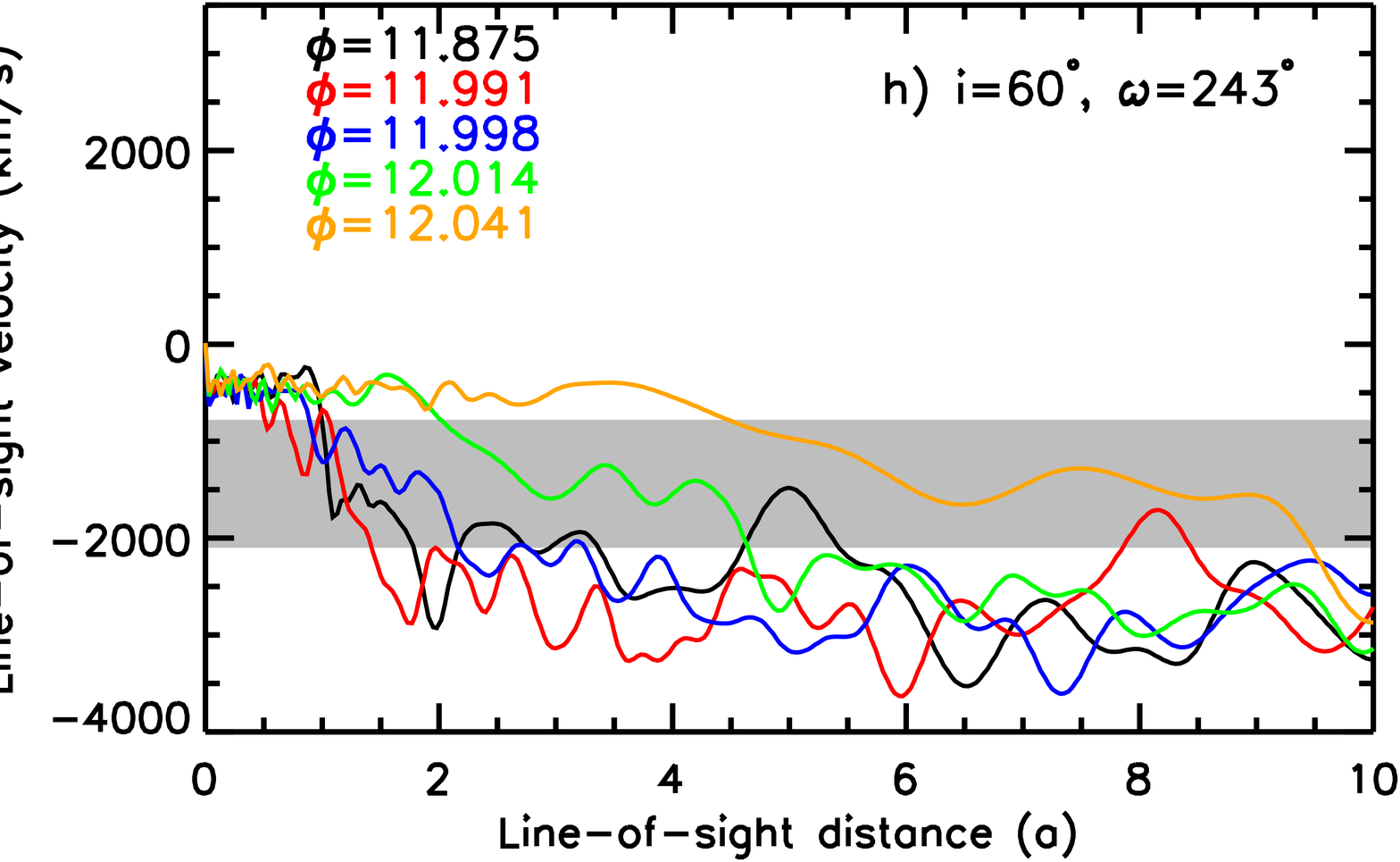}}
\resizebox{0.33\hsize}{!}{\includegraphics{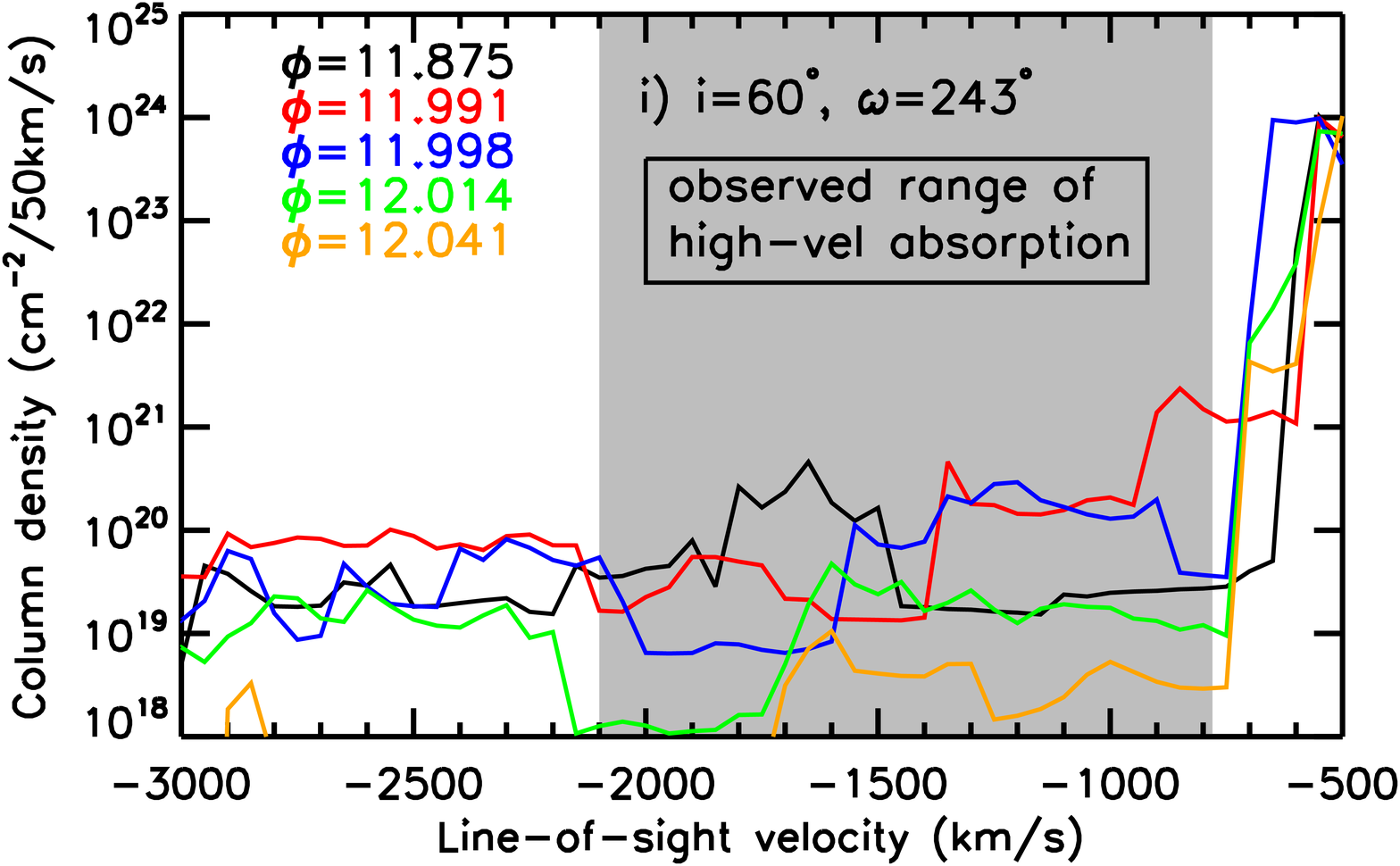}}\\

\resizebox{0.33\hsize}{!}{\includegraphics{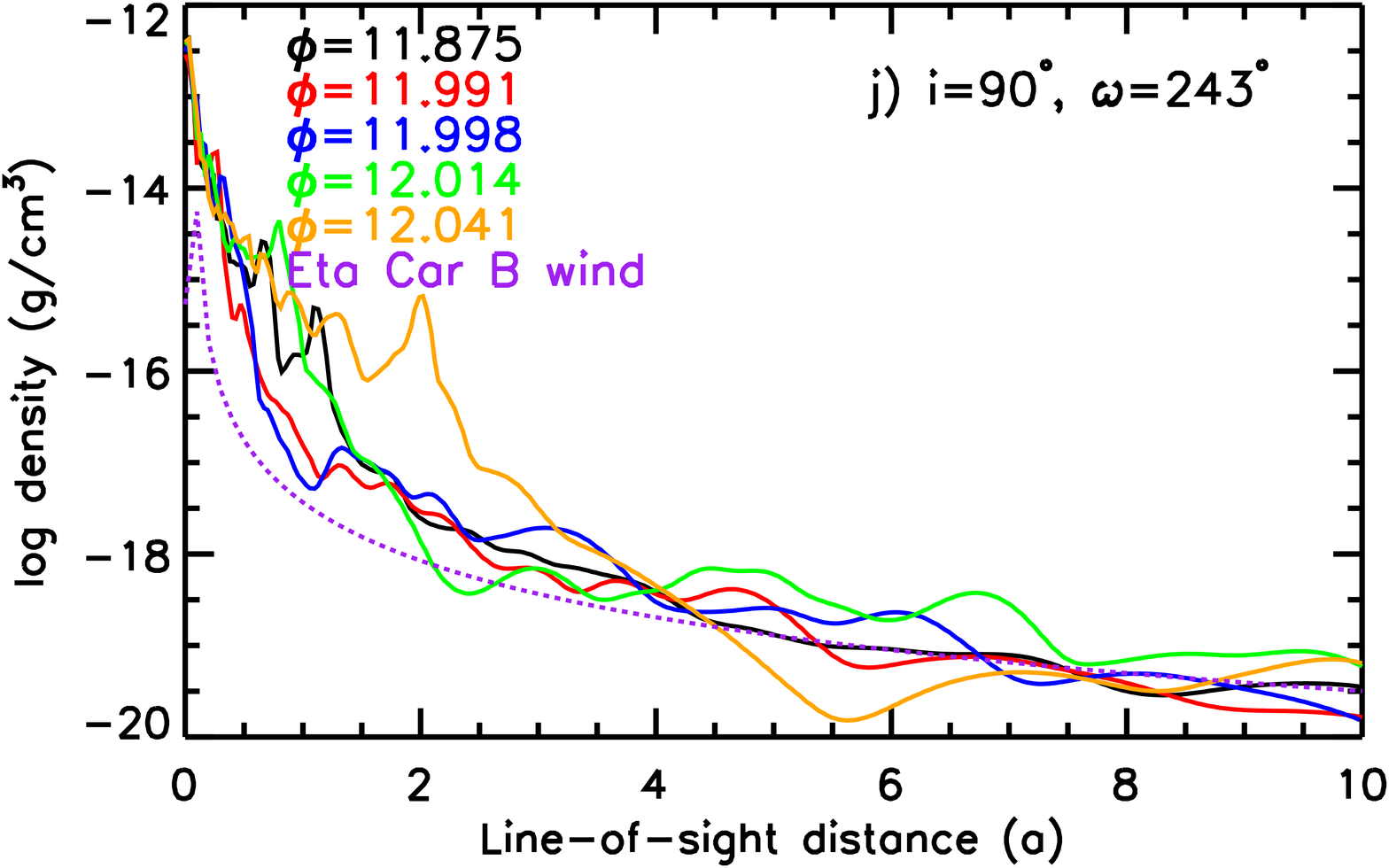}}
\resizebox{0.33\hsize}{!}{\includegraphics{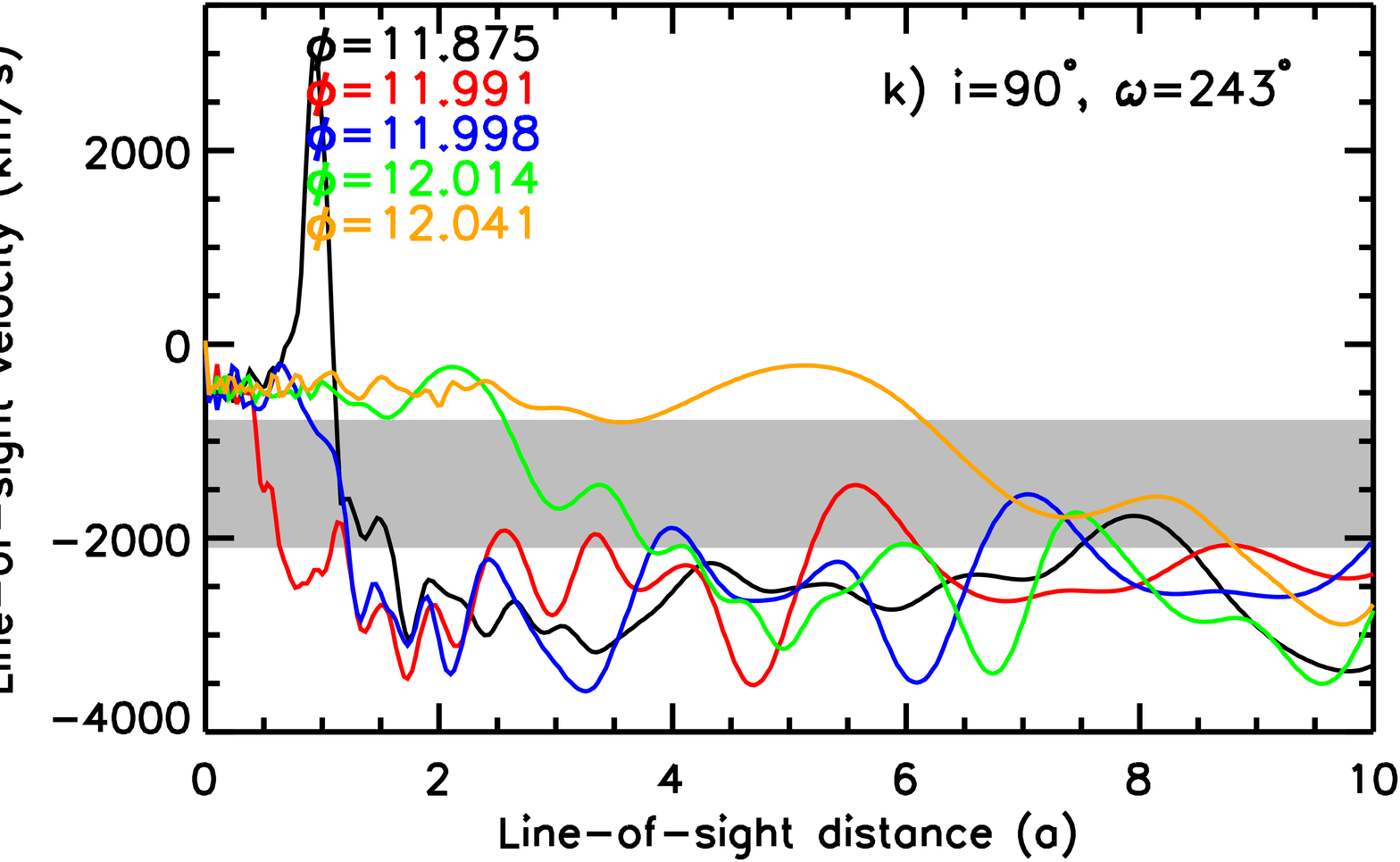}}
\resizebox{0.33\hsize}{!}{\includegraphics{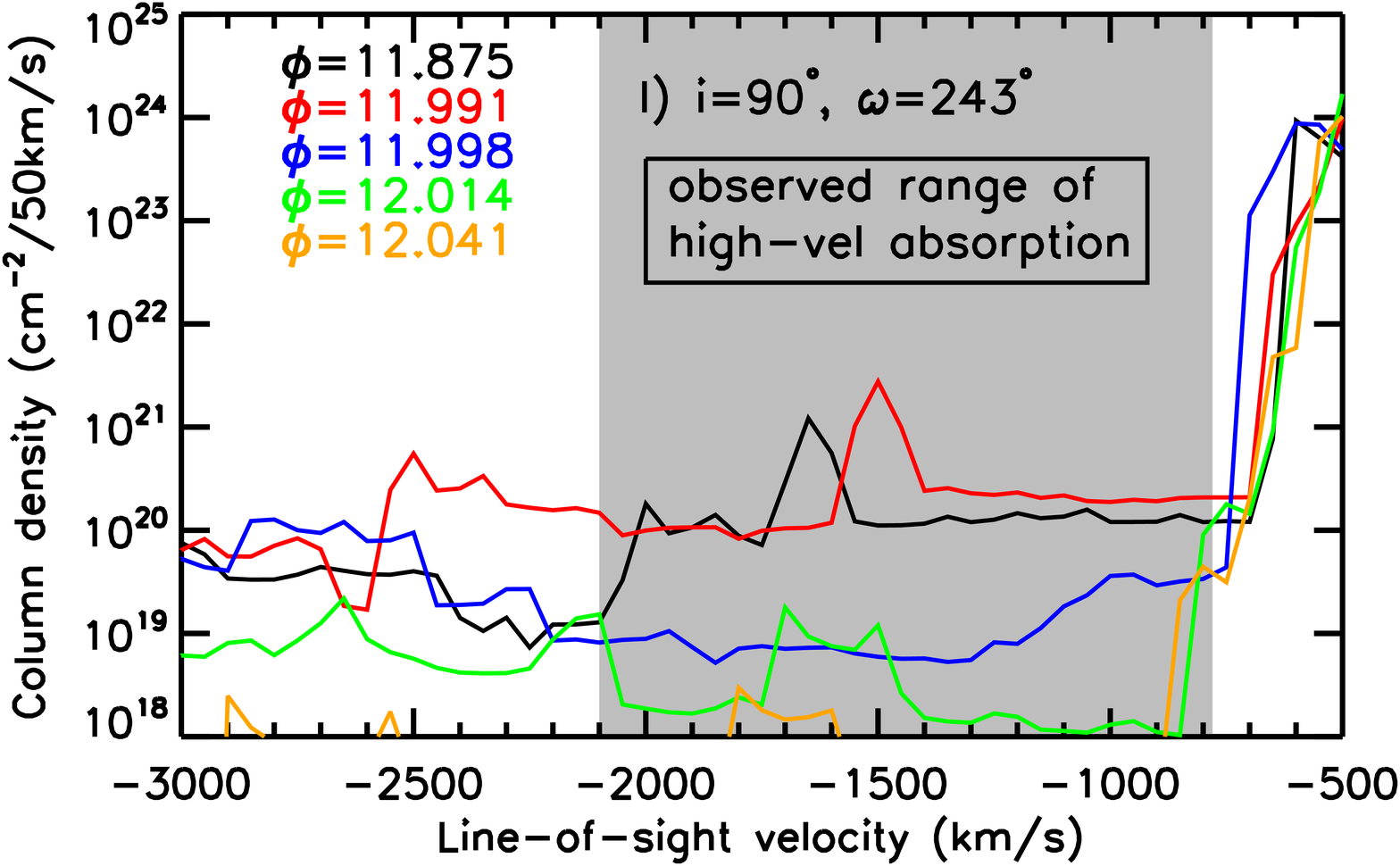}}\\

\caption{\label{sph_tom2}  Similar to Fig. \ref{sph_tom1} but, from top to bottom, the following line-of-sights are shown: $i=60\degr$ and $\omega=50\degr$, $i=90\degr$ and $\omega=50\degr$, $i=60\degr$ and $\omega=243\degr$, and $i=90\degr$ and $\omega=243\degr$, respectively.}
\end{figure*}

Recent works have suggested that the orbital plane of the Eta Car binary system might not be aligned with the Homunculus equatorial plane \citep{abraham07,abraham09,okazaki08,henley08,diego09}. To verify whether our observations support a tilted orbital axis scenario, here we investigate whether the 3-D SPH models show a significant column density of high-velocity gas at other orbital inclination and at the orbital phases corresponding to when the high-velocity absorption was observed. Figure \ref{sph_tom2} presents, similar to Figure \ref{sph_tom1}, the one-dimensional density, velocity, and column density structure of the gas for different lines-of-sight to Eta Car~A, but inclined at $i=60\degr$ and $90 \degr$ from the plane of the sky.

For longitudes of periastron $\omega \sim 50\degr-90 \degr$, similar to the aligned case, most of the material in the line-of-sight towards Eta Car~A is from its own wind at roughly $-500~\kms$, except during a brief period after periastron when the ``wind-eclipse'' scenario occurs (Fig. \ref{sph_tom2}a--f). For inclination angles higher than $i=41\degr$, such as $60\degr$ and $90\degr$, a higher fraction of the wind of Eta Car~B crosses the line-of-sight to Eta Car~A and, thus, higher column densities are obtained at higher velocities after periastron. However, that occurs only briefly and it is still in disagreement with our observations for the same reasons discussed in Sect. \ref{windeclipse}. Hence, our data are not in agreement with the orbital parameters derived by \citet{abraham07} and \citet{diego09} if the high-velocity absorption originates in the wind-wind collision zone.s

If the orbital plane is tilted relative to the Homunculus, the hydrodynamics of the 3-D simulations still require that the binary system is seen under 
$\omega\sim240-270\degr$, otherwise no dense, extended, high-velocity material from the wind-wind collision zone is in line-of-sight during the observed  amount of time. The temporal behavior of the column density of the high-velocity gas for $i=60\degr$ and $\omega = 243\degr$ (Fig. \ref{sph_tom2}j--l) is very similar to that obtained for $i=41\degr$ and $\omega = 243\degr$ (Fig. \ref{sph_tom1}g--i). Indeed, for $\omega=243\degr$, the overall relative increase (decrease) in the column density of the high-velocity material before (after) periastron is even higher in the case of $i=60\degr$ than for $i=41\degr$. For $i=90\degr$, there is no significant increase in the column density of the high-velocity material before periastron and, actually, a strong decrease is seen at $\phi=11.998$ (Fig. \ref{sph_tom2}g--i). Strong ionization effects would be necessary to explain the high-velocity component if indeed $i=90\degr$, which seems unlikely. We can also argue that relatively less strong ionization effects would be needed to explain the appearance and disappearance of high-velocity material during periastron if $i=60\degr$ compared to the $i=41\degr$ case. However, since the exact amount of ionization effects due to Eta Car~B is unclear, we conclude that our observations can be explained by the 3-D hydrodynamical models with both $i=41\degr$ and $i=60\degr$ for $\omega=243\degr$, which is consistent with previous X-ray analyses \citep{okazaki08,henley08,parkin09}.  

\section{Concluding remarks}

VLT/CRIRES observations of Eta Car provide definitive evidence that high-velocity material, up to $\sim-1900~\kms$, was present in the system during the 2009.0 periastron passage. The broad, high-velocity absorption is seen in \ion{He}{i} $\lambda$10833 in the VLT/CRIRES dataset only in the spectrum obtained at phase $\phi=11.991$ to 11.998, showing that it is connected to the spectroscopic event. Near-infrared observations obtained at OPD/LNA from 1992 to 2009 indeed show that the high-velocity absorption in \ion{He}{i} $\lambda$10833 is periodic, and tightly connected with phase zero of the spectroscopic cycle as well. Based on the OPD/LNA dataset, we constrained the timescale of detection of the high-velocity gas from 95 to $206~\mathrm{d}$  (0.047 to 0.102 in phase) around phase zero. We analyzed archival {\it HST}/STIS ultraviolet data, showing that the \ion{Si}{iv} $\lambda$1393, 1402 resonance line also presented a high-velocity absorption component up to $-2100~\kms$. 

We presented several reasons why the high-velocity absorption is unlikely to be due to a transitory high-velocity wind of Eta Car~A, or due to a wind eclipse of Eta Car~B. We suggest that our observations provide direct detection of shocked, high-velocity material flowing from the wind-wind collision zone around the binary system. Using detailed 3-dimensional hydrodynamical simulations of the wind-wind collision zone of Eta Car, we found that dense high-velocity gas is in the line-of-sight to the primary star only if the binary system is oriented in the sky such that the companion is behind the primary star during periastron, corresponding to a longitude of periastron of $\omega \sim 240\degr-270\degr$. Our data is broadly consistent with an orbital inclination in the range $i=40\degr-60\degr$. We derived that the high-velocity gas is located at distances of 15 to 45~AU in the line-of-sight to Eta Car~A. More importantly, we can rule out orbital orientations in the range $\omega \sim 0\degr-180\degr$ for all inclination angles, since these do not produce a significant column density of high-velocity gas in our line-of-sight to match our observations of the high-velocity absorption component.

The current 3-D SPH simulations used in this paper do not account for radiative cooling, which makes it difficult to estimate the ionization stage of the high-velocity material in the wind-wind collision zone. In addition to the increase in column density of the high-velocity gas, ionization effects due to the close presence of Eta Car B likely play an important role in explaining the amount of high-velocity absorption seen during periastron. Time-dependent, multi-dimensional radiative transfer modeling of the outflowing gas from the wind-wind collision zone of the Eta Car binary system is urgently warranted, and will allow us to better understand the influence of Eta Car~B on the wind of Eta Car~A across periastron. Ultimately, this will provide constraints on the masses of the stars and on the wind parameters of the Eta Car binary system.

\begin{acknowledgements}

We wish to thank the kind allocation of ESO Director Discretionary Time that was crucial for the completion of this project, and the ESO staff at Garching and Paranal, in particular Hugues Sana and Andrea Ahumada, for carrying out the VLT/CRIRES observations. We appreciate many discussions and comments on the manuscript from Michael Corcoran and Nathan Smith. We thank an anonymous referee for the suggestions to improve the original manuscript. JHG thanks the Max-Planck-Gesellschaft for financial support for this work. AD and MT thanks the FAPESP foundation for continuous support. TIM is supported by a NASA GSRP fellowship. The HST observations were accomplished through STIS GTO, HST GO and HST Eta Car Treasury Team programmes. HST/STIS data were obtained through the HST Eta Car Treasury archive hosted at University of Minnesota.

\end{acknowledgements}

\Online


\begin{thebibliography}{76}


\bibitem[{{Abraham} \& {Falceta-Gon{\c c}alves}(2007)}]{abraham07}
{Abraham}, Z. \& {Falceta-Gon{\c c}alves}, D. 2007, \mnras, 378, 309

\bibitem[{{Abraham} \& {Falceta-Gon{\c c}alves}(2009)}]{abraham09}
{Abraham}, Z. \& {Falceta-Gon{\c c}alves}, D. 2009, \mnras, 1574

\bibitem[{{Abraham} {et~al.}(2005){Abraham}, {Falceta-Gon{\c c}alves},
  {Dominici}, {Nyman}, {Durouchoux}, {McAuliffe}, {Caproni}, \&
  {Jatenco-Pereira}}]{abraham05a}
{Abraham}, Z., {Falceta-Gon{\c c}alves}, D., {Dominici}, T.~P., {et~al.} 2005,
  \aap, 437, 977

\bibitem[{{Behar} {et~al.}(2007){Behar}, {Nordon}, \& {Soker}}]{behar07}
{Behar}, E., {Nordon}, R., \& {Soker}, N. 2007, \apjl, 666, L97

\bibitem[{{Chesneau} {et~al.}(2005){Chesneau}, {Min}, {Herbst}, {Waters},
  {Hillier}, {Leinert}, {de Koter}, {Pascucci}, {Jaffe}, {K{\"o}hler},
  {Alvarez}, {van Boekel}, {Brandner}, {Graser}, {Lagrange}, {Lenzen}, {Morel},
  \& {Sch{\"o}ller}}]{chesneau05}
{Chesneau}, O., {Min}, M., {Herbst}, T., {et~al.} 2005, \aap, 435, 1043

\bibitem[{{Corcoran}(2005)}]{corcoran05}
{Corcoran}, M.~F. 2005, \aj, 129, 2018

\bibitem[{{Corcoran} {et~al.}(1997){Corcoran}, {Ishibashi}, {Davidson},
  {Swank}, {Petre}, \& {Schmitt}}]{corcoran97}
{Corcoran}, M.~F., {Ishibashi}, K., {Davidson}, K., {et~al.} 1997, \nat, 390,
  587

\bibitem[{{Corcoran} {et~al.}(2001){Corcoran}, {Ishibashi}, {Swank}, \&
  {Petre}}]{corcoran01}
{Corcoran}, M.~F., {Ishibashi}, K., {Swank}, J.~H., \& {Petre}, R. 2001, \apj,
  547, 1034

\bibitem[{{Damineli}(1996)}]{damineli96}
{Damineli}, A. 1996, \apjl, 460, L49

\bibitem[{{Damineli} {et~al.}(1997){Damineli}, {Conti}, \&
  {Lopes}}]{damineli97}
{Damineli}, A., {Conti}, P.~S., \& {Lopes}, D.~F. 1997, New Astronomy, 2, 107

\bibitem[{{Damineli} {et~al.}(2008{\natexlab{a}}){Damineli}, {Hillier},
  {Corcoran}, {Stahl}, {Groh}, {Arias}, {Teodoro}, {Morrell}, {Gamen},
  {Gonzalez}, {Leister}, {Levato}, {Levenhagen}, {Grosso}, {Colombo}, \&
  {Wallerstein}}]{damineli08_multi}
{Damineli}, A., {Hillier}, D.~J., {Corcoran}, M.~F., {et~al.}
  2008{\natexlab{a}}, \mnras, 386, 2330

\bibitem[{{Damineli} {et~al.}(2008{\natexlab{b}}){Damineli}, {Hillier},
  {Corcoran}, {Stahl}, {Levenhagen}, {Leister}, {Groh}, {Teodoro}, {Albacete
  Colombo}, {Gonzalez}, {Arias}, {Levato}, {Grosso}, {Morrell}, {Gamen},
  {Wallerstein}, \& {Niemela}}]{damineli08_period}
{Damineli}, A., {Hillier}, D.~J., {Corcoran}, M.~F., {et~al.}
  2008{\natexlab{b}}, \mnras, 384, 1649

\bibitem[{{Damineli} {et~al.}(2000){Damineli}, {Kaufer}, {Wolf}, {Stahl},
  {Lopes}, \& {de Ara{\'u}jo}}]{damineli00}
{Damineli}, A., {Kaufer}, A., {Wolf}, B., {et~al.} 2000, \apjl, 528, L101

\bibitem[{{Damineli} {et~al.}(1998){Damineli}, {Stahl}, {Kaufer}, {Wolf},
  {Quast}, \& {Lopes}}]{damineli98}
{Damineli}, A., {Stahl}, O., {Kaufer}, A., {et~al.} 1998, \aaps, 133, 299

\bibitem[{{Davidson} {et~al.}(1995){Davidson}, {Ebbets}, {Weigelt},
  {Humphreys}, {Hajian}, {Walborn}, \& {Rosa}}]{davidson1995}
{Davidson}, K., {Ebbets}, D., {Weigelt}, G., {et~al.} 1995, \aj, 109, 1784

\bibitem[{{Davidson} \& {Humphreys}(1997)}]{dh97}
{Davidson}, K. \& {Humphreys}, R.~M. 1997, \araa, 35, 1

\bibitem[{{Davidson} {et~al.}(2005){Davidson}, {Martin}, {Humphreys},
  {Ishibashi}, {Gull}, {Stahl}, {Weis}, {Hillier}, {Damineli}, {Corcoran}, \&
  {Hamann}}]{davidson05}
{Davidson}, K., {Martin}, J., {Humphreys}, R.~M., {et~al.} 2005, \aj, 129, 900

\bibitem[{{Duncan} \& {White}(2003)}]{duncan03}
{Duncan}, R.~A. \& {White}, S.~M. 2003, \mnras, 338, 425

\bibitem[{{Falceta-Gon{\c c}alves} \& {Abraham}(2009)}]{diego09}
{Falceta-Gon{\c c}alves}, D. \& {Abraham}, Z. 2009, \mnras, 399, 1441

\bibitem[{{Falceta-Gon{\c c}alves} {et~al.}(2005){Falceta-Gon{\c c}alves},
  {Jatenco-Pereira}, \& {Abraham}}]{diego05}
{Falceta-Gon{\c c}alves}, D., {Jatenco-Pereira}, V., \& {Abraham}, Z. 2005,
  \mnras, 357, 895

\bibitem[{{Feast} {et~al.}(2001){Feast}, {Whitelock}, \& {Marang}}]{feast01}
{Feast}, M., {Whitelock}, P., \& {Marang}, F. 2001, \mnras, 322, 741

\bibitem[{{Fern{\'a}ndez-Laj{\'u}s} {et~al.}(2010){Fern{\'a}ndez-Laj{\'u}s},
  {Fari{\~n}a}, {Calder{\'o}n}, {Salerno}, {Torres}, {Schwartz}, {von Essen},
  {Giudici}, \& {Bareilles}}]{lajus10}
{Fern{\'a}ndez-Laj{\'u}s}, E., {Fari{\~n}a}, C., {Calder{\'o}n}, J.~P.,
  {et~al.} 2010, New Astronomy, 15, 108

\bibitem[{{Fern{\'a}ndez-Laj{\'u}s} {et~al.}(2009){Fern{\'a}ndez-Laj{\'u}s},
  {Fari{\~n}a}, {Torres}, {Schwartz}, {Salerno}, {Calder{\'o}n}, {von Essen},
  {Calcaferro}, {Giudici}, {Llinares}, \& {Niemela}}]{lajus09}
{Fern{\'a}ndez-Laj{\'u}s}, E., {Fari{\~n}a}, C., {Torres}, A.~F., {et~al.}
  2009, \aap, 493, 1093

\bibitem[{{Fern{\'a}ndez-Laj{\'u}s} {et~al.}(2003){Fern{\'a}ndez-Laj{\'u}s},
  {Gamen}, {Schwartz}, {Salerno}, {Llinares}, {Farina}, {Amor{\'{\i}}n}, \&
  {Niemela}}]{lajus03}
{Fern{\'a}ndez-Laj{\'u}s}, E., {Gamen}, R., {Schwartz}, M., {et~al.} 2003,
  Information Bulletin on Variable Stars, 5477, 1

\bibitem[{{Gaviola}(1953)}]{gaviola53}
{Gaviola}, E. 1953, \apj, 118, 234

\bibitem[{{Groh} \& {Damineli}(2004)}]{gd04}
{Groh}, J.~H. \& {Damineli}, A. 2004, Information Bulletin on Variable Stars,
  5492, 1

\bibitem[{{Groh} {et~al.}(2007){Groh}, {Damineli}, \& {Jablonski}}]{gdj07}
{Groh}, J.~H., {Damineli}, A., \& {Jablonski}, F. 2007, \aap, 465, 993

\bibitem[{{Gull} {et~al.}(2009){Gull}, {Nielsen}, {Corcoran}, {Madura},
  {Owocki}, {Russell}, {Hillier}, {Hamaguchi}, {Kober}, {Weis}, {Stahl}, \&
  {Okazaki}}]{gull09}
{Gull}, T.~R., {Nielsen}, K.~E., {Corcoran}, M.~F., {et~al.} 2009, \mnras, 396,
  1308

\bibitem[{{Hamaguchi} {et~al.}(2007){Hamaguchi}, {Corcoran}, {Gull},
  {Ishibashi}, {Pittard}, {Hillier}, {Damineli}, {Davidson}, {Nielsen}, \&
  {Kober}}]{hamaguchi07}
{Hamaguchi}, K., {Corcoran}, M.~F., {Gull}, T., {et~al.} 2007, \apj, 663, 522

\bibitem[{{Hartman} {et~al.}(2005){Hartman}, {Damineli}, {Johansson}, \&
  {Letokhov}}]{hartman05}
{Hartman}, H., {Damineli}, A., {Johansson}, S., \& {Letokhov}, V.~S. 2005,
  \aap, 436, 945

\bibitem[{{Henley} {et~al.}(2008){Henley}, {Corcoran}, {Pittard}, {Stevens},
  {Hamaguchi}, \& {Gull}}]{henley08}
{Henley}, D.~B., {Corcoran}, M.~F., {Pittard}, J.~M., {et~al.} 2008, \apj, 680,
  705

\bibitem[{{Hillier} \& {Allen}(1992)}]{hillier92}
{Hillier}, D.~J. \& {Allen}, D.~A. 1992, \aap, 262, 153

\bibitem[{{Hillier} {et~al.}(2001){Hillier}, {Davidson}, {Ishibashi}, \&
  {Gull}}]{hillier01}
{Hillier}, D.~J., {Davidson}, K., {Ishibashi}, K., \& {Gull}, T. 2001, \apj,
  553, 837

\bibitem[{{Hillier} {et~al.}(2006){Hillier}, {Gull}, {Nielsen}, {Sonneborn},
  {Iping}, {Smith}, {Corcoran}, {Damineli}, {Hamann}, {Martin}, \&
  {Weis}}]{hillier06}
{Hillier}, D.~J., {Gull}, T., {Nielsen}, K., {et~al.} 2006, \apj, 642, 1098

\bibitem[{{Hofmann} \& {Weigelt}(1988)}]{hofmann88}
{Hofmann}, K. \& {Weigelt}, G. 1988, \aap, 203, L21

\bibitem[{{Humphreys} {et~al.}(2008){Humphreys}, {Davidson}, \&
  {Koppelman}}]{humphreys08}
{Humphreys}, R.~M., {Davidson}, K., \& {Koppelman}, M. 2008, \aj, 135, 1249

\bibitem[{{Iping} {et~al.}(2005){Iping}, {Sonneborn}, {Gull}, {Massa}, \&
  {Hillier}}]{iping05}
{Iping}, R.~C., {Sonneborn}, G., {Gull}, T.~R., {Massa}, D.~L., \& {Hillier},
  D.~J. 2005, \apjl, 633, L37

\bibitem[{{Ishibashi} {et~al.}(1999){Ishibashi}, {Corcoran}, {Davidson},
  {Swank}, {Petre}, {Drake}, {Damineli}, \& {White}}]{ishibashi99}
{Ishibashi}, K., {Corcoran}, M.~F., {Davidson}, K., {et~al.} 1999, \apj, 524,
  983

\bibitem[{{Kaeufl} {et~al.}(2004){Kaeufl}, {Ballester}, {Biereichel},
  {Delabre}, {Donaldson}, {Dorn}, {Fedrigo}, {Finger}, {Fischer}, {Franza},
  {Gojak}, {Huster}, {Jung}, {Lizon}, {Mehrgan}, {Meyer}, {Moorwood}, {Pirard},
  {Paufique}, {Pozna}, {Siebenmorgen}, {Silber}, {Stegmeier}, \&
  {Wegerer}}]{kaufl04}
{Kaeufl}, H.-U., {Ballester}, P., {Biereichel}, P., {et~al.} 2004, in Proc. of
  the SPIE, ed. A.~F.~M. {Moorwood} \& M.~{Iye}, Vol. 5492, 1218

\bibitem[{{Kashi} \& {Soker}(2007)}]{kashi07b}
{Kashi}, A. \& {Soker}, N. 2007, New Astronomy, 12, 590

\bibitem[{{Kashi} \& {Soker}(2008{\natexlab{a}})}]{kashi08b}
{Kashi}, A. \& {Soker}, N. 2008{\natexlab{a}}, New Astronomy, 13, 569

\bibitem[{{Kashi} \& {Soker}(2008{\natexlab{b}})}]{kashi08a}
{Kashi}, A. \& {Soker}, N. 2008{\natexlab{b}}, \mnras, 390, 1751

\bibitem[{{Kashi} \& {Soker}(2009)}]{kashi09b}
{Kashi}, A. \& {Soker}, N. 2009, \mnras, 394, 923

\bibitem[{{Kerber} {et~al.}(2008){Kerber}, {Nave}, \& {Sansonetti}}]{kerber08}
{Kerber}, F., {Nave}, G., \& {Sansonetti}, C.~J. 2008, \apjs, 178, 374

\bibitem[{{Mehner} {et~al.}(2010){Mehner}, {Davidson}, {Ferland}, \&
  {Humphreys}}]{mehner10}
{Mehner}, A., {Davidson}, K., {Ferland}, G.~J., \& {Humphreys}, R.~M. 2010,
  \apj, 710, 729

\bibitem[{{Nielsen} {et~al.}(2007){Nielsen}, {Corcoran}, {Gull}, {Hillier},
  {Hamaguchi}, {Ivarsson}, \& {Lindler}}]{nielsen07}
{Nielsen}, K.~E., {Corcoran}, M.~F., {Gull}, T.~R., {et~al.} 2007, \apj, 660,
  669

\bibitem[{{Nielsen} {et~al.}(2005){Nielsen}, {Gull}, \& {Vieira
  Kober}}]{nielsen05}
{Nielsen}, K.~E., {Gull}, T.~R., \& {Vieira Kober}, G. 2005, \apjs, 157, 138

\bibitem[{{Okazaki} {et~al.}(2008){Okazaki}, {Owocki}, {Russell}, \&
  {Corcoran}}]{okazaki08}
{Okazaki}, A.~T., {Owocki}, S.~P., {Russell}, C.~M.~P., \& {Corcoran}, M.~F.
  2008, \mnras, 388, L39

\bibitem[{{Parkin} {et~al.}(2009){Parkin}, {Pittard}, {Corcoran}, {Hamaguchi},
  \& {Stevens}}]{parkin09}
{Parkin}, E.~R., {Pittard}, J.~M., {Corcoran}, M.~F., {Hamaguchi}, K., \&
  {Stevens}, I.~R. 2009, \mnras, 394, 1758

\bibitem[{{Pittard} \& {Corcoran}(2002)}]{pc02}
{Pittard}, J.~M. \& {Corcoran}, M.~F. 2002, \aap, 383, 636

\bibitem[{{Pittard} {et~al.}(1998){Pittard}, {Stevens}, {Corcoran}, \&
  {Ishibashi}}]{pittard98}
{Pittard}, J.~M., {Stevens}, I.~R., {Corcoran}, M.~F., \& {Ishibashi}, K. 1998,
  \mnras, 299, L5

\bibitem[{{Smith}(2002)}]{smith02}
{Smith}, N. 2002, \mnras, 337, 1252

\bibitem[{{Smith}(2004)}]{smith04_vel}
{Smith}, N. 2004, \mnras, 351, L15

\bibitem[{{Smith}(2006)}]{smith06}
{Smith}, N. 2006, \apj, 644, 1151

\bibitem[{{Smith}(2008)}]{smith08b}
{Smith}, N. 2008, \nat, 455, 201

\bibitem[{{Smith}(2010)}]{smith10}
{Smith}, N. 2010, \mnras, 402, 145

\bibitem[{{Smith} {et~al.}(2003{\natexlab{a}}){Smith}, {Davidson}, {Gull},
  {Ishibashi}, \& {Hillier}}]{smith03}
{Smith}, N., {Davidson}, K., {Gull}, T.~R., {Ishibashi}, K., \& {Hillier},
  D.~J. 2003{\natexlab{a}}, \apj, 586, 432

\bibitem[{{Smith} {et~al.}(2003{\natexlab{b}}){Smith}, {Gehrz}, {Hinz},
  {Hoffmann}, {Hora}, {Mamajek}, \& {Meyer}}]{smith03b}
{Smith}, N., {Gehrz}, R.~D., {Hinz}, P.~M., {et~al.} 2003{\natexlab{b}}, \aj,
  125, 1458

\bibitem[{{Smith} {et~al.}(2004){Smith}, {Morse}, {Collins}, \&
  {Gull}}]{smith04}
{Smith}, N., {Morse}, J.~A., {Collins}, N.~R., \& {Gull}, T.~R. 2004, \apjl,
  610, L105

\bibitem[{{Stahl} {et~al.}(2005){Stahl}, {Weis}, {Bomans}, {Davidson}, {Gull},
  \& {Humphreys}}]{stahl05}
{Stahl}, O., {Weis}, K., {Bomans}, D.~J., {et~al.} 2005, \aap, 435, 303

\bibitem[{{Steiner} \& {Damineli}(2004)}]{sd04}
{Steiner}, J.~E. \& {Damineli}, A. 2004, \apjl, 612, L133

\bibitem[{{Teodoro} {et~al.}(2008){Teodoro}, {Damineli}, {Sharp}, {Groh}, \&
  {Barbosa}}]{teodoro08}
{Teodoro}, M., {Damineli}, A., {Sharp}, R.~G., {Groh}, J.~H., \& {Barbosa},
  C.~L. 2008, \mnras, 387, 564

\bibitem[{{Thackeray}(1953)}]{thackeray53}
{Thackeray}, A.~D. 1953, \mnras, 113, 211

\bibitem[{{van Boekel} {et~al.}(2003){van Boekel}, {Kervella}, {Sch{\"o}ller},
  {Herbst}, {Brandner}, {de Koter}, {Waters}, {Hillier}, {Paresce}, {Lenzen},
  \& {Lagrange}}]{vb03}
{van Boekel}, R., {Kervella}, P., {Sch{\"o}ller}, M., {et~al.} 2003, \aap, 410,
  L37

\bibitem[{{van Genderen} {et~al.}(2003){van Genderen}, {Sterken}, {Allen}, \&
  {Liller}}]{vg03}
{van Genderen}, A.~M., {Sterken}, C., {Allen}, W.~H., \& {Liller}, W. 2003,
  \aap, 412, L25

\bibitem[{{van Genderen} {et~al.}(2006){van Genderen}, {Sterken}, {Allen}, \&
  {Walker}}]{vg06}
{van Genderen}, A.~M., {Sterken}, C., {Allen}, W.~H., \& {Walker}, W.~S.~G.
  2006, Journal of Astronomical Data, 12, 3

\bibitem[{{Verner} {et~al.}(2005){Verner}, {Bruhweiler}, \& {Gull}}]{verner05}
{Verner}, E., {Bruhweiler}, F., \& {Gull}, T. 2005, \apj, 624, 973

\bibitem[{{Viotti} {et~al.}(1989){Viotti}, {Rossi}, {Cassatella}, {Altamore},
  \& {Baratta}}]{viotti89}
{Viotti}, R., {Rossi}, L., {Cassatella}, A., {Altamore}, A., \& {Baratta},
  G.~B. 1989, \apjs, 71, 983

\bibitem[{{Walborn}(1973)}]{walborn73}
{Walborn}, N.~R. 1973, \apj, 179, 517

\bibitem[{{Weigelt} \& {Ebersberger}(1986)}]{weigelt86}
{Weigelt}, G. \& {Ebersberger}, J. 1986, \aap, 163, L5

\bibitem[{{Weigelt} {et~al.}(2007){Weigelt}, {Kraus}, {Driebe}, {Petrov},
  {Hofmann}, {Millour}, {Chesneau}, {Schertl}, {Malbet}, {Hillier}, {Gull},
  {Davidson}, {Domiciano de Souza}, {Antonelli}, {Beckmann}, {Bresson},
  {Chelli}, {Dugu{\'e}}, {Duvert}, {Gennari}, {Gl{\"u}ck}, {Kern}, {Lagarde},
  {Le Coarer}, {Lisi}, {Perraut}, {Puget}, {Rantakyr{\"o}}, {Robbe-Dubois},
  {Roussel}, {Tatulli}, {Zins}, {Accardo}, {Acke}, {Agabi}, {Altariba},
  {Arezki}, {Aristidi}, {Baffa}, {Behrend}, {Bl{\"o}cker}, {Bonhomme},
  {Busoni}, {Cassaing}, {Clausse}, {Colin}, {Connot}, {Delboulb{\'e}},
  {Feautrier}, {Ferruzzi}, {Forveille}, {Fossat}, {Foy}, {Fraix-Burnet},
  {Gallardo}, {Giani}, {Gil}, {Glentzlin}, {Heiden}, {Heininger}, {Hernandez
  Utrera}, {Kamm}, {Kiekebusch}, {Le Contel}, {Le Contel}, {Lesourd}, {Lopez},
  {Lopez}, {Magnard}, {Marconi}, {Mars}, {Martinot-Lagarde}, {Mathias},
  {M{\`e}ge}, {Monin}, {Mouillet}, {Mourard}, {Nussbaum}, {Ohnaka}, {Pacheco},
  {Perrier}, {Rabbia}, {Rebattu}, {Reynaud}, {Richichi}, {Robini},
  {Sacchettini}, {Sch{\"o}ller}, {Solscheid}, {Spang}, {Stee}, {Stefanini},
  {Tallon}, {Tallon-Bosc}, {Tasso}, {Testi}, {Vakili}, {von der L{\"u}he},
  {Valtier}, {Vannier}, {Ventura}, {Weis}, \& {Wittkowski}}]{weigelt07}
{Weigelt}, G., {Kraus}, S., {Driebe}, T., {et~al.} 2007, \aap, 464, 87

\bibitem[{{Weis} {et~al.}(2005){Weis}, {Stahl}, {Bomans}, {Davidson}, {Gull},
  \& {Humphreys}}]{weis05}
{Weis}, K., {Stahl}, O., {Bomans}, D.~J., {et~al.} 2005, \aj, 129, 1694

\bibitem[{{Whitelock} {et~al.}(1983){Whitelock}, {Carter}, {Roberts},
  {Whittet}, \& {Baines}}]{whitelock83}
{Whitelock}, P.~A., {Carter}, B.~S., {Roberts}, G., {Whittet}, D.~C.~B., \&
  {Baines}, D.~W.~T. 1983, \mnras, 205, 577

\bibitem[{{Whitelock} {et~al.}(2004){Whitelock}, {Feast}, {Marang}, \&
  {Breedt}}]{whitelock04}
{Whitelock}, P.~A., {Feast}, M.~W., {Marang}, F., \& {Breedt}, E. 2004, \mnras,
  352, 447

\bibitem[{{Zanella} {et~al.}(1984){Zanella}, {Wolf}, \& {Stahl}}]{zanella84}
{Zanella}, R., {Wolf}, B., \& {Stahl}, O. 1984, \aap, 137, 79

\bibitem[{{Zethson}(2001)}]{zethson01}
{Zethson}, T. 2001, PhD thesis, Lunds Universitet (Sweden)

\end{thebibliography}
\end{document}